\documentclass[10pt,english,11pt,aps,letter,superscriptaddress,nofootinbib,pre]{revtex4}
\usepackage[T1]{fontenc}
\usepackage[latin9]{inputenc}
\usepackage{color}
\usepackage{babel}
\usepackage{verbatim}
\usepackage{amsmath}
\usepackage{amssymb}
\usepackage{graphicx}
\usepackage{esint}
\usepackage[unicode=true,pdfusetitle,
 bookmarks=true,bookmarksnumbered=false,bookmarksopen=false,
 breaklinks=false,pdfborder={0 0 1},backref=false,colorlinks=false]
 {hyperref}

\makeatletter

\providecommand{\tabularnewline}{\\}

\@ifundefined{textcolor}{}
{%
 \definecolor{BLACK}{gray}{0}
 \definecolor{WHITE}{gray}{1}
 \definecolor{RED}{rgb}{1,0,0}
 \definecolor{GREEN}{rgb}{0,1,0}
 \definecolor{BLUE}{rgb}{0,0,1}
 \definecolor{CYAN}{cmyk}{1,0,0,0}
 \definecolor{MAGENTA}{cmyk}{0,1,0,0}
 \definecolor{YELLOW}{cmyk}{0,0,1,0}
}

\usepackage{ae,aecompl}

\newcommand{\sfrac}[2]{\mathchoice
  {\kern0em\raise.5ex\hbox{\the\scriptfont0 #1}\kern-.15em/
   \kern-.15em\lower.25ex\hbox{\the\scriptfont0 #2}}
  {\kern0em\raise.5ex\hbox{\the\scriptfont0 #1}\kern-.15em/
   \kern-.15em\lower.25ex\hbox{\the\scriptfont0 #2}}
  {\kern0em\raise.5ex\hbox{\the\scriptscriptfont0 #1}\kern-.2em/
   \kern-.15em\lower.25ex\hbox{\the\scriptscriptfont0 #2}}
  {#1\!/#2}}

\makeatletter
\DeclareMathSizes{\@xipt}{10}{6}{5}
\makeatother

\makeatother

\begin{document}

\title{Inertial Coupling Method for particles in an incompressible fluctuating
fluid}

\author{Florencio Balboa Usabiaga}

\affiliation{Departamento de Física Teórica de la Materia Condensada, Condensed
Matter Physics Center (IFIMAC), Univeridad Autónoma de Madrid, Madrid
28049, Spain}

\author{Rafael Delgado-Buscalioni}

\affiliation{Departamento de Física Teórica de la Materia Condensada, Condensed
Matter Physics Center (IFIMAC), Univeridad Autónoma de Madrid, Madrid
28049, Spain}

\author{Boyce E. Griffith}

\affiliation{Leon H. Charney Division of Cardiology, Department of Medicine, New
York University School of Medicine, New York, NY 10016}

\affiliation{Courant Institute of Mathematical Sciences, New York University,
New York, NY 10012}

\author{Aleksandar Donev}

\email{donev@courant.nyu.edu}

\affiliation{Courant Institute of Mathematical Sciences, New York University,
New York, NY 10012}
\begin{abstract}
We develop an inertial coupling method for modeling the dynamics of
point-like ``blob'' particles immersed in an incompressible fluid,
generalizing previous work for compressible fluids {[}\emph{F. Balboa
Usabiaga, I. Pagonabarraga, and R. Delgado-Buscalioni, J. Comp. Phys.,
235:701-722, 2013}{]}. The coupling consistently includes excess (positive
or negative) inertia of the particles relative to the displaced fluid,
and accounts for thermal fluctuations in the fluid momentum equation.
The coupling between the fluid and the blob is based on a no-slip
constraint equating the particle velocity with the local average of
the fluid velocity, and conserves momentum and energy. We demonstrate
that the formulation obeys a fluctuation-dissipation balance, owing
to the non-dissipative nature of the no-slip coupling. We develop
a spatio-temporal discretization that preserves, as best as possible,
these properties of the continuum formulation. In the spatial discretization,
the local averaging and spreading operations are accomplished using
compact kernels commonly used in immersed boundary methods. We find
that the special properties of these kernels allow the blob to provide
an effective model of a particle; specifically, the volume, mass,
and hydrodynamic properties of the blob are remarkably grid-independent.
We develop a second-order semi-implicit temporal integrator that maintains
discrete fluctuation-dissipation balance, and is not limited in stability
by viscosity. Furthermore, the temporal scheme requires only constant-coefficient
Poisson and Helmholtz linear solvers, enabling a very efficient and
simple FFT-based implementation on GPUs. We numerically investigate
the performance of the method on several standard test problems. In
the deterministic setting, we find the blob to be a remarkably robust
approximation to a rigid sphere, at both low and high Reynolds numbers.
In the stochastic setting, we study in detail the short and long-time
behavior of the velocity autocorrelation function and observe agreement
with all of the known behavior for rigid sphere immersed in a fluctuating
fluid. The proposed inertial coupling method provides a low-cost coarse-grained
(minimal resolution) model of particulate flows over a wide range
of time-scales ranging from Brownian to convection-driven motion.
\end{abstract}
\maketitle

\global\long\def\V#1{\boldsymbol{#1}}
\global\long\def\M#1{\boldsymbol{#1}}
\global\long\def\Set#1{\mathbb{#1}}

\global\long\def\D#1{\Delta#1}
\global\long\def\d#1{\delta#1}

\global\long\def\norm#1{\left\Vert #1\right\Vert }
\global\long\def\abs#1{\left|#1\right|}

\global\long\def\grad{\M{\nabla}}
\global\long\def\avv#1{\langle#1\rangle}
\global\long\def\av#1{\left\langle #1\right\rangle }

\global\long\def\P{\mathcal{P}}

\global\long\def\ki{k}
\global\long\def\wi{\omega}

\global\long\def\pRe{\mathrm{Re}_{P}}
 \global\long\def\fRe{\mathrm{Re}_{F}}
 \global\long\def\bu{{\bf u}}
 \global\long\def\bv{{\bf v}}
 \global\long\def\br{{\bf r}}

\setlength{\abovedisplayskip}{0.35ex}\setlength{\belowdisplayskip}{0.35ex}

\section{Introduction}

The dynamics of small particles immersed in a fluid is key to many
applications involving disparate length and time scales \cite{KimKarrila}:
from the dynamics of millimeter particles (dust) in turbulent flow,
to multiphase flow with micron and nanoscopic colloidal molecules
in quiescent, laminar \cite{SRD_Review,LB_SoftMatter_Review,MigrationFractination},
or turbulent regimes \cite{Eaton2009,Tanaka2010}. In many engineering
applications colloidal particles are exposed to disparate dynamic
regimes coexisting in different subdomains of the same chamber \cite{Rhodes_book}.
Such processes demand fast computational methods able to efficiently
resolve the motion of many ($O\left(10^{5}\right)$) colloidal particles
in quite different dynamics ranging from diffusive to inertial dynamics.
Such scenarios are paradigmatic of what one might call \emph{multi-regime}
systems.

A group of methods such as smooth particle hydrodynamics (SPH) \cite{Bian2012},
smoothed dissipative particle dynamics \cite{SDPD_Scaling}, and stochastic
rotation dynamics (SRD) \cite{SRD_Review} resolve both the particle
and fluid phase using similar discrete Lagrangian descriptions, and
as such, seem to be natural candidates to become multi-regime solvers
\cite{Horbach2006,Bian2012}. Particle-particle methods allow for
an easy treatment of complex boundary conditions, and offer a natural
way to couple moving boundaries or immersed particles to the fluid.
However, particle-particle methods have important drawbacks when compared
with standard solvers for discretized Computational Fluid Dynamics
(CFD). In particular, they offer limited control over the fluid properties
and require relatively small time steps compared to, for example,
semi-implicit CFD schemes. Moreover, they cannot be adapted to efficiently
treat the natural time scales governing the different dynamical regimes
(e.g., the Brownian or overdamped limit). Similar advantages and drawbacks
also apply to the lattice Boltzmann (LB) method \cite{LB_SoftMatter_Review},
although the LB approach has proven to be a rather flexible framework
\textbf{\cite{Melchionna2010}}.

Many other approaches use CFD for the solvent flow and couple its
dynamics with that of the immersed particles. In the realm of CFD
one can still distinguish two large subgroups of methods. The first
group of methods involves a Lagrangian description of the computational
mesh which self-adapts to follow the particle \cite{Hu2001}. The
second group uses a fixed (Eulerian) grid and requires converting
the particle boundary conditions into body forces or some interaction
equations \cite{Yamamoto2007,Luo2009,Melchionna2010}. The present
work focuses on this second group, sometimes called mixed Eulerian-Lagrangian
methods. These schemes are particularly suited to attack the ``multi-regime''
problem, because they are faster, more flexible and can work with
minimal resolution models (pointwise particle descriptions).

In their seminal work, Maxey and Riley \cite{Maxey1983} decomposed
the fluid velocity as $\bv(\br,t)=\bv_{0}(\br,t)+\bv_{1}(\br,t)$,
where $\bv_{0}$ is the undisturbed flow (which would result if the
boundary conditions at the particle surface were not applied), and
$\bv_{1}$ is the perturbative component created by the fluid-particle
interaction. In the bulk flow, convection (advection) becomes relevant
for $\mathrm{Re}_{F}=v_{0}L\rho/\eta>1$; where the \emph{fluid} Reynolds
number $\fRe$ is defined in terms of the typical flow speed $v_{0}$,
the fluid density $\rho$, the dynamic viscosity $\eta=\rho\nu$,
and a characteristic length $L$ for velocity variation in the flow.
The fluid force on the particle arises from the local fluid inertia
(proportional to the local material derivative of $\bv_{0}$) and
also from the local stress created by the particle disturbance. The
\emph{relaxational} part of the particle inertia is a consequence
of its mass resistance to instantaneously follow the velocity of the
surrounding fluid; the fluid drag damps the particle velocity towards
the local fluid velocity within an inertial time $\tau_{P}\sim(\rho_{P}-\rho)R^{2}/\eta$
which increases with the density contrast $\rho_{P}-\rho$ and with
the particle radius $R$.

By contrast, \emph{convective} inertia arises from non-linear interactions
between the particle dynamics and its perturbative flow \cite{Lomholt2001}.
The \emph{particle} Reynolds number $\mathrm{Re}_{P}=2wR/\nu$, defined
with the particle-fluid relative speed $w$ \cite{Maxey1983}, determines
the relative strength of advection by the perturbative flow relative
to viscous dissipation. The importance of convective inertia is indicated
by the ratio $\fRe(R/L)^{2}$ between the characteristic times associated
with Stokes drag and convection \cite{Maxey1983,Lomholt2001}. At
finite values of the non-dimensional groups $\mathrm{Re}_{P}$ and
$\fRe(R/L)^{2}$ inertia effects due to particle mass and particle
size are not similar anymore, especially in the turbulent regime,
where non-linear interactions between the mean flow and the particle
perturbative field become relevant \cite{Monchaux2012,Eaton2009}.
Non-linear interaction between particle advection and thermal fluctuations
are also possible at small Reynolds number. Some examples are the
change in the mobility of colloidal particles ($R\sim10^{-[5-8]}\,\mbox{m}$)
over the Stokes limit at low values of the Schmidt number \cite{Falck2004,Ripoll2005},
and inertial effects in directional locking, a process to separate
nanoparticles at very small $\pRe$ \cite{Balbin2009}.

Computational approaches are usually tailored to tackle some specific
dynamical regime and they can be naturally classified according to
the range of $\mathrm{Re}_{F}$, $\mathrm{Re}_{P}$ and $R/L$ they
can be safely applied to. In the creeping flow limit, $\fRe\rightarrow0$
and $\pRe\rightarrow0$, the perturbative flow $\bv_{1}$ has a negligible
effect on the unperturbed field, which is a priori fixed. The perturbative
field created by a collection of particles is the linear superposition
of the Stokes fields and it determines the multi-body hydrodynamic
forces on the particle ensemble. Analytical expressions for these
forces are embedded in the mobility matrix of Brownian hydrodynamics
(BD) \cite{Ermak1978,BrownianDynamics_DNA} and Stokesian dynamics
(SD) \cite{Banchio2003}. In addition to the stokeslet (monopole)
terms, in SD one can include higher terms of the multipole expansion
of the perturbative stress \cite{KimKarrila}. The zero-Reynolds regime
resolves the long-time diffusive (Smoluchowski) limit of colloidal
motion, in which fluctuations make an important ($O(1)$) contribution.
Direct implementation of the fluctuation dissipation (FD) relation
between the friction and noise matrices requires O$(N^{3})$ operations,
where $N$ is the number of particles. Sophisticated and technically-complex
techniques such as the accelerated Stokesian dynamics \cite{BrownianDynamics_OrderNlogN},
and the general geometry Ewald-like method for confined geometries
\cite{BrownianDynamics_OrderN}, reduce the large raw cost to $O(N\,\ln N)$
operations, albeit with large multiplicative prefactors.

As an alternative to BD and SD methods, two-way coupling algorithms
using a Stokes frictional force were developed for mixed Eulerian-Lagrangian
dynamics \cite{LB_SoftMatter_Review,FluctuatingHydroMD_Coveney,SELM}.
The idea is to deploy a relative simple and efficient fluid solver
to explicitly resolve the perturbative flow responsible for the hydrodynamic
coupling between particles. The particle cost is dominated by neighbour
searching and scales (almost) linearly with $N$, while the (added)
fluid solver cost scales like the system volume. The Eulerian-Lagrangian
mixed approach permits to work at finite $\fRe$. However, the Stokes
(i.e. frictional) coupling assumption limits the scheme to $\pRe<1$
and only resolves far-field hydrodynamics ($R/L<1$). The Stokes coupling
consistently neglects convective inertia and only includes relaxational
particle inertia in an approximate way, with a finite particle response
time $\tau_{P}$ introduced by a \emph{phenomenological} friction
coefficient. Frictional coupling is obviously dissipative and requires
introducing an additional noise term in the particle equation, different
from that of the fluctuating fluid stress tensor \cite{LB_SoftMatter_Review,SELM}.

Other methods for finite $\fRe$ have been restricted to $\pRe=0$,
where the particle inertia is absent. Two relevant examples are the
stochastic Immersed Boundary(IB) method \cite{StochasticImmersedBoundary,SELM}
commonly used for fluid-structure interaction at $R/L=O(1)$, and
the Force Coupling method (FCM) \cite{ForceCoupling_Stokes,ForceCoupling_Monopole,Lomholt2001},
where each particle is represented by a low-order expansion of force
multipoles ($R/L<1$). Very recently, subsequent to the competition
and submission of this paper, an extension of the FCM that includes
fluctuations in the overdamped or inertia-less limit has been developed
\cite{ForceCoupling_Fluctuations}. For $\pRe=0$ the relative fluid-particle
acceleration is zero and the particle velocity just follows the local
fluid velocity. The hydrodynamic force due to the particle-fluid interaction
is then equal to the total force exerted on the particle by sources
other than the fluid. This permits a fluid-only formulation whereby
the net non-hydrodynamic particle force is \emph{spread} from the
particle to the surrounding fluid using some compact kernel. This
important spreading operation differs substantially from method to
method. In FCM two different Gaussian kernels are used to spread the
force monopole and force dipole moments (stresslet); their widths
are fitted in the \emph{continuum} model to recover the Stokes drag
and linear Faxen terms \cite{ForceCoupling_Stokes}. By contrast,
the IBM kernels are specifically designed to minimize the effects
of the \emph{discrete} Eulerian mesh in the spreading of Lagrangian
point forces (monopole terms) \cite{IBM_PeskinReview}.

For $\pRe\ll1$ and $R/L\ll1$ advection of the perturbative flow
can be neglected leading to the (analytically solvable) unsteady Stokes
equation for the perturbative field \cite{Maxey1983}. The fluid-particle
force can be expressed as some function of the relative velocity field
$\bu-\bv_{0}$ \emph{interpolated} at the particle site. This forms
the basis of \emph{one-way-coupling} schemes for point-particle dynamics
frequently used in turbulence research, $\fRe\gg1$ \cite{Lomholt2001}.
Generalizations to $\pRe\sim1$ have been also derived (see e.g. \cite{Lovalenti1993})
but, even in the simpler $\pRe\ll1$ limit, the evaluation of the
fluid-particle force involves cumbersome expressions which require
interpolations of the displaced fluid acceleration (the added mass
effect), second order spatial derivatives of $\V v_{0}$ (Faxen terms),
and time-convolved integrals which recast the history of vorticity
diffusion around the particle (Basset memory). For a sphere moving
with velocity $\bu$ at $\pRe\ll1$, the leading term is the steady
Stokes force $F_{\text{Stokes}}=6\pi\eta R(\bu-\bv_{0})$, which,
due to its simple form, has been overused in \emph{two-way} point-particle
approximations of turbulent ($\fRe\gg1$) and pre-turbulent regimes
\cite{Monchaux2012,Eaton2009}. Although the point-particle approach
can probably describe the relaxational inertia of very small ($R/L\ll1$)
heavy particles in a light fluid (e.g. aerosol), it has the serious
limitation of neglecting the convective inertia arising from the particle
finite size \cite{Botto2012}. Even at low $\pRe$, convection of
perturbative flow is known to alter the Basset memory and the long
time particle dynamics \cite{Lomholt2001}. And vice versa, recent
works show that micron-size particles can alter the turbulent spectra
at moderate $\pRe$ and $R\sim L$ (with $L$ the Kolmogorov length)
\cite{Tanaka2010,Eaton2009} due to energy dissipation and vorticity
production in the particles wake \cite{Botto2012}.

Several Eulerian-Lagrangian methods have appeared in recent years
to allow for a fully consistent treatment of the coupled particle
and fluid inertia. A key issue is the spatial resolution of the particle.
In the ``direct forcing'' method \cite{Uhlmann2008}, and related
extensions to fluctuating hydrodynamics \cite{FluctuatingHydro_FluidOnly,FIMAT_Patankar,Melchionna2010,ImmersedFEM_Patankar},
the fluid force on the particle is obtained by imposing the no-slip
constraint on a well-resolved particle surface, and in some cases,
to ensure rigid body motion, also in the particle interior \cite{ParticleLaden_Proteus}.
High spatial resolution requires substantial computational effort;
the largest simulations so far reached $O(10^{3})$ particles \cite{Lucci2010,Uhlmann2008,ParticleLaden_Heat}.
The smoothed particle method (SPM) \cite{Yamamoto2007,Luo2009} works
with a mixed (particle-fluid) velocity field constructed with a smooth
characteristic function which discriminates particle and fluid cells.
This permits an intermediate resolution with a typical particle radius
$R\simeq5h$ (here $h$ is the mesh size) requiring $O(10^{3})$ fluid
cells per particle. These fully or partially resolved methods are
quite far from a point-particle approach, which can require as few
as $13$ cells to perform a fourth order orthogonal Lagrangian interpolation
\cite{Dejoan2011}. ``Blob'' particle descriptions, more appropriately
termed \emph{minimally-resolved models}, offer a way to explore finite
particle effects at moderate computational cost. In this work each
particle (blob) is described by an unique kernel of small support
(27 cells) which is shown to provide a consistent set of particle
physical properties (volume, mass, hydrodynamic radius). It has to
be noted that due to the long-ranged nature of hydrodynamic interactions,
sufficiently large boxes are required to reduce (to an acceptable
extent) the finite size effects; hence a reduction in the linear size
of the particle description means a large (cubic) reduction in the
overall fluid solver cost.

In a preceding paper \cite{DirectForcing_Balboa}, some of us proposed
an inertial coupling method that directly couples a compressible finite-volume
fluctuating hydrodynamic solver \cite{LLNS_Staggered} to blob particles.
A distinguishing feature of the coupling methodology is that it includes
the effect of the particle and fluid inertia in the dynamics, while
still consistently including thermal fluctuations even in non-trivial
geometries. It was numerically demonstrated that the inertial coupling
method can reproduce ultrasound forces on colloidal particles, taking
place at much faster rates than viscous friction \cite{Wang2011}.
In this previous work \cite{DirectForcing_Balboa}, a compressible
solver was used because one of the focus applications was the interaction
between ultrasound and colloidal particles \cite{Wang2011}. In many
applications sonic effects can be ignored and the essential hydrodynamic
interactions can be captured by using the \emph{isothermal incompressible}
Navier-Stokes equations instead of the compressible equations. This
eliminates the fast sound waves and allows for a much larger time
step size in the fluid solver. Here we develop an \emph{inertial coupling
method} (ICM) that directly couples an incompressible finite-volume
solver for the fluctuating Navier-Stokes equations \cite{LLNS_Staggered}
with suspended particles, which do not necessarily have the same density
as the fluid. We demonstrate that the coupling obeys a continuum and
a discrete fluctuation-dissipation balance and study the performance
of our algorithm. The ICM is a coarse-grained model for particle hydrodynamics
which aims to capture hydrodynamic effects (unsteady forcing, viscous
friction and advection) over a broad range of time scales and $\mathrm{Re}_{P}$:
from Brownian motion to convection-driven regimes. In the ICM, the
coupling between the particle and the fluid is not assumed to have
any functional form (e.g. Stokes drag) but \emph{naturally} arises
from the no-slip constraint averaged over the particle (or ``blob'')
domain. The present results (see also Ref. \cite{DirectForcing_Balboa})
indicate that this type of (non-linear) coupling permits to take into
account both fluid \emph{and} particle inertia beyond the Stokes limit,
where advective interactions take place.

It is useful to point out the main similarities and differences between
our work and closely related work by others. First, at the level of
the formulation, the idea of using a minimally-resolved description
of particles immersed in a fluid has a long history in both deterministic
and fluctuating hydrodynamics. In the deterministic setting, Maxey
and collaborators have explored in extensive detail the coupling between
a minimally-resolved spherical particle and fluid flow within the
context of the FCM \cite{ForceCoupling_Stokes,ForceCoupling_Monopole,Lomholt2001}.
In the context of fluctuating hydrodynamics, a blob description of
particles has been used frequently to couple a Lattice-Boltzmann (LB)
fluid solver to immersed particles \cite{LB_SoftMatter_Review,FluctuatingHydroMD_Coveney}.
There are two distinguishing features of our method from that described
in extensive detail in an excellent survey by Dünweg and Ladd \cite{LB_SoftMatter_Review}.
The first difference, inherited from our previous work \cite{DirectForcing_Balboa},
is the fact that we employ a \emph{direct} frictionless (conservative)
coupling that instantaneously enforces a no-slip constraint. This
eliminates an artificial frictional time scale from the problem and
allows us to obtain physically-accurate short-time dynamics, as we
demonstrate in detail in Section \ref{sec:Results} by examining the
velocity autocorrelation function of the blobs for very short times.
In principle, a more direct coupling could be done in the formulation
of Dünweg and Ladd by taking a very large frictional constant, as
explained in more detail by Atzberger \cite{SELM}. However, the resulting
dynamics are stiff and doing this numerically requires using small
time steps sizes. The second important distinguishing feature of the
work presented here is the use of a semi-implicit incompressible fluid
solver instead of the compressible explicit LB solver, as also done
in the stochastic IB method \cite{StochasticImmersedBoundary,SELM}.
This allows us to maintain stability at much larger time step size
than used in typical LB simulations, and, more importantly, allows
us to approach the Brownian or Stokesian (overdamped) dynamics limit
without any \emph{uncontrolled} approximations. A final distinguishing
feature of our work from all work we are aware of is the consistent
inclusion of inertial effects in both the formulation and the numerical
algorithm. Specifically, inertial effects (excess particle mass) are
not included in the FCM or the stochastic IB method. Recently, the
very strong coupling limit of the frictional formulation described
in Refs. \cite{LB_SoftMatter_Review,SELM} has been considered theoretically
by Tabak and Atzberger \cite{SELM_Reduction}. The equations obtained
by them are identical to the ones we derive here based on physical
guiding principles. Finally, we point out that our numerical algorithm,
which includes both the spatial and a \emph{second-order semi-implicit}
temporal discretizations of the fluid-particle equations, is distinct
from any other work we are aware of. The algorithm is a generalization
of the first order coupling introduced by the direct forcing method
\cite{DirectForcing_Uhlmann,DirectForcing_Balboa,ParticleLaden_Proteus}.
Our scheme achieves second-order accuracy, avoids pressure-velocity
splitting \cite{bellColellaGlaz:1989}, as required for low Reynolds
number flows, and can be extended to other immersed-boundary fluid-structure
coupling methods as well.

It is important to point out that in this work we do not consider
thermal (heat) transfer between the particles and the fluid and instead
use an isothermal description, as most relevant to microfluidic applications.
Heat transfer is important in many large-scale particulate flows,
and has been included in Direct Numerical Simulation (DNS) algorithms
\cite{ParticleLaden_Heat}. It is possible to include additional transport
processes in the minimally-resolved approach we employ here, however,
the specifics of how to express the surface boundary conditions to
a volumetric blob condition are very problem specific and need to
be carefully constructed on a case by case basis. In this respect,
recently, reaction-diffusion processes have been included in the type
of method studied here \cite{ReactiveBlobs}, and compressible blobs
have been considered and found to adequately describe coupling between
ultrasound waves and small particles \cite{CompressibleBlobs}.

In the remainder of this Introduction we introduce some notation and
fundamental concepts. In Section \ref{sec:Continuum} we discuss the
continuum equations of the incompressible inertial coupling method.
We present both a constrained and a constraint-free formulation, and
demonstrate momentum and energy conservation, as well as fluctuation-dissipation
balance. In Section \ref{sec:Discretization} we present a second-order
semi-implicit spatio-temporal discretization of the continuum equations,
and demonstrate second-order temporal accuracy in the deterministic
setting. In Section \ref{sec:Results} we test and apply the algorithm
to a collection of standard test problems. We demonstrate that the
algorithm correctly reproduces static equilibrium properties such
as the radial distribution function in a suspension of soft spheres,
and also reproduces known features of single and pair hydrodynamic
interactions at small Reynolds numbers. We also study in detail the
short and long-time behavior of the velocity autocorrelation function
of a freely diffusing particle. Finally, we study the behavior of
the blob particle in high Reynolds number flow. In Section \ref{sec:Conclusions}
we offer some conclusions and thoughts on possible extensions of the
method and algorithm. Several more technical calculations and algorithmic
details are presented in a collection of Appendices.

\subsection{Notation and Basic Concepts}

In the beginning, we focus on the continuum formulation of the fluid-particle
coupling. However, it is important to point out that most of the notation
and conclusions can directly be adopted in the discrete formulation
by simply replacing spatial integrals with sums over grid points.
We will return to the spatially-discrete formulation in Section \ref{sec:Discretization}.

Let us consider a particle of physical mass $m$ and size (e.g., radius)
$a$ immersed in a fluid with density $\rho$. In real problems there
will be many particles $i=1,\dots,N_{p}$ that interact with each
other, for example, in microfluidic applications involving polymers
each particle could represent a bead in a bead-spring or bead-link
polymer model \cite{BrownianDynamics_DNA}. Unless otherwise indicated
it is straightforward to extend the proposed formulation to a collection
of interacting particles by simply adding a summation over the different
particles. Therefore, for simplicity of notation, we will typically
focus on a single particle and omit the particle index.

The position of the particle is denoted with $\V q(t)$ and its velocity
with $\V u=\dot{\V q}$. The shape of the particle and its effective
interaction with the fluid is captured through a smooth kernel function
$\delta_{a}\left(\V r\right)$ that integrates to unity and whose
support is localized in a region of size $a$. For example, one may
choose any one-dimensional ``bell-shaped'' curve $\delta_{a}\left(r\right)$
with half-width of order $a$, and define a spherically-symmetric
$\delta_{a}\left(\V r\right)=\delta_{a}\left(r\right)$; alternatively,
in $d$ dimensions one may define a tensor-product
\begin{equation}
\delta_{a}\left(\V r\right)=\prod_{\alpha=1}^{d}\delta_{a}\left(r_{\alpha}\right).\label{eq:delta_tensor}
\end{equation}
In immersed-boundary methods \cite{IBM_PeskinReview}, the kernel
function $\delta_{a}$ is considered to be an approximation of the
Dirac delta function of purely numerical origin and has the tensor-product
form (\ref{eq:delta_tensor}). By contrast, in the force-coupling
method \cite{ForceCoupling_Monopole,ForceCoupling_Stokes}, the shape
of the kernel function is chosen to be a spherically-symmetric Gaussian
whose width is related to the physical size and properties of the
actual particle. We adopt an approach that is intermediate between
these two extremes and choose the shape of the function based on numerical
considerations, but relate its shape to the physical properties of
the particle.

The fluid velocity field is denoted with $\V v(\V r,t)$ and is assumed
to extend over the whole domain, including the particle interior.
The interaction between the fluid and particle is mediated via the
kernel function through two crucial local operations. The\emph{ local
averaging }linear operator\textbf{ $\M J(\V q)$} averages the fluid
velocity inside the particle to estimate a local fluid velocity
\[
\V v_{\V q}\left(t\right)=\M J\V v(\V r,t)=\int\delta_{a}\left(\V q-\V r\right)\V v\left(\V r,t\right)\, d\V r.
\]
The reverse of local averaging is accomplished using the \emph{local
spreading }linear operator $\M S(\V q)$ which takes a force $\V F$
applied to the particle and spreads it over the extent of the kernel
function to return a smooth force density field,
\[
\V f\left(\V r,t\right)=\M S\V F\left(t\right)=\V F\left(t\right)\,\delta_{a}\left(\V q-\V r\right).
\]
Note that the local spreading operator $\M S$ has dimensions of inverse
volume. For notational simplicity we will slightly abuse notation
and assume that the local spreading and interpolation operators can
be applied to a scalar, a vector, or a tensor field, with the interpretation
that the same local averaging or spreading operation is applied to
each component independently. This sort of block-diagonal form of
the spreading and interpolation operators is not strictly required
for the mathematical formulation \cite{SELM}, but applies to the
specific Peskin forms of the operators we use in practice \cite{IBM_PeskinReview}.

The physical volume of the particle $\D V$ is related to the shape
and width of the kernel function via $\M J\M S=\D V^{-1}\,\M I$,
that is,
\begin{equation}
\D V=\left(\M J\M S\,1\right)^{-1}=\left[\int\delta_{a}^{2}\left(\V r\right)d\V r\right]^{-1}.\label{eq:dV_JS}
\end{equation}
Therefore, even though the particle is represented only by the position
of its centroid, it is not appropriate to consider it a ``point''
particle. Rather, it can be thought of as a diffuse sphere that has
some physical extent and interacts with the fluid in its interior.
For lack of better terminology, we will refer to such a diffuse particle
as a ``blob''. In fluctuating hydrodynamics the fluid velocity is
a distribution and cannot be evaluated pointwise, therefore, to obtain
well-defined fluctuating equations spatial averaging must be used
and a physical volume associated to each blob.

Because fluid permeates the interior of the particle, the effective
inertia of the particle is enlarged by $\rho\D V$ giving the physical
mass
\[
m=m_{e}+\rho\D V=m_{e}+m_{f},
\]
where $m_{e}$ is the \emph{excess mass} of the particle over the
mass of the entrained fluid $m_{f}=\rho\D V$. In particular, $m_{e}=0$
corresponds to a neutrally-buoyant particle, meaning that the inertia
of the fluid is unchanged by the presence of the particle. It is a
crucial property that $\D V$ is a constant that only depends on the
shape of the kernel function and \emph{not} on the position of the
particle. This ensures that the mass of the particle $m$ is constant
and can be given a well-defined physical interpretation. Preserving
this translational invariance of the physical properties of the blob
in the spatially-discrete setting requires using special discrete
averaging and spreading operations.

One could alternatively use the dimensionless operator $\widetilde{\M S}=\M S\D V$,
as done in Ref. \cite{DirectForcing_Balboa}, with the property that
$\M J\widetilde{\M S}=\M I$. We prefer to use the dimensional version
because the averaging and spreading operators are \emph{adjoint},
$\M S=\M J^{\star}$, i.e., the natural dot products in the particle
(Lagrangian) and fluid (Eulerian) domains are related via \cite{SELM}
\begin{equation}
\left(\M J\V v\right)\cdot\V u=\int\V v\cdot\left(\M S\V u\right)d\V r=\int\delta_{a}\left(\V q-\V r\right)\left(\V v\cdot\V u\right)d\V r\label{eq:adjoint_cond}
\end{equation}
for any $\V u$ and $\V v$. This adjoint property follows from the
fact that the same kernel function is used is used for both averaging
and spreading, and is crucial in maintaining energy conservation and
fluctuation-dissipation balance. This adjoint condition will also
be preserved by the discrete local averaging and spreading operators.

\subsection{Fluctuating Incompressible Navier-Stokes Equation}

In this work we assume that the fluid can be described via the fluctuating
Navier-Stokes equation \cite{FluctHydroNonEq_Book}. Specifically,
we model the dynamics of the fluid velocity field $\V v(\V r,t)$
assuming an isothermal incompressible Newtonian fluid, $\grad\cdot\V v=0$,
\begin{equation}
\rho\left(\partial_{t}\V v+\V v\cdot\grad\V v\right)=-\grad\pi+\grad\cdot\M{\sigma}+\V f=-\grad\pi+\eta\grad^{2}\V v+\grad\cdot\left[\left(k_{B}T\eta\right)^{\frac{1}{2}}\left(\M{\mathcal{W}}+\M{\mathcal{W}}^{T}\right)\right]+\V f,\label{eq:LLNS_incomp_v}
\end{equation}
where the stress tensor $\M{\sigma}$ includes the viscous $\eta\left(\grad\V v+\grad^{T}\V v\right)$
and fluctuating contributions, $\pi$ is the non-thermodynamic pressure,
$\rho$ is the (constant) fluid density, $\eta=\rho\nu$ is the (constant)
fluid shear viscosity, $\nu$ is the kinematic viscosity, and $\V f\left(\V r,t\right)$
is an additional force density such as gravity or the force exerted
by the particles on the fluid. Note that we prefer to use the standard
physics notation instead of the differential notation more common
in the mathematics literature since there is no difference between
the Ito and Stratonovich interpretations of stochastic integrals for
additive noise.

In the momentum conservation law (\ref{eq:LLNS_incomp_v}), the stochastic
momentum flux is modeled using a white-noise random Gaussian tensor
field $\M{\mathcal{W}}\left(\V r,t\right)$, that is, a tensor field
whose components are independent (space-time) white noise processes,
\[
\av{\mathcal{W}_{ij}(\V r,t)\mathcal{W}_{kl}(\V r^{\prime},t^{\prime})}=\left(\delta_{ik}\delta_{jl}\right)\delta(t-t^{\prime})\delta(\V r-\V r^{\prime}).
\]
The form of the stochastic forcing term ensures fluctuation-dissipation
balance between the random forcing and the viscous dissipation and
gives the correct spectrum for the thermally-induced velocity fluctuations.
The symmetrized form of the fluctuating stress $\left(k_{B}T\eta\right)^{1/2}\left(\M{\mathcal{W}}+\M{\mathcal{W}}^{T}\right)$
mimics the symmetry of the viscous stress tensor, ensuring fluctuation-dissipation
balance even for variable viscosity flows \cite{DFDB,FluctHydroNonEq_Book}.
The discretization and numerical solution of (\ref{eq:LLNS_incomp_v})
is discussed in more detail in Refs. \cite{LLNS_Staggered,DFDB}.

It is important to emphasize here that the non-linear fluctuating
Navier-Stokes equation forced with white-noise fluxes is ill-defined
because the solution should be a distribution rather than a function
and the nonlinear term $\V v\cdot\grad\V v$ cannot be interpreted
in the sense of distributions. This term needs to be regularized in
order to give a precise meaning to (\ref{eq:LLNS_incomp_v}). Such
a regularization has the physical interpretation that the fluctuating
fields are only defined from the underlying microscopic dynamics via
spatial coarse-graining with a characteristic mesoscopic length $\sigma$,
as discussed at length in Ref. \cite{DiscreteLLNS_Espanol}. In the
continuum setting, one can replace the term $\V v\cdot\grad\V v$
with $\tilde{\V v}\cdot\grad\V v+\grad\tilde{\V v}^{T}\cdot\V v$,
where $\tilde{\V v}$ is a smoothed velocity in which features in
$\V v$ at scales smaller than $\sigma$ are filtered %
\footnote{In the $\alpha$-Navier-Stokes equations \cite{alphaNS_Titi} the
smoothing is chosen to be an inverse Helmholtz operator, $\V v=\M u-\sigma^{2}\grad^{2}\V u$,
with boundary conditions chosen such that $\V u$ is divergence free
in the whole domain of interest \cite{alphaNS_BCs}.%
}, following the $\alpha$-Navier-Stokes model \cite{alphaNS_Titi}
in turbulence. An alternative is to filter the stochastic forcing
$\M{\mathcal{W}}$ with a smoothing kernel of width $\sigma$ \cite{StokesLaw}
(see also Appendix in Ref. \cite{LowMachExplicit}). We are not aware
of any careful studies of what regularization is the most appropriate
(i.e., produces the best match with molecular dynamics), and we do
not attempt to address this complex issue in this work.

In finite-volume or finite-element spatial discretizations, both the
nonlinear terms and the stochastic forcing are naturally regularized
by the discretization or coarse-graining length scale (grid spacing).
In our spatial discretization, the advective term $\V v\cdot\grad\V v$
is discretized using a skew-adjoint (conservative in both momentum
and energy) discrete advection operator, as explained in detail in
Ref. \cite{DFDB}. This ensures that the inclusion of that term does
not alter the equilibrium Gibbs-Boltzmann distribution for the fluctuating
velocity field. In practice, we have not observed any measurable effect
of the nonlinearity on the results presented here, as tested by simply
omitting advective fluxes in the velocity equation. This is consistent
with the notion that as long as there are sufficiently many molecules
per hydrodynamic cell the fluctuations will be small and the behavior
of the nonlinear equations will closely follow that of the \emph{linearized}
equations of fluctuating hydrodynamics, which can be given a precise
meaning \cite{DaPratoBook}. Specifically, advective terms such as
$\V v\cdot\grad\V v$ or $\V u\cdot\grad\V v$ (see, for example,
Eq. (\ref{eq:a_J_def})) scale like the \emph{square} of the magnitude
of the fluctuations, and in practice we observe they give unmeasurably
small corrections when the blobs are much larger than the fluid molecules.

\subsection{No-Slip Condition}

Coupling of a continuum (fluctuating) fluid with point-like (blob)
particles has been considered by other researchers. In particular,
in Lattice-Boltzmann methods \cite{Ladd:93,LB_SoftMatter_Review,VACF_LBM_Chineese}
a Stokes \emph{frictional} force between the particle and the fluid
is postulated. Specifically, the motion of the particle is described
by a Langevin equation in which a phenomenological Stokes frictional
force between the particle and the fluid is postulated, proportional
to the difference $\V u-\M J\V v$ between the particle and the locally-averaged
fluid velocity. A corresponding force is added to the fluid equations
to ensure momentum and energy conservation and fluctuation-dissipation
balance in the fluid-particle system \cite{LB_SoftMatter_Review,SELM}.

An important downside of the inertial Stokes coupling is the imposition
of an artificial friction parameter and an associated delay with the
response of the particle to changes in the flow. Such a delay is often
not physically acceptable unless a very large friction constant is
imposed, leading to numerical stiffness. Instead, following Ref. \cite{DirectForcing_Balboa},
we impose an \emph{instantaneous} coupling between the fluid and the
particle in the form of a \emph{no-slip constraint},
\begin{equation}
\V u=\dot{\V q}=\M J\V v=\int\delta_{a}\left(\V q-\V r\right)\V v\left(\V r,t\right)\, d\V r,\label{eq:no-slip}
\end{equation}
The no-slip condition simply states that the velocity of the particle
is equal to a local average of the fluid velocity. This is a constraint
that formally eliminates the particle velocity from the formulation
and leaves only the fluid degrees of freedom. We now demonstrate that
the imposition of (\ref{eq:no-slip}) leads to a physically-consistent
coarse-grained model of the coupled fluid-particle system. Notably,
the fluid-particle coupling conserves momentum, energy, and obeys
a fluctuation-dissipation principle.

It is important to point out that due to the finite extent of the
kernel $\delta_{a}$, the particle velocity (\ref{eq:no-slip}) differs
from that of a point tracer even for a smooth fluid velocity field.
As noted by Maxey and Patel \cite{ForceCoupling_Monopole}, to second-order
in the gradients of $\V v$, the particle velocity obeys a Faxen relation
\cite{FaxenRelations},
\begin{eqnarray*}
\V u & \approx & \int\delta_{a}\left(\V q-\V r\right)\left\{ \V v\left(\V q,t\right)+\grad\V v\left(\V q,t\right)\cdot\left(\V r-\V q\right)+\frac{1}{2}\grad\grad\V v\left(\V q,t\right):\left[\left(\V r-\V q\right)\left(\V r-\V q\right)^{T}\right]\right\} d\V r=\\
 & = & \V v\left(\V q,t\right)+\frac{1}{2}\grad\grad\V v\left(\V q,t\right):\int\delta_{a}\left(\V r\right)\V r\V r^{T}\, d\V r.
\end{eqnarray*}
If the kernel $\delta_{a}\left(\V r\right)$ is spherically-symmetric,
\begin{equation}
\V u=\V v\left(\V q,t\right)+\left[\int\frac{r_{x}^{2}}{2}\delta_{a}\left(r\right)d\V r\right]\grad^{2}\V v\left(\V q,t\right)=\V v\left(\V q,t\right)+\frac{a_{F}^{2}}{6}\grad^{2}\V v\left(\V q,t\right),\label{eq:Faxen_blob}
\end{equation}
where $a_{F}$ can be termed the ``Faxen'' radius of the blob \cite{ForceCoupling_Monopole},
in general different from the hydrodynamic radius (unlike for a fully-resolved
rigid sphere). The same formula applies for the case of a tensor-product
kernel (\ref{eq:delta_tensor}).

The particle acceleration is
\begin{equation}
\dot{\V u}=\frac{d}{dt}\left[\M J\left(\V q\right)\V v\right]=\M J\left(\partial_{t}\V v\right)+\left(\V u\cdot\frac{\partial}{\partial\V q}\M J\right)\V v,\label{eq:approx_adv_der}
\end{equation}
where for our choice of interpolation operator we have the explicit
form:
\[
\left(\V u\cdot\frac{\partial}{\partial\V q}\M J\right)\V v=\int\left[\V u\cdot\frac{\partial}{\partial\V q}\delta_{a}\left(\V q-\V r\right)\right]\V v\left(\V r,t\right)d\V r.
\]
Observe that in the limit of a ``point particle'', $a\rightarrow0$,
the kernel function approaches a Dirac delta function and one can
identify (\ref{eq:approx_adv_der}) with the advective derivative,
\[
\frac{d}{dt}\left(\M J\V v\right)\approx\frac{d}{dt}\V v\left(\V q(t),t\right)=D_{t}\V v=\partial_{t}\V v+\left(\V v\cdot\grad\right)\V v,
\]
which is expected since in this limit the particle becomes a Lagrangian
marker. In Ref. \cite{IBM_Implicit_Peskin}, the term $\frac{d}{dt}\left(\M J\V v\right)$
is replaced with the interpolated Navier-Stokes advective derivative
$\M J\left(D_{t}\V v\right)$, thus avoiding the need to differentiate
the kernel function. For a blob particle with finite size, however,
in general, the relative fluid-particle acceleration is non-zero,
\begin{equation}
\M a_{\M J}=\frac{d}{dt}\left(\M J\V v\right)-\M J\left(D_{t}\V v\right)=\left(\V u\cdot\frac{\partial}{\partial\V q}\M J\right)\V v-\M J\V v\cdot\grad\V v\neq\V 0.\label{eq:a_J_def}
\end{equation}

\section{\label{sec:Continuum}Incompressible Inertial Coupling Method}

Following the discussion in the Introduction and the derivation in
Section 2 of Ref. \cite{DirectForcing_Balboa} we take the equations
of motion for a single particle coupled to a fluctuating fluid to
be
\begin{eqnarray}
\rho\left(\partial_{t}\V v+\V v\cdot\grad\V v\right)=\rho D_{t}\V v & = & -\grad\pi+\grad\cdot\M{\sigma}-\M S\left(\V q\right)\V{\lambda}\label{eq:v_t}\\
m_{e}\dot{\V u} & = & \V F\left(\V q\right)+\V{\lambda}\label{eq:u_t}\\
\mbox{s.t. } & \V u= & \M J\left(\V q\right)\V v,\label{eq:no_slip}
\end{eqnarray}
where the fluid-particle force $\V{\lambda}$ is a Lagrange multiplier
that enforces the constraint (\ref{eq:no_slip}) and $\V F\left(\V q\right)$
is the external force applied to the particle. Observe that the total
particle-fluid momentum
\[
\M P=m_{e}\V u+\int\rho\V v\left(\V r,t\right)d\V r
\]
is conserved because Newton's third law is enforced; the opposite
total force is exerted on the fluid by the particle as is exerted
on the particle by the fluid. When there is more than one particle
one simply adds the forces from all the particles in the fluid equation.

Note that similar equations apply for both compressible and incompressible
fluids. In the compressible case \cite{DirectForcing_Balboa}, a density
equation is added to the system (\ref{eq:v_t},\ref{eq:u_t},\ref{eq:no_slip})
and the pressure $\pi\left(\rho\right)$ obtained from the equation
of state. In the incompressible case the divergence-free condition
$\grad\cdot\V v=0$ is used instead to determine the (non-thermodynamic)
pressure as a Lagrange multiplier.

For now, we will silently ignore the fact that the fluctuating equations
include a non-smooth white noise component that must be handled with
care, and return to a discussion of the stochastic equations later
on. For a neutrally-buoyant particle, $m_{e}=0$, $\V{\lambda}=-\V F$,
and the fluid equation is the standard Navier-Stokes with the force
on the particle spread back to the fluid as a force density $\M S\V F$
\cite{IBM_PeskinReview,ForceCoupling_Stokes}. In this case our formulation
is equivalent to the Stochastic Immersed Boundary Method \cite{StochasticImmersedBoundary,SELM},
and in the deterministic context, it is equivalent to the initial
(``monopole'') version of the Force Coupling Method \cite{ForceCoupling_Monopole}.

In the determinstic setting, several extensions of the deterministic
IB method to include inertial effects have already been developed
by Peskin and collaborators \cite{InertialIBM_nonuniform,InertialIBM_Penalty,InertialIBM_Density,IBM_Implicit_Peskin},
as well as by Uhlmann in the context of the direct-forcing method
\cite{DirectForcing_Uhlmann}, to which our method is closely related.
In the penalty method of Kim and Peskin \cite{InertialIBM_Penalty},
a particle (Lagrangian marker in the context of IBM) is represented
as a pair of particles, a (neutrally-buoyant) passive tracer that
follows the flow, $\dot{\V q}=\M J\left(\V q\right)\V v$, and an
inertial particle of mass $m_{e}$ connected to the tracer via an
elastic spring. In the limit of an infinitely stiff spring (infinite
penalty parameter) the spring force $\V{\lambda}$ becomes a Lagrange
multiplier enforcing the no-slip constraint. An advantage of our constrained
formulation is that it does not include the fast dynamics associated
with the stiff elastic springs and thereby avoids the time step size
restrictions associated with an explicit penalty method \cite{InertialIBM_Penalty}.

\subsection{Primitive-variable Formulation}

In this section we study the properties of (\ref{eq:v_t},\ref{eq:u_t},\ref{eq:no_slip})
in order to better understand the physics of the fluid-particle coupling.
Using (\ref{eq:u_t}) to eliminate $\V{\lambda}=m_{e}\dot{\V u}-\M F$
and (\ref{eq:a_J_def}) to eliminate $\dot{\V u}$, the fluid equation
(\ref{eq:v_t}) becomes,
\begin{eqnarray}
\rho D_{t}\V v=\rho\left(\partial_{t}\V v+\V v\cdot\grad\V v\right) & = & -m_{e}\M S\M J\left(D_{t}\V v\right)-\grad\pi+\grad\cdot\M{\sigma}-m_{e}\M S\M a_{\M J}+\M S\M F.\label{eq:v_t_no_u}
\end{eqnarray}
This gives the effective fluid equation
\begin{equation}
\left(\rho+m_{e}\M S\V J\right)\partial_{t}\V v=-\left[\rho\left(\V v\cdot\grad\right)+m_{e}\M S\left(\V u\cdot\frac{\partial}{\partial\V q}\M J\right)\right]\V v-\grad\pi+\grad\cdot\M{\sigma}+\M S\M F,\label{eq:fluid_only}
\end{equation}
in which the effective fluid inertia is given by the operator $\rho+m_{e}\M S\V J$,
and the kinetic stress term $\rho\V v\cdot\grad\V v$ includes an
additional term due to the excess inertia of the particle. When there
are many interacting particles one simply adds a summation over all
particles in front of all terms involving particle quantities in (\ref{eq:fluid_only}).
Note that for a neutrally-buoyant particle $m_{e}=0$ and one obtains
the constant-density Navier-Stokes equation with external forcing
$\M S\M F$.

Similarly, by eliminating $\V{\lambda}$ from (\ref{eq:u_t}) we obtain
the effective particle equation (see also Section 2 of Ref. \cite{DirectForcing_Balboa}),
\begin{equation}
m\dot{\V u}=\D V\,\M J\left(-\grad\pi+\grad\cdot\M{\sigma}\right)+\V F+m_{f}\M a_{\M J},\label{eq:particle_only}
\end{equation}
where $m_{f}=\rho\D V$ is the mass of the fluid dragged with the
particle. This equation makes it clear why $m=m_{e}+m_{f}$ has the
physical interpretation of particle mass (inertia). If the particle
were a rigid sphere, the force exerted by the fluid on the particle
would be the surface average of the stress tensor. It is sensible
that for a blob particle this is replaced by the locally averaged
divergence of the stress tensor (first term on right hand side). The
last term in the particle equation $m_{f}\M a_{\M J}$ has a less-clear
physical interpretation and comes because the fluid is allowed to
have a local acceleration different from the particle. It is expected
that at small Reynolds numbers the velocity field will be smooth at
the scale of the particle size and thus $\M a_{\M J}\approx\V 0$
\cite{balboa2011}. Nevertheless, we will retain the terms involving
$\M a_{\M J}$ to ensure a consistent formulation, see Appendix \ref{sec:AppendixLangevin}.

\subsection{Momentum Formulation}

Let us define a \emph{momentum field} as the sum of the fluid momentum
and the spreading of the particle momentum,
\begin{equation}
\V p\left(\V r,t\right)=\rho\V v+m_{e}\M S\V u=\left(\rho+m_{e}\M S\M J\right)\V v.\label{eq:v_to_p}
\end{equation}
The total momentum is $\M P\left(t\right)=\int\V p\left(\V r,t\right)d\V r$
and therefore a local conservation law for $\V p\left(\V r,t\right)$
implies conservation of the total momentum.

By adding the fluid and particle equations (\ref{eq:v_t},\ref{eq:u_t})
together we can obtain the dynamics of the momentum field,
\begin{eqnarray}
\partial_{t}\V p & = & \rho\left(\partial_{t}\V v\right)+m_{e}\M S\dot{\V u}+m_{e}\left(\V u\cdot\frac{\partial}{\partial\V q}\M S\right)\V u\nonumber \\
 & = & -\grad\pi+\grad\cdot\M{\sigma}-\grad\cdot\left[\rho\V v\V v^{T}+m_{e}\M S\left(\V u\V u^{T}\right)\right]+\M S\V F,\label{eq:p_t}
\end{eqnarray}
where we used the fact that $\M S$ depends on the difference $\left(\V q-\V r\right)$
only, and not on $\V q$ and $\V r$ separately. In the absence of
applied external forces we can write the right hand side as a divergence
of a total stress tensor $-\pi\M I+\M{\sigma}+\M{\sigma}_{\text{kin}}$,
where the kinetic stress tensor includes a contribution from the inertia
of the particle,
\begin{equation}
\M{\sigma}_{\text{kin}}=-\rho\V v\V v^{T}-m_{e}\M S\left(\V u\V u^{T}\right).\label{eq:sigma_kin}
\end{equation}
This means that the momentum field obeys a local conservation law,
as expected for short-ranged interactions between the particle and
the fluid molecules.

The formulation (\ref{eq:p_t}) is not only informative from a physical
perspective, but was also found very useful in performing adiabatic
elimination in the case of frictional coupling in Refs. \cite{SELM,SELM_Reduction}.

\subsection{Pressure-Free Formulation}

The equations we wrote so far contain the $\grad\pi$ term and can
easily be generalized to the case of a compressible fluid \cite{DirectForcing_Balboa}.
For analysis purposes, in the incompressible case it is useful to
eliminate the pressure from the equations using a projection operator
formalism. This well-known procedure \cite{Chorin68} can be understood
as follows. The fluid equation (\ref{eq:v_t}) is of the form
\[
\partial_{t}\V v+\rho^{-1}\grad\pi=\V g,\quad\grad\cdot\V v=0.
\]
By taking the divergence of the evolution equation, we get
\[
\partial_{t}\left(\grad\cdot\V v\right)+\rho^{-1}\grad^{2}\pi=\rho^{-1}\grad^{2}\pi=\grad\cdot\V g,
\]
which is a Poisson equation for the pressure whose solution can be
formally written as $\pi=\rho\grad^{-2}\left(\grad\cdot\V g\right)$.
This means that
\[
\partial_{t}\V v=\V g-\rho^{-1}\grad\pi=\V g-\grad\left[\grad^{-2}\left(\grad\cdot\V g\right)\right]=\M{\mathcal{P}}\V g,
\]
where $\M{\mathcal{P}}$ is a \emph{projection operator} that projects
the right-hand side $\V g$ or a given velocity field onto the space
of divergence-free vector fields. Note that the boundary conditions
are implicit in the definitions of the gradient, divergence, and Laplacian
operators. For periodic boundary conditions, the projection can most
easily be implemented using a spatial Fourier transform. Specifically,
in Fourier space the projection operator is simply a multiplication
by the $d\times d$ matrix $\widehat{\M{\mathcal{P}}}=\M I-k^{-2}(\V k\V k^{T})$,
where $d$ is the dimensionality, $d=2$ or $d=3$, and $\V k$ is
the wavenumber.

By using the projection operator, we can eliminate the pressure from
the equations of motion (\ref{eq:v_t},\ref{eq:u_t}), to obtain
\[
\rho\partial_{t}\V v=\M{\mathcal{P}}\left[-\rho\V v\cdot\grad\V v+\grad\cdot\M{\sigma}-m_{e}\M S\dot{\V u}+\M S\M F\right].
\]
If we now use (\ref{eq:approx_adv_der}) to eliminate $\dot{\V u}$
we obtain the fluid equation 
\[
\rho\partial_{t}\V v+m_{e}\M{\mathcal{P}}\M S\M J\M{\mathcal{P}}\left(\partial_{t}\V v\right)=\M{\mathcal{P}}\left[-\rho\V v\cdot\grad\V v-m_{e}\M S\left(\V u\cdot\frac{\partial}{\partial\V q}\M J\right)\V v+\grad\cdot\M{\sigma}+\M S\M F\right],
\]
where we used the fact that $\M{\mathcal{P}}\V v=\V v$ since $\grad\cdot\V v=0$,
and we added a $\M{\mathcal{P}}$ in front of the second term for
symmetry purposes. This shows that the pressure-free form of the fluid-only
equation (\ref{eq:fluid_only}) is 
\begin{equation}
\M{\rho}_{\text{eff}}\partial_{t}\V v=\M{\mathcal{P}}\left\{ -\left[\rho\left(\V v\cdot\grad\right)+m_{e}\M S\M J\left(\V v\cdot\frac{\partial}{\partial\V q}\M J\right)\right]\V v+\grad\cdot\M{\sigma}\right\} +\M{\mathcal{P}}\M S\M F=\M{\mathcal{P}}\V f+\M{\mathcal{P}}\M S\M F,\label{eq:fluid_only_no_p}
\end{equation}
where the force density $\V f$ contains the advective, viscous and
stochastic contributions to the fluid dynamics. This form of the equation
of motion can be shown to be identical to the limiting equation for
velocity obtained by Tabak and Atzberger, with the exception of the
advective term $\V v\cdot\grad\V v$ which is omitted in Ref. \cite{SELM_Reduction}.
An important feature of this formulation is that the density $\rho$
in the usual Navier-Stokes equation is now replaced by the effective
density \emph{operator}
\begin{equation}
\M{\rho}_{\text{eff}}=\rho\M I+m_{e}\M{\mathcal{P}}\M S\V J\M{\mathcal{P}},\label{eq:rho_eff_one_particle}
\end{equation}
where $\M I$ is the identity operator or matrix. Notice that the
effective density operator for incompressible flow is not $\rho\M I+m_{e}\M S\M J$
as one might naively expect based on (\ref{eq:fluid_only}). The distinction
between $\M S\M J$ and $\M{\mathcal{P}}\M S\V J\M{\mathcal{P}}$
is important, and leads to a well-known but surprising difference
between the short-time motion of a particle immersed in a compressible
versus an incompressible fluid \cite{BrownianCompressibility_Zwanzig}.
When there are many particles present, the effective inertia tensor
is generalized straightforwardly by summing the added inertia over
the particles,
\begin{equation}
\M{\rho}_{\text{eff}}=\rho\M I+\sum_{i}\left(m_{e}\right)_{i}\M{\mathcal{P}}\M S_{i}\V J_{i}\M{\mathcal{P}}.\label{eq:rho_eff_many_particles}
\end{equation}

Equation (\ref{eq:fluid_only_no_p}) together with the no-slip condition
$\dot{\V q}=\M J\V v$ gives a closed set of equations for $\V v$
and $\V q$ without any constraints. We can use this unconstrained
formulation to simplify analysis of the properties of the coupled
fluid-particle problem. In particular, in Appendix \ref{sec:AppendixLangevin},
the constraint-free form is used for showing fluctuation-dissipation
balance in the stochastic setting. The formal solution of (\ref{eq:fluid_only_no_p}),
\[
\partial_{t}\V v=\M{\rho}_{\text{eff}}^{-1}\M{\mathcal{P}}\left(\V f+\M S\M F\right),
\]
involves the cumbersome operator $\M{\mathcal{P}}\M S\V J\M{\mathcal{P}}$
via the inverse of the effective inertia $\M{\rho}_{\text{eff}}$.
In principle this makes the pressure-free formulation difficult to
use in numerical methods. However, as we explain in Appendix \ref{AppendixPeriodicBCs},
with periodic boundary conditions it is possible to efficiently invert
the operator $\M{\rho}_{\text{eff}}$ using Fourier transforms and
thus obtain a closed-form equation (\ref{eq:fluid_only_periodic})
for $\partial_{t}\V v$ suitable for numerical implementations. This
relies on the fact that for a large $d$-dimensional periodic system
$\M J\M{\mathcal{P}}\M S$ is a constant multiple of the $d\times d$
identity matrix.

\subsection{\label{sub:EnergyConservation}Energy Conservation}

A crucial property of no-slip coupling (unlike frictional-coupling)
between the particles and the flow is that it is non-dissipative,
and therefore all dissipation (drag) comes from viscous dissipation.
Specifically, in the absence of viscous dissipation, the equations
of motion (\ref{eq:v_t},\ref{eq:u_t},\ref{eq:no_slip}) conserve
a coarse-grained Hamiltonian \cite{GrabertBook,CoarseGraining_Pep}
given by the sum of potential energy and the kinetic energy of the
particle and the fluid,
\begin{equation}
H\left(\V v,\V u,\V q\right)=\rho\int\frac{v^{2}}{2}\, d\V r+m_{e}\frac{u^{2}}{2}+U\left(\V q\right),\label{eq:Hamiltonian_u}
\end{equation}
where $U\left(\V q\right)$ is the interaction potential of the particle
with external sources and other particles, with an associated conservative
force 
\[
\M F\left(\V q\right)=-\frac{\partial U}{\partial\V q}=-\frac{\partial H}{\partial\V q}.
\]
For compressible flow one needs to include the (density-dependent)
internal energy of the fluid in the Hamiltonian as well \cite{HamiltonianFluid}.

To demonstrate energy conservation, we calculate the rate of change
\[
\frac{dH}{dt}=-\M F\cdot\V u+m_{e}\V u\cdot\dot{\V u}+\int\rho\V v\cdot\left(\partial_{t}\V v\right)\, d\V r
\]
in the absence of viscous and stochastic fluxes. Using the equations
of motion (\ref{eq:v_t},\ref{eq:u_t}) we get
\begin{eqnarray*}
\frac{dH}{dt} & = & -\M F\cdot\V u+\V u\cdot\left(\V F+\V{\lambda}\right)-\int\V v\cdot\left(\M S\V{\lambda}\right)\, d\V r\\
 &  & -\int\V v\cdot\grad\pi\, d\V r-\rho\int\V v\cdot\left(\V v\cdot\grad\V v\right)\, d\V r\\
 & = & \left(\V u-\M J\V v\right)\cdot\V{\lambda}+\int\pi\left(\grad\cdot\V v\right)\, d\V r\\
 &  & -\frac{\rho}{2}\int\V v\cdot\grad\left(v^{2}\right)\, d\V r+\rho\int\V v\cdot\left[\left(\grad\times\V v\right)\times\V v\right]\, d\V r,
\end{eqnarray*}
where integration by parts and the adjoint property (\ref{eq:adjoint_cond})
were used for the first two terms, and a vector identity was used
to express $\V v\cdot\grad\V v$ in terms of the vorticity $\grad\times\V v$.
The first term vanishes due to the no-slip constraint $\V u=\M J\V v$.
The second and third terms vanish for incompressible flow $\grad\cdot\V v$,
and the last term vanishes because of the basic properties of the
cross product. This demonstrates that $dH/dt=0$ in the absence of
viscous dissipation, that is, the non-dissipative terms in the equation
strictly conserve the coarse-grained free energy.

\subsection{Fluctuation-Dissipation Balance}

So far, we considered the equations of motion for the fluid-particle
system ignoring thermal fluctuations. In Appendix \ref{sec:AppendixLangevin}
we formally demonstrate that in order to account for thermal fluctuations
in a manner that preserves fluctuation-dissipation balance it is sufficient
to add the usual Landau-Lifshitz stochastic stress $\left(k_{B}T\eta\right)^{1/2}\left(\M{\mathcal{W}}+\M{\mathcal{W}}^{T}\right)$
to the viscous stress tensor in $\M{\sigma}$, without adding any
stochastic forces on the particle. The key physical insight is that
the fluid-particle coupling is non-dissipative, as demonstrated in
Section \ref{sub:EnergyConservation}, and the only dissipation comes
from the viscous terms.

Fluctuation-dissipation balance here means that at thermodynamic equilibrium
the particle-fluid system is ergodic and time-reversible with respect
to the Gibbs-Boltzmann distribution $Z^{-1}\exp\left(-H/k_{B}T\right)$,
where the ``Hamiltonian'' $H$ given in (\ref{eq:Hamiltonian_u})
is to be interpreted as a coarse-grained free energy. Since $\V u=\M J\V v$
is not an independent degree of freedom, we can formally write the
Hamiltonian in terms of the degrees of freedom of the system as a
sum of potential and kinetic energy. The total kinetic energy includes,
in addition to the kinetic energy of the fluid $\int\left(\rho/2\right)v^{2}\, d\V r$,
a kinetic energy contribution due to the motion of the particle,
\[
E_{p}=m_{e}\frac{u^{2}}{2}=m_{e}\frac{\left(\M J\V v\right)\cdot\V u}{2}=\frac{m_{e}}{2}\int\V v\cdot\left(\M S\V u\right)\, d\V r=\frac{m_{e}}{2}\int\V v^{T}\M S\M J\V v\, d\V r,
\]
where use was made of the adjoint condition (\ref{eq:adjoint_cond}).
This leads to the coarse-grained Hamiltonian 
\begin{equation}
H\left(\V v,\V q\right)=m_{e}\int\frac{\V v^{T}\M S\M J\V v}{2}\, d\V r+\rho\int\frac{v^{2}}{2}\, d\V r+U\left(\V q\right)=\frac{1}{2}\int\M v^{T}\M{\rho}_{\text{eff}}\V v\, d\V r+U\left(\V q\right).\label{eq:H_u_q}
\end{equation}
Note that it is not necessary here to include an entropic contribution
to the coarse-grained free energy because our formulation is isothermal,
and we assume that the particles do not have internal structure.

We emphasize that the form of $H\left(\V q,\V v\right)$ in (\ref{eq:H_u_q})
is postulated based on physical reasoning rather than derived from
a more refined model. Atzberger \emph{et al.} \cite{SELM,SELM_Reduction}
provide a careful and detailed discussion of how one might eliminate
the particle velocity $\V u$ by performing an adiabatic elimination,
starting from a frictional coupling model in which the particle velocity
is independent from the fluid velocity. The starting frictional coupling
model, however, as we discussed earlier, involves an arbitrary frictional
force parameter that is yet to be given a microscopic interpretation.
We believe that the consistency of our inertial coupling model with
general thermodynamic principles and deterministic hydrodynamics is
sufficient to adopt the inertial coupling model as a consistent coarse-grained
model without having to justify it from ``first principles'' \cite{OttingerBook}.
We also note that the equations obtained in Ref. \cite{SELM_Reduction}
are similar in structure to the ones we have presented here.

The fact that fluid-particle coupling conserves the Hamiltonian (see
Section \ref{sub:EnergyConservation}) and is therefore non-dissipative
is a crucial component of fluctuation-dissipation balance. However,
this is not sufficient on its own. An important additional requirement
is that the phase space dynamics should be incompressible, which means
that the dynamics preserves not just phase-space functions of $H$,
but also preserves phase-space measures such as the Gibbs-Boltzmann
distribution. As discussed by Atzberger \cite{SELM}, even for the
case of a neutrally-buoyant particle an additional ``Ito'' or ``thermal''
drift term needs to be added to the velocity equation to ensure fluctuation-dissipation
balance. This term has the form of an additional contribution to the
stress tensor
\begin{equation}
\M{\sigma}_{\text{th}}=-\left[\M S\left(k_{B}T\right)\right]\M I.\label{eq:sigma_th}
\end{equation}
The physical origin of this term is the Kirkwood kinetic stress due
to the thermal motion of the particle lost when eliminating $\V u$
as a degree of freedom. Another way to interpret this term is that
it adds a particle contribution of $\M S\left(k_{B}T\right)$ to the
pressure. For incompressible flow, this simply changes the pressure
but does not change the dynamics of the velocity field since the projection
$\M{\mathcal{P}}$ eliminates the scalar gradient term $\grad\cdot\M{\sigma}_{\text{th}}=-\grad\M S\left(k_{B}T\right)$.
In Appendix \ref{sec:AppendixLangevin} we argue that for non-neutrally
buoyant particles there is also no need to include an additional thermal
drift term for periodic boundary conditions. Note, however, that the
above calculations rely on continuum identities that fail to be strictly
obeyed discretely. As we explain in Section \ref{sub:ThermalDrift}
of the Appendix in more detail, ensuring strict \emph{discrete} fluctuation-dissipation
balance requires keeping the contribution $-\grad\M S\left(k_{B}T\right)$
in the momentum equation in \emph{both} the compressible and incompressible
settings.

\subsection{\label{sub:Equipartition}Equipartition of Energy}

For a single particle immersed in a periodic incompressible fluid
in $d$ dimensions, translational invariance implies that there is
no dependence of expectation values on $\V q$, and therefore we can
keep $\V q$ fixed when calculating expectation values. The fact that
the Hamiltonian (\ref{eq:H_u_q}) is quadratic in $\V v$ means that
the fluctuations of velocity are Gaussian with covariance $\av{\V v\V v^{\star}}=\left(k_{B}T\right)\M{\rho}_{\text{eff}}^{-1}$.
The fluctuations of the particle velocity have variance
\[
\av{u^{2}}=\mbox{Trace}\left[\M J\av{\V v\V v^{\star}}\M S\right]=\left(k_{B}T\right)\mbox{Trace}\left[\M J\M{\rho}_{\text{eff}}^{-1}\M S\right].
\]
Using the relations (\ref{eq:rho_eff_inv_incomp}) and (\ref{eq:JPS})
derived in the Appendix, we can simplify
\begin{equation}
\av{u^{2}}=\frac{k_{B}T}{\rho}\mbox{Trace}\left[\M J\left(\M I-\frac{m_{e}\D{\widetilde{V}}}{\tilde{m}}\M{\mathcal{P}}\M S\V J\M{\mathcal{P}}\right)\M S\right]=d\frac{k_{B}T}{\tilde{m}},\label{eq:m_eff_incomp}
\end{equation}
where $\D{\widetilde{V}}=d\,\D V/\left(d-1\right)$ and $\tilde{m}=m_{e}+d\, m_{f}/\left(d-1\right)$.

The result $\av{u^{2}}=d\left(k_{B}T\right)/\tilde{m}$ should be
compared to the corresponding result for a compressible fluid \cite{DirectForcing_Balboa},
$\av{u^{2}}=d\left(k_{B}T\right)/m$, which follows from the usual
equipartition principle of statistical mechanics. When incompressibility
is accounted for, a fraction of the equilibrium kinetic energy is
carried in the unresolved sound waves, and therefore the apparent
mass of the particle is $\tilde{m}$ and not $m=m_{e}+m_{f}$, as
we verify numerically in Section \ref{sub:VACF_short}. It is reassuring
that our model equations reproduce the result for rigid particles.
This suggests that the model introduced here can be used to study
more complicated questions such as the effect of multi-particle interactions
on $\av{u^{2}}$ in semi-dilute to dense colloidal suspensions \cite{VACF_Suspension}.

\section{\label{sec:Discretization}Spatio-Temporal Discretization}

In this section we describe our second-order spatio-temporal discretization
of the equations of motion (\ref{eq:v_t},\ref{eq:u_t},\ref{eq:no_slip}).
Our spatio-temporal discretization is based on the deterministic Immersed
Boundary Method (IBM), and in particular, on the deterministic second-order
temporal integrator presented in Ref. \cite{IBM_FEM_Boyce}. For the
fluctuating fluid solver, we use the second-order staggered-grid spatial
discretization of the fluctuating Navier-Stokes equations described
in detail in Ref. \cite{LLNS_Staggered}. A notable feature of the
fluid solver we employ is that it handles the viscous terms semi-implicitly
and is stable for large time steps. Furthermore, for the fluctuating
Stokes equations, our fluid solver produces the correct spectrum of
the velocity fluctuations for any time step size \cite{LLNS_Staggered}.

There are two key novel features in our incompressible inertial coupling
algorithm from those previously developed. Firstly, our algorithm
includes the effects of particle excess inertia in a manner that strictly
conserves momentum and is second-order deterministically for smooth
problems. Secondly, we focus our initial development on systems with
periodic boundaries only, allowing the use of the Fast Fourier Transform
(FFT) as a linear solver for the time-dependent Stokes equations.
This greatly simplifies the implementation of the algorithm and allows
us to use Graphics Processing Units (GPUs) for very efficient parallelization
of the algorithm.

For neutrally-buoyant particles our temporal discretization is exactly
that described in Ref. \cite{IBM_FEM_Boyce} with the fluid solver
replaced by that described in Ref. \cite{LLNS_Staggered}. This simplified
algorithm is already implemented in the IBAMR software framework \cite{IBAMR},
an open-source library for developing fluid-structure interaction
models that use the immersed boundary method. Note that IBAMR can
handle non-periodic boundary conditions using a preconditioned iterative
solver for the time-dependent Stokes equations \cite{NonProjection_Griffith}.
In this paper we focus on describing the additional steps required
to handle excess inertia and to use FFTs as a linear solver.

The majority of our presentation will focus on a single particle coupled
to a fluctuating fluid. Only small changes are required to handle
multiple particles by simply summing the single-particle term over
the different particles. As we explain in more detail shortly, the
error introduced by superposing the single particle solutions to solve
the multi-particle system is small if the kernels of the different
particles are not overlapping, which in practice means that there
are at least 3 grid cells between the centroids of the particles.

\subsection{Spatial Discretization}

Our second-order spatial discretization of the equations (\ref{eq:v_t},\ref{eq:u_t},\ref{eq:no_slip})
is based on standard techniques for incompressible flow and the immersed
boundary method \cite{IBM_PeskinReview}, as described in more detail
in, for example, Ref. \cite{IBM_FEM_Boyce}. In the spatially-discretized
equations, the same equations as for the continuum apply, but with
the interpretation that the velocity $\V v$ is not a (random) vector
field but rather a finite-volume discretization of that field \cite{DiscreteLLNS_Espanol,DiscreteDiffusion_Espanol}.
We use a uniform Cartesian staggered-grid spatial discretization of
the incompressible Navier-Stokes equations, as described in more detail
in Ref. \cite{LLNS_Staggered}. In the staggered discretization the
control volume grid associated to each component of velocity is shifted
by half a grid spacing along the corresponding dimension relative
to the pressure grid. In the discrete setting, the various continuum
operators acting on vector fields become matrices. The spatial discretization
of the differential operators, notably the discrete gradient, divergence
and Laplacian operators, is described in detail in Ref. \cite{LLNS_Staggered}.

\subsubsection{Discrete Interpolation and Spreading Operators}

Application of the local averaging operator $\M J$, which is a convolution
operator in the continuum setting, becomes a discrete summation over
the grid points that are near the particle,
\[
\M J\V v\equiv\sum_{\V k\in\text{grid}}\phi_{a}\left(\V q-\V r_{k}\right)\V v_{k},
\]
where $\V r_{k}$ denotes the center of the control volume with which
$\V v_{k}$ is associated, and $\phi_{a}$ is a function that takes
the role of the kernel function $\delta_{a}$. We follow the traditional
choice \cite{IBM_PeskinReview} and do the local averaging independently
along each dimension,
\[
\phi_{a}\left(\V q-\V r_{k}\right)=\prod_{\alpha=1}^{d}\phi_{a}\left[q_{\alpha}-\left(r_{k}\right)_{\alpha}\right],
\]
which improves the isotropy of the spatial discretization (but note
that the local averaging is not rotationally invariant). As a matrix,
the local spreading operator $\M S=\left(\D V_{f}\right)^{-1}\M J^{\star}$
is a weighted transpose of $\M J$,
\[
\left(\M S\V F\right)_{k}=\left(\D V_{f}\right)^{-1}\phi_{a}\left(\V q-\V r_{k}\right)\V F,
\]
where $\D V_{f}=\D x\D y\D z$ is the volume of the hydrodynamic cell
\footnote{The cell volume $\D V_{f}$ is introduced here because the fluid kinetic
energy appearing in the discrete Hamiltonian is $\D V_{f}\,\sum_{\V k\in\text{grid}}\rho v_{k}^{2}/2$
and therefore $\partial H/\partial\V v$ is not the functional derivative
$\rho\V v$ as in the continuum (see Appendix \ref{sec:AppendixLangevin})
but rather the partial derivative $\D V_{f}\,\rho\V v$.%
}.

The discrete kernel function $\phi_{a}$ was constructed by Peskin
\cite{IBM_PeskinReview} to yield translationally-invariant zeroth-
and first-order moment conditions, along with a quadratic condition,
\begin{eqnarray}
\sum_{\V k\in\text{grid}}\phi_{a}\left(\V q-\V r_{k}\right) & = & 1\nonumber \\
\sum_{\V k\in\text{grid}}\left(\V q-\V r_{k}\right)\phi_{a}\left(\V q-\V r_{k}\right) & = & 0\nonumber \\
\sum_{\V k\in\text{grid}}\phi_{a}^{2}\left(\V q-\V r_{k}\right) & = & \D V^{-1}=\mbox{const.},\label{eq:JS_invariance}
\end{eqnarray}
independent of the position of the particle $\V q$ relative to the
underlying (fixed) velocity grid. Ensuring these properties requires
relating the support of the kernel function to the grid spacing, that
is, making $a\sim\D x$ (more specifically, typically the width of
the function $\phi_{a}$ has to be an integer multiple of the grid
spacing). This means that the size and shape of the particles is directly
tied to the discretization of the fluid equations, and the two cannot
be varied independently, for example, simulating the motion of a ``spherical''
particle requires choosing the same grid spacing along each dimension,
$\D x=\D y=\D z$. This is a shortcoming of our method, but, at the
same time, it is physically unrealistic to resolve the fluid flow
and, in particular, the fluctuations in fluid velocity, with different
levels of resolution for different particles or dimensions.

The physical size of the particle (hydrodynamic radius) can be varied
over a certain range independently of the grid resolution by using
modified discrete kernel functions. This can be very useful when simulating
polydisperse suspensions with mild polydispersity. A simple approach
is to use shifted or \emph{split kernels},
\[
\phi_{a,s}\left(\V q-\V r_{k}\right)=\frac{1}{2^{d}}\,\prod_{\alpha=1}^{d}\left\{ \phi_{a}\left[q_{\alpha}-\left(r_{k}\right)_{\alpha}-\frac{s}{2}\right]+\phi_{a}\left[q_{\alpha}-\left(r_{k}\right)_{\alpha}+\frac{s}{2}\right]\right\} ,
\]
where $s$ denotes a shift that parametrizes the kernel. By varying
$s$ in a certain range, for example, $0\leq s\leq\D x$, one can
smoothly increase the support of the kernel and thus increase the
hydrodynamic radius of the blob by as much as a factor of two. While
in principle one can increase the width of the kernels arbitrarily,
if the support of the kernel grows beyond 5-6 cells it better to abandon
the minimally-resolved blob approach and employ a more resolved representation
of the particle \cite{ParticleLaden_Proteus,Luo2009}. We do not use
split kernels in this work but have found them to work as well as
the unshifted kernels, while allowing increased flexibility in varying
the grid spacing relative to the hydrodynamic radius of the particles.

The last condition (\ref{eq:JS_invariance}), was imposed by Peskin
\cite{IBM_PeskinReview} as a way of approximating independence under
shifts of order of the grid spacing. This property is especially important
in our context since it implies that the particle volume $\D V=\left(\M J\M S\,1\right)^{-1}$
will remain constant and independent of the position of the blob relative
to the underlying grid. The function with minimal support that satisfies
(\ref{eq:JS_invariance}) is uniquely determined \cite{IBM_PeskinReview}.
In our numerical experiments we employ this \emph{three-point} discrete
kernel function, which means that the support of $\phi$ extends to
only three grid points along each dimension (i.e., $3^{d}$ discrete
velocities are involved in the averaging and spreading operations
in $d$ dimensions), see Ref. \cite{DirectForcing_Balboa} for details.
This particular choice gives $\D V=8\D V_{f}$ in three dimensions
and $\D V=4\D V_{f}$ in two dimensions. The narrow kernel improves
the computational efficiency on the bandwidth-limited GPU, as detailed
in Ref. \cite{DirectForcing_Balboa}.

\subsubsection{\label{sub:TranslationalInvariance}Translational Invariance}

In the continuum derivation, obtaining the closed-form pressure-free
velocity equation (\ref{eq:fluid_only_periodic}) relied sensitively
on the fact that for a large continuum system
\begin{equation}
\M J\M{\mathcal{P}}\M S=\frac{d-1}{d}\D V^{-1}\,\M I,\label{eq:JPS_cont}
\end{equation}
see Eq. (\ref{eq:JPS}) in Appendix \ref{AppendixPeriodicBCs}. Ideally,
we would like the spatial discretization to have the additional property
that, for periodic boundary conditions, $\M J\M{\P}\M S$ should be
invariant under translations of the particle relative to the underlying
fluid grid. This is \emph{not} ensured by the Peskin operators, which
are constructed without any reference to the fluid equations and the
particular form of the discrete projection or the discrete viscous
dissipation. In fact, in the traditional immersed boundary method
\cite{IBM_PeskinReview}, a centered discretization of the velocity
was used, which implies a very different form for the discrete projection
operator $\M{\P}$ and required the introduction of an additional
``odd-even'' moment condition not strictly necessary with a staggered
discretization.

Numerical experiments suggest that for the staggered grid discretization
that we employ, the continuum identity (\ref{eq:JPS_cont}) is obeyed
to within a maximal deviation of a few percent,
\begin{equation}
\D{\widetilde{\V V}}^{-1}\left(\V q\right)=\M J\M{\P}\M S\approx\D{\widetilde{V}}^{-1}\,\M I\label{eq:dV_tilde_inv}
\end{equation}
for a sufficiently large periodic system, where $\D{\widetilde{V}}=d\D V/\left(d-1\right)$
is a modified volume of the blob. For a periodic three-dimensional
system of $N_{x}\times N_{y}\times N_{z}$ hydrodynamic cells, we
expect to see deviations from (\ref{eq:dV_tilde_inv}) if one of the
grid dimensions becomes of the order of the kernel width (which is
3 cells for our spatial discretization). This is because a particle
then becomes affected by its nearby periodic images. For a grid size
$N_{x}\times N_{y}\times1$ cells we expect to obtain two-dimensional
behavior {[}see Eq. (\ref{eq:dV_2D}){]}.

\begin{figure*}
\centering{}\includegraphics[width=0.49\columnwidth]{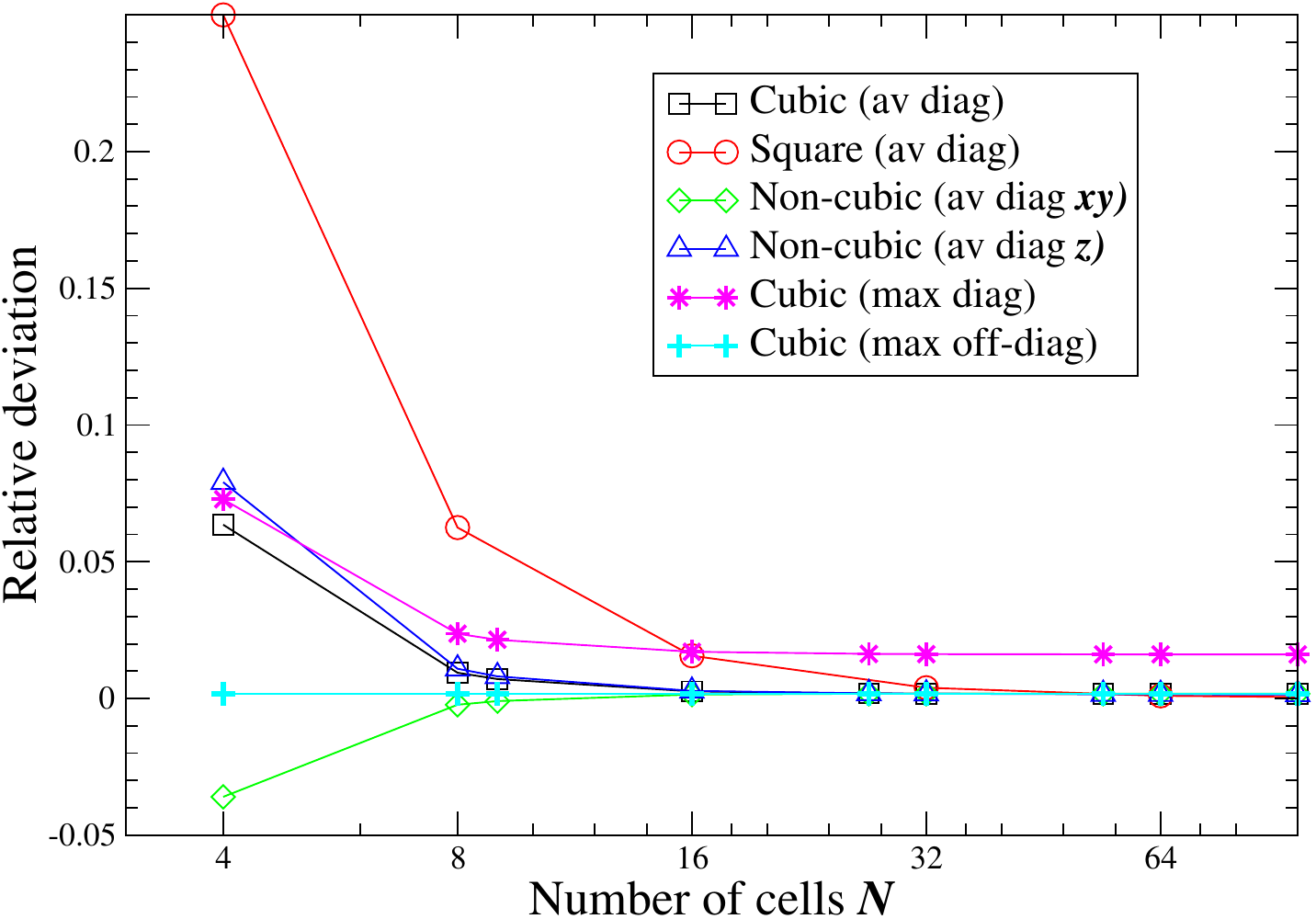}\includegraphics[width=0.49\columnwidth]{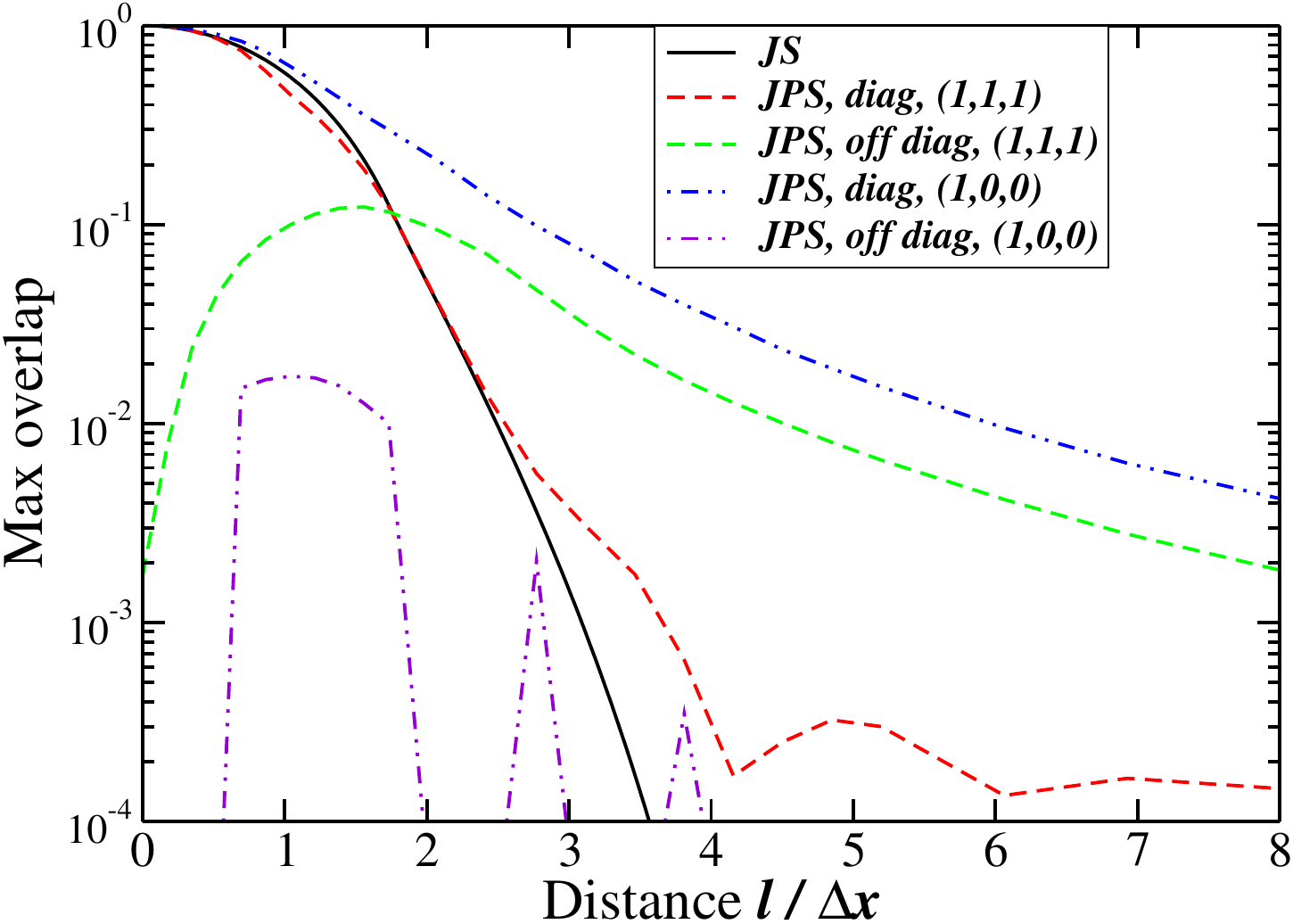}
\caption{\label{fig:JPS}(\emph{Left panel}) Translational invariance of the
approximation $d\D V\left(\M J\M{\P}\M S\right)/\left(d-1\right)\approx\M I$
over a set of $10^{3}$ positions of the particle relative to the
underlying fluid grid. For a periodic three-dimensional system of
$N^{3}$ cells, we show the average diagonal value of $\d{\M I}$
(black squares), the maximum diagonal element of $\d{\M I}$ (magenta
stars), as well as the maximum off-diagonal element of $\d{\M I}$
(cyan pluses). We also show the average diagonal value of $\d{\M I}$
for a two-dimensional system of $N^{2}$ cells (red circles), as well
as the average value $(\d{\M I}_{xx}+\d{\M I}_{yy})/2$ (green diamonds)
and $\d{\M I}_{zz}$ (blue triangles) for a three-dimensional system
of $32\times32\times N$ cells. (\emph{Right panel}) Maximum diagonal
and off-diagonal value of $3\D V\left(\M J_{i}\M{\P}\M S_{j}\right)/2$
and the maximum diagonal value of $\D V\left(\M J_{i}\M S_{j}\right)$
for two particles $i$ and $j$ a distance $l$ apart along the $\left(1,0,0\right)$
or $\left(1,1,1\right)$ direction, in a three-dimensional periodic
box of $32^{3}$ cells.}
\end{figure*}

In the left panel of Fig. \ref{fig:JPS} we show numerical results
for the average and maximum deviation of $\D{\widetilde{V}}\left(\M J\M{\P}\M S\right)$
from the $d\times d$ identity matrix,
\[
\d{\M I}\left(\V q\right)=\D{\widetilde{V}}^{-1}\left(\M J\M{\P}\M S\right)-\M I,
\]
as we vary the position of the particle $\V q$ relative to a cubic
Eulerian grid of size $N_{x}=N_{y}=N_{z}=N$ cells. Specifically,
we show the average diagonal value, the maximum diagonal element,
and the maximum off-diagonal element of $\d{\M I}\left(\V q\right)$.
For all but the smallest systems the diagonal elements are smaller
than $5\%$ and the off-diagonal elements are on the order of $0.1\%$.
For smaller system sizes there are visible finite-size effects due
to interactions with periodic images. For comparison, we also show
the corresponding two-dimensional results for a square grid of $N_{x}=N_{x}=N$
cells. The finite-size effects are more pronounced in two dimensions
due to the slower decay of the Green's function for the Poisson equation,
but for systems larger than $N=32$ cells we find (\ref{eq:JPS_cont})
to hold to a percent or so. In the left panel of Fig. \ref{fig:JPS}
we also show the average diagonal values $(\d{\M I}_{xx}+\d{\M I}_{yy})/2$
and $\d{\M I}_{zz}$ for non-cubic systems, illustrating the change
from three-dimensional to the two-dimensional behavior as $N_{z}\rightarrow1$.

Given these numerical observations, we will make an approximation
and assume that $\M J\M{\P}\M S$ is a constant multiple of the $d\times d$
identity matrix independent of $\V q$, and (\ref{eq:dV_tilde_inv})
holds as an equality. While this is, in principle, an uncontrolled
approximation, we will correct for it in order to ensure strict momentum
conservation and thus strict adherence to fundamental physical laws.
We also note that the small approximations we make when computing
$\M{\rho}_{\text{eff}}^{-1}$ only affect the short-time inertial
dynamics and do \emph{not} affect static properties or the long-time
dynamics, as deduced from (\ref{eq:fluid_only_no_p}). The approximation
(\ref{eq:dV_tilde_inv}) allows us to write the discrete equivalent
of (\ref{eq:rho_eff_inv_incomp}) 
\begin{equation}
\M{\rho}_{\text{eff}}^{-1}=\left(\rho\V I+m_{e}\M{\P}\M S\M J\M{\P}\right)^{-1}\approx\rho^{-1}\left(\M I-\frac{m_{e}\D{\widetilde{V}}}{\tilde{m}}\M{\P}\M S\V J\M{\P}\right).\label{eq:rho_eff_inv_discrete}
\end{equation}

When there are multiple particles in the system we simply sum the
second term on the right hand side over all particles, see Eq. (\ref{eq:rho_eff_inv_approx}),
which is an approximation even in the continuum setting. This approximation
relies on the assumption that $\M J_{i}\M{\P}\M S_{j}\approx\M 0$
for two particles $i$ and $j$ that are far away from each other.
To test the validity of this approximation, in the right panel of
Fig. \ref{fig:JPS} we show the maximum diagonal and off-diagonal
value for $\M J_{i}\M{\P}\M S_{j}$ for two particles a distance $l$
away from each other, for both $\V q_{i}-\V q_{j}=l\left(1,0,0\right)$
and $\V q_{i}-\V q_{j}=l\left(\sqrt{3},\sqrt{3},\sqrt{3}\right)/3$,
in three dimensions. If $l=0$ then $\M J_{i}\M{\P}\M S_{j}$ approaches
$2\D V^{-1}\,\M I/3$, so we normalize the value of $\M J_{i}\M{\P}\M S_{j}$
by a factor of $3\D V/2$. We see that for $l\gtrsim5\D x$, the maximum
normalized value of $\M J_{i}\M{\P}\M S_{j}$ is less than $0.01$.
As seen in the figure, $\M J_{i}\M S_{j}$ vanishes identically if
the kernels of particles $i$ and $j$ are disjoint, which is always
the case when the distance between the blobs is larger than $3\D x$
along at least one direction.

Another operator that is important for the long-time diffusive (Brownian)
motion of the particle, as explained in Refs. \cite{LB_SoftMatter_Review,SELM},
is the discrete mobility tensor for a single particle,
\[
\M{\mu}\left(\V q\right)=-\M J\M{\mathcal{L}^{-1}}\M S,
\]
where $\M{\mathcal{L}}^{-1}$ denotes the discrete Stokes solution
operator. For a single particle in a periodic domain, we numerically
find that $\M{\mu}\left(\V q\right)$ is approximately translationally-invariant
to within a few percent,
\[
\M{\mu}\left(\V q\right)\approx\mu\,\M I=\frac{1}{6\pi\eta R_{H}}\,\M I,
\]
where the effective hydrodynamic radius for an infinite system is
numerically extrapolated to be $R_{H}\approx0.91\D x$ for a uniform
grid spacing $\D x$ and the three-point kernel \cite{DirectForcing_Balboa}.
This is consistent with previous results for a cell-centered discretization
of the Navier-Stokes equations \cite{IBM_Sphere} and also with the
results obtained using a Lattice-Boltzmann fluid solver in \cite{LB_SoftMatter_Review}.
By using the Peskin four-point kernel \cite{IBM_PeskinReview} instead
of the three-point discrete kernel function the translational invariance
of the spatial discretization can be improved, however, at a potentially
significant increase in computational cost. A more systematic investigation
of different choices for the discrete kernel functions has been performed
by Mori \cite{IBMKernels_Mori}, however, these types of investigations
have yet to be carried out within the ``blob'' particle approach.
We have found the inexpensive three-point function to perform quite
well in our tests and use it exclusively in this work.

For non-periodic systems, one must generalize the definition of $\M J$
and $\M S$ in the case when the particle overlaps a physical boundary
\cite{IBMDelta_Boundary}. Even if a particle does not overlap a boundary,
however, it will feel the boundary hydrodynamically and therefore
both $\M J\M{\P}\M S$ and $\M J\M{\mathcal{L}^{-1}}\M S$ will depend
on the proximity of the particle to physical boundaries. Implementing
our algorithm in such cases may require first pre-tabulating the values
of these $d\times d$ matrices for different positions of the particle
relative to the boundaries.

\subsection{Temporal Discretization}

In this section, we describe how to integrate the spatially-discretized
equations in time and update the fluid and particle velocities and
particle position from time $n\D t$ to time $\left(n+1\right)\D t$,
where $\D t$ is the time step size, which can in principle be adjusted
dynamically but we will assume it is kept fixed. We will use a superscript
to denote the time level at which a given quantity is evaluated, for
example, $\V q^{n+\frac{1}{2}}$ will denote a mid-point estimate
for the position of the particle at time $\left(n+\frac{1}{2}\right)\D t$.
Similarly, $\V F^{n+\frac{1}{2}}=\V F\left(\V q^{n+\frac{1}{2}}\right)$
will denote the force on the particle (due to external sources or
other particles) evaluated at the position $\V q^{n+\frac{1}{2}}$
and, implicitly, at time $\left(n+\frac{1}{2}\right)\D t$ in the
case of a time dependent force.

For a neutrally-buoyant particle, $m_{e}=0$, the fluid momentum equation
is coupled to the particle position only through the forcing term
$\M S\V F$. In the deterministic setting, Griffith and Luo \cite{IBM_FEM_Boyce}
have developed a second-order splitting scheme for integrating the
spatially-discretized equations in time. The fluid solver in this
scheme is very similar to the predictor-corrector scheme employed
in the stochastic setting in Ref. \cite{LLNS_Staggered}. The temporal
discretization that we present next is based on replacing the fluid
solver in Ref. \cite{IBM_FEM_Boyce} with that in Ref. \cite{LLNS_Staggered},
at least for the simpler case $m_{e}=0$. The main difficulty is including
the additional inertia from the particles in the fluid momentum update
in a computationally-efficient manner.

In Ref. \cite{DirectForcing_Balboa} a first-order splitting algorithm
was developed for the case of a compressible fluid. This type of algorithm
is similar to the original projection algorithm of Chorin \cite{Chorin:1968}
for incompressible flow and can be summarized as follows. Update the
fluid first without accounting for the force $\V{\lambda}$ exerted
by the particle. Then, solve for the value of $\V{\lambda}$ that,
when applied as a correction to the fluid update, exactly imposes
the no-slip condition. Extending this type of approach to be higher
than first order accurate is known to be difficult from the literature
on incompressible flow \cite{bellColellaGlaz:1989}, due to the fact
that the splitting introduces a commutator error. Here we follow a
different, though related approach, which allows to construct a more
accurate algorithm for viscous-dominated flows.

Our temporal scheme will be based on the following approach, which
can be shown to be second-order deterministically by a Taylor series
expansion of the temporal local truncation-error:
\begin{enumerate}
\item Estimate the position of the particle at the midpoint to leading order,
\begin{equation}
\V q^{n+\frac{1}{2}}=\V q^{n}+\frac{\D t}{2}\M J^{n}\V v^{n}.\label{eq:particle_pred-2}
\end{equation}

\item \label{enu:momentum_update}Update the fluid velocity based on (\ref{eq:fluid_only})
using a second-order algorithm, while keeping the particle positions
fixed at the midpoint estimates,
\begin{eqnarray}
\left(\rho\M I+m_{e}\M S^{n+\frac{1}{2}}\M J^{n+\frac{1}{2}}\right)\frac{\V v^{n+1}-\V v^{n}}{\D t}+\grad\pi^{n+\frac{1}{2}} & = & -\grad\cdot\left(\rho\V v\V v^{T}-\M{\sigma}\right)^{n+\frac{1}{2}}+\M S^{n+\frac{1}{2}}\V F^{n+\frac{1}{2}},\nonumber \\
 & - & \left[m_{e}\M S\M J\left(\V v\cdot\frac{\partial}{\partial\V q}\M J\right)\V v\right]^{n+\frac{1}{2}}\label{eq:velocity_update}
\end{eqnarray}
subject to $\grad\cdot\V v^{n+1}=0$. Here a second order Runge-Kutta
\cite{DFDB} or Adams-Bashforth \cite{IBM_FEM_Boyce} scheme can be
used to evaluate the fluid momentum fluxes to at least second-order
accuracy, denoted generically here by superscript $n+\frac{1}{2}$.
\item Update the particle position using a second-order midpoint estimate
of the velocity,
\begin{equation}
\V q^{n+1}=\V q^{n}+\frac{\D t}{2}\M J^{n+\frac{1}{2}}\left(\V v^{n+1}+\V v^{n}\right).\label{eq:update_q-2}
\end{equation}

\end{enumerate}
Observe that the above scheme never actually uses the particle velocity
$\V u$, although one can and should keep track of the particle excess
momentum $m_{e}\V u$ and update it whenever the fluid momentum is
updated, to ensure strict conservation of momentum. Also observe that
during the fluid update we fix the particle at its midpoint position
$\V q^{n+\frac{1}{2}}$.

\subsubsection{Velocity Update}

The most difficult step in the time stepping algorithm summarized
above is the momentum (velocity) update, step \ref{enu:momentum_update}.
In order to update the velocity of the fluid we need to calculate
the fluid momentum change due to viscosity and thermal fluctuations
and also the momentum exchange with the particle, all to second order
in time. Our scheme is based on solving for the values of the Lagrange
multipliers $\pi^{n+\frac{1}{2}}$ and $\V{\lambda}^{n+\frac{1}{2}}$
such that at the end of the time step both the incompressibility and
the no-slip constraints are satisfied, 
\begin{eqnarray}
\rho\frac{\V v^{n+1}-\V v^{n}}{\D t}+\grad\pi^{n+\frac{1}{2}} & = & -\grad\cdot\left(\rho\V v\V v^{T}-\M{\sigma}\right)^{n+\frac{1}{2}}-\M S^{n+\frac{1}{2}}\V{\lambda}^{n+\frac{1}{2}}\nonumber \\
m_{e}\V u^{n+1} & = & m_{e}\V u^{n}+\D t\,\V F^{n+\frac{1}{2}}+\D t\,\V{\lambda}^{n+\frac{1}{2}}\nonumber \\
\grad\cdot\V v^{n+1} & = & 0\nonumber \\
\V u^{n+1} & = & \M J^{n+\frac{1}{2}}\V v^{n+1}+\D{\V u}^{n+\frac{1}{2}}.\label{eq:uv_update}
\end{eqnarray}
A correction $\D{\V u}^{n+\frac{1}{2}}$ is included to account for
the fact that the no-slip condition is not correctly centered since
$\M J$ and $\V v$ are evaluated at different points in time, as
we explain shortly. 

If we multiply the particle velocity update by $\M S^{n+\frac{1}{2}}$
and add it to the fluid equation, and use the no-slip constraint including
$\V u^{n}=\M J^{n-\frac{1}{2}}\V v^{n}+\D{\V u}^{n-\frac{1}{2}}$,
we get (\ref{eq:velocity_update}) with the kinetic term in the last
line approximated as
\begin{equation}
\left[m_{e}\M S\left(\V u\cdot\frac{\partial}{\partial\V q}\M J\right)\V v\right]^{n+\frac{1}{2}}=m_{e}\M S^{n+\frac{1}{2}}\left[\left(\frac{\M J^{n+\frac{1}{2}}-\M J^{n-\frac{1}{2}}}{\D t}\right)\V v^{n}+\frac{\D{\V u}^{n+\frac{1}{2}}-\D{\V u}^{n-\frac{1}{2}}}{\D t}\right].\label{eq:kinetic_term_lagged}
\end{equation}
The penultimate term in the above equation can be seen as a discretization
of the kinetic term 
\[
\left(\frac{\M J^{n+\frac{1}{2}}-\M J^{n-\frac{1}{2}}}{\D t}\right)\V v^{n}\approx\left[\left(\V u\cdot\frac{\partial}{\partial\V q}\M J\right)\V v\right]^{n}.
\]
This is consistent with the continuum equations but it yields only
first-order and \emph{not} second-order accuracy in $\D t$ because
it is not centered at time level $n+\frac{1}{2}$. This reduction
of the accuracy comes because the no-slip constraint in (\ref{eq:uv_update})
uses the midpoint instead of the endpoint position of the particle.

The above discussion shows that setting $\D{\V u}^{n+\frac{1}{2}}=\V 0$
results in a first-order scheme if $m_{e}\neq0$. To get second-order
accuracy, we need to apply a nonzero correction to the no-slip constraint.
Imposing the no-slip constraint at the end of the time step, $\V u^{n+1}=\M J^{n+1}\V v^{n+1}$,
leads to a formulation that is implicit in both $\V q^{n+1}$ and
$\V v^{n+1}$, which is difficult to implement in practice. Instead,
we can center the no-slip constraint as
\[
\frac{1}{2}\M J^{n+\frac{1}{2}}\left(\V v^{n+1}+\V v^{n}\right)=\frac{1}{2}\left(\V u^{n+1}+\M J^{n}\V v^{n}\right)=\frac{1}{2}\left(\M J^{n+\frac{1}{2}}\V v^{n+1}+\D{\V u}^{n+\frac{1}{2}}+\M J^{n}\V v^{n}\right),
\]
which gives the no-slip centering correction
\begin{equation}
\D{\V u}^{n+\frac{1}{2}}=\left(\M J^{n+\frac{1}{2}}-\M J^{n}\right)\V v^{n}.\label{eq:du_2nd}
\end{equation}
This correction for the no-slip constraint is simple to implement
with only one additional local averaging operation to evaluate $\M J^{n}\V v^{n}$.
Note that we purposely used $\M J^{n}\V v^{n}$ instead of $\V u^{n}$
here since in our formulation, and also in our algorithm, $\V u^{n+1}$
is only used as an intermediate variable. A Taylor series analysis
shows that using (\ref{eq:du_2nd}) makes (\ref{eq:kinetic_term_lagged})
a centered second-order approximation of the kinetic term
\[
\left(\frac{\M J^{n+\frac{1}{2}}-\M J^{n-\frac{1}{2}}}{\D t}\right)\V v^{n}+\frac{\D{\V u}^{n+\frac{1}{2}}-\D{\V u}^{n-\frac{1}{2}}}{\D t}=\left[\left(\V u\cdot\frac{\partial}{\partial\V q}\M J\right)\V v\right]^{n+\frac{1}{2}}+O\left(\D t^{3}\right).
\]
A Taylor series analysis confirms that using the no-slip correction
(\ref{eq:du_2nd}) leads to a second-order algorithm for updating
the position of the particle and the velocity of the fluid.

In order to avoid one more additional local averaging operation (which
requires an irregular memory access pattern and is thus costly, especially
in a GPU-based implementation) we can set $\D{\V u}^{n+\frac{1}{2}}=\V 0$.
We are primarily concerned with viscous-dominated (low Reynolds number)
flows, for which the kinetic term $\left(\V u\cdot\partial\M J/\partial\V q\right)\V v$
is small (quadratic in $\V v$, just like the advective term $\V v\cdot\grad\V v$)
and can be approximated to first order without a significant reduction
in the overall accuracy of the method. As we explain in Appendix \ref{sec:VelocityCorrection},
in our scheme $\D{\V u}^{n+\frac{1}{2}}$ contains an additional higher-order
correction ($O(\D t^{3}$) for smooth flows) that arises solely due
to the implicit handling of viscosity.

\subsubsection{Semi-Implicit Discretization of Viscous Terms}

During the fluid update the particle position remains fixed at $\V q^{n+\frac{1}{2}}$.
For notational simplicity, in the remainder of this paper we will
sometimes drop the time step index from $\M J$ and $\M S$; unless
otherwise indicated, they are always evaluated at $\V q^{n+\frac{1}{2}}$.

Following Ref. \cite{LLNS_Staggered}, our second-order implementation
of the velocity update (\ref{eq:velocity_update}) treats the viscous
term semi-implicitly and the remaining terms explicitly,
\[
-\grad\cdot\left(\rho\V v\V v^{T}-\M{\sigma}\right)^{n+\frac{1}{2}}=\frac{\eta}{2}\grad^{2}\left(\V v^{n+1}+\V v^{n}\right)+\grad\cdot\M{\Sigma}^{n}-\grad\cdot\left(\rho\V v\V v^{T}\right)^{n+\frac{1}{2}}.
\]
The spatial discretization of the stochastic flux is \cite{LLNS_Staggered}
\[
\M{\Sigma}^{n}=\left(\frac{k_{B}T\eta}{\D V_{f}\,\D t}\right)^{\frac{1}{2}}\left[\M W^{n}+\left(\M W^{n}\right)^{T}\right],
\]
where $\D V_{f}=\D x\D y\D z$ is the volume of the hydrodynamic cells,
and $\M W^{n}$ is a collection of i.i.d. unit normal variates, generated
independently at each time step on the faces of the staggered momentum
grid. To approximate the advective fluxes to second order in $\D t$,
one can use either the predictor-corrector method described in Ref.
\cite{LLNS_Staggered} or, more efficiently, one can use the classical
(time lagged) Adams-Bashforth method \cite{IBM_FEM_Boyce}
\begin{equation}
\grad\cdot\left(\rho\V v\V v^{T}\right)^{n+\frac{1}{2}}=\frac{3}{2}\grad\cdot\left(\rho\V v\V v^{T}\right)^{n}-\frac{1}{2}\grad\cdot\left(\rho\V v\V v^{T}\right)^{n-1}.\label{eq:Adams_Bashforth}
\end{equation}
For viscous-dominated (small Reynolds number) flows, one can also
approximate the advective terms to first order only without a significant
reduction of the overall accuracy for reasonably large time steps.

Referring back to Eq. (\ref{eq:velocity_update}), we see that updating
the fluid momentum semi-implicitly requires solving the linear system
\begin{eqnarray}
\left[\left(\frac{\rho\M I+m_{e}\M S^{n+\frac{1}{2}}\M J^{n+\frac{1}{2}}}{\D t}\right)-\frac{\eta}{2}\grad^{2}\right]\V v^{n+1}+\grad\pi^{n+\frac{1}{2}} & = & \left[\left(\frac{\rho\M I+m_{e}\M S^{n+\frac{1}{2}}\M J^{n+\frac{1}{2}}}{\D t}\right)+\frac{\eta}{2}\grad^{2}\right]\V v^{n}\nonumber \\
-\grad\cdot\left(\rho\V v\V v^{T}\right)^{n+\frac{1}{2}}+\grad\cdot\M{\Sigma}^{n} & + & \M S^{n+\frac{1}{2}}\V F^{n+\frac{1}{2}}-\left[m_{e}\M S\M J\left(\V v\cdot\frac{\partial}{\partial\V q}\M J\right)\V v\right]^{n+\frac{1}{2}}.\label{eq:v_implicit_system}
\end{eqnarray}
If $m_{e}=0$, we can solve the linear system (\ref{eq:v_implicit_system})
for the unknowns $\V v^{n+1}$ and $\pi^{n+\frac{1}{2}}$ using a
preconditioned iterative solver \cite{NonProjection_Griffith}, as
explained in more detail in Ref. \cite{LLNS_Staggered}. For periodic
systems the system (\ref{eq:v_implicit_system}) can be solved easily
by using a projection method together with FFT-based velocity and
pressure linear solvers. For non-neutrally-buoyant particles, however,
solving (\ref{eq:v_implicit_system}) requires developing a specialized
preconditioned Krylov method. Here we develop an approximate solver
for (\ref{eq:uv_update}) that only requires a few FFTs, and will
be shown in Section \ref{sub:Deterministic} to give nearly second-order
accuracy for a wide range of relevant time step sizes.

Our approach consists of splitting the velocity solver into two steps.
In the first step, we ignore the inertia of the particle, i.e., delete
the $m_{e}\M S\M J$ term in (\ref{eq:v_implicit_system}), and solve
for a provisional velocity $\tilde{\V v}^{n+1}$and pressure $\tilde{\pi}^{n+\frac{1}{2}}$,
\begin{equation}
\left(\frac{\rho}{\D t}\V I-\frac{\eta}{2}\grad^{2}\right)\tilde{\V v}^{n+1}+\grad\tilde{\pi}^{n+\frac{1}{2}}=\left(\frac{\rho}{\D t}\V I+\frac{\eta}{2}\grad^{2}\right)\V v^{n}+\grad\cdot\M{\Sigma}^{n}+\M S\V F^{n+\frac{1}{2}}-\grad\cdot\left(\rho\V v\V v^{T}\right)^{n+\frac{1}{2}},\label{eq:unperturbed_v}
\end{equation}
subject to $\grad\cdot\tilde{\V v}^{n+1}=0$. If $m_{e}=0$, this
completes the fluid solve and setting $\V v^{n+1}=\tilde{\V v}^{n+1}$
gives us second-order accuracy for the viscous and stochastic terms
\cite{DFDB}. If $m_{e}\neq0$, we need to find a velocity correction
$\D{\V v}^{n+\frac{1}{2}}=\V v^{n+1}-\tilde{\V v}^{n+1}$ and pressure
correction $\D{\pi}^{n+\frac{1}{2}}=\pi^{n+\frac{1}{2}}-\tilde{\pi}^{n+\frac{1}{2}}$
that takes into account the inertia of the particle. We do this by
splitting the linear system (\ref{eq:uv_update}) into two equations,
(\ref{eq:unperturbed_v}) for the unperturbed velocity field, and
\begin{eqnarray}
\left(\frac{\rho}{\D t}\V I-\frac{\eta}{2}\grad^{2}\right)\D{\V v}^{n+\frac{1}{2}}+\grad\left(\D{\pi}^{n+\frac{1}{2}}\right) & = & -\M S\left(\V{\lambda}^{n+\frac{1}{2}}+\V F^{n+\frac{1}{2}}\right)\label{eq:sys_dv_eq}\\
\frac{m_{e}}{\D t}\V u^{n+1} & = & \frac{m_{e}}{\D t}\V u^{n}+\left(\V{\lambda}^{n+\frac{1}{2}}+\V F^{n+\frac{1}{2}}\right)\label{eq:u_np1_eq}\\
\V u^{n+1} & = & \M J\left(\tilde{\V v}^{n+1}+\D{\V v}^{n+\frac{1}{2}}\right)+\D{\V u}^{n+\frac{1}{2}}\label{eq:no_slip_split}\\
\grad\cdot\left(\D{\V v}^{n+\frac{1}{2}}\right) & = & 0,\nonumber 
\end{eqnarray}
for the perturbed field. This gives a linear system of equations for
the unknowns $\tilde{\V v}^{n+1}$, $\V u^{n+1}$, $\D{\V v}^{n+\frac{1}{2}}$,
$\tilde{\pi}^{n+\frac{1}{2}}$, $\D{\pi}^{n+\frac{1}{2}}$, and $\V{\lambda}^{n+\frac{1}{2}}$.
We explain how we solve this linear system of equations in Appendix
\ref{sec:VelocityCorrection} for periodic boundaries using Fourier
Transform techniques. Here we simply summarize the resulting algorithm,
as implemented in our code. In Appendix \ref{AppendixCompressible}
we give a summary of a similar algorithm for compressible flow, which
our code also implements.

\subsection{\label{sub:SummaryIncompressible}Summary of Algorithm}
\begin{enumerate}
\item Estimate the position of the particle at the midpoint,
\begin{equation}
\V q^{n+\frac{1}{2}}=\V q^{n}+\frac{\D t}{2}\M J^{n}\V v^{n},\label{eq:particle_pred}
\end{equation}
and evaluate the external or interparticle forces $\V F^{n+\frac{1}{2}}\left(\V q^{n+\frac{1}{2}}\right)$.
\item Solve the unperturbed fluid equation
\begin{eqnarray}
\rho\frac{\tilde{\V v}^{n+1}-\V v^{n}}{\D t}+\grad\tilde{\pi}^{n+\frac{1}{2}} & = & \frac{\eta}{2}\grad^{2}\left(\tilde{\V v}^{n+1}+\V v^{n}\right)+\grad\cdot\left[\left(\frac{k_{B}T\eta}{\D V_{f}\,\D t}\right)^{1/2}\left(\M W^{n}+\left(\M W^{n}\right)^{T}\right)\right]\label{eq:provisional_step-1}\\
 & - & \left[\frac{3}{2}\grad\cdot\left(\rho\V v\V v^{T}\right)^{n}-\frac{1}{2}\grad\cdot\left(\rho\V v\V v^{T}\right)^{n-1}\right]+\M S^{n+\frac{1}{2}}\V F^{n+\frac{1}{2}},\nonumber \\
\grad\cdot\tilde{\V v}^{n+1} & = & 0,
\end{eqnarray}
using a projection algorithm and FFTs to diagonalize the Laplacian
operator.
\item If $m_{e}=0$, set $\V v^{n+1}=\tilde{\V v}^{n+1}$ and skip to step
\ref{enu:update_q}.
\item Evaluate the slip correction
\begin{equation}
\d{\V u}^{n+\frac{1}{2}}=\left(\M J^{n+\frac{1}{2}}-\M J^{n}\right)\V v^{n}+\frac{\nu\D t}{2}\M J^{n-\frac{1}{2}}\grad^{2}\left(\D{\V v}^{n-\frac{1}{2}}\right)\label{eq:delta_u_nph}
\end{equation}
and the change of the particle excess momentum
\[
\D{\V p}=m_{e}\left(\V u^{n}-\M J^{n+\frac{1}{2}}\tilde{\V v}^{n+1}-\d{\V u}^{n+\frac{1}{2}}\right).
\]

\item Calculate the fluid velocity perturbation due to the excess inertia
of the particle
\[
\D{\tilde{\V v}}=\frac{\tilde{m}_{f}}{\rho\left(\tilde{m}_{f}+m_{e}\right)}\M{\P}\M S\D{\V p},
\]
 using FFTs to implement the discrete projection $\M{\P}$, where
$\tilde{m}_{f}=d\,\rho\D V/\left(d-1\right)$.
\item Account for the viscous contribution to the velocity perturbation
by solving the system 
\begin{eqnarray}
\left(\rho\V I-\frac{\D t}{2}\eta\grad^{2}\right)\D{\V v}^{n+\frac{1}{2}}+\D t\,\grad\left(\D{\pi}^{n+\frac{1}{2}}\right) & = & \M S^{n+\frac{1}{2}}\left(\D{\V p}-m_{e}\M J^{n+\frac{1}{2}}\D{\tilde{\V v}}\right),\label{eq:dv_eq_modified-1}\\
\grad\cdot\left(\D{\V v}^{n+\frac{1}{2}}\right) & = & 0
\end{eqnarray}
using a projection algorithm and FFTs to diagonalize the Laplacian
operator.
\item Update the fluid velocity
\begin{equation}
\V v^{n+1}=\tilde{\V v}^{n+1}+\D{\V v}^{n+\frac{1}{2}}.\label{eq:update_v}
\end{equation}

\item Update the particle velocity in a momentum-conserving manner,
\begin{equation}
\V u^{n+1}=\M J^{n+\frac{1}{2}}\left(\tilde{\V v}^{n+1}+\D{\tilde{\V v}}\right)+\d{\V u}^{n+\frac{1}{2}}.\label{eq:update_u}
\end{equation}

\item \label{enu:update_q}Update the particle position,
\begin{equation}
\V q^{n+1}=\V q^{n}+\frac{\D t}{2}\M J^{n+\frac{1}{2}}\left(\V v^{n+1}+\V v^{n}\right).\label{eq:update_q}
\end{equation}

\end{enumerate}
We note that the full slip correction (\ref{eq:delta_u_nph}) is only
required if $m_{e}/m_{f}$ is large and the Reynolds number is large.
For sufficiently small Reynolds numbers (viscous-dominated flows)
we can neglect the quadratic advective term and only keep the linear
term, and set
\begin{equation}
\d{\V u}^{n+\frac{1}{2}}=\frac{\nu\D t}{2}\M J^{n-\frac{1}{2}}\grad^{2}\left(\D{\V v}^{n-\frac{1}{2}}\right).\label{eq:delta_u_nph_visc}
\end{equation}
We can also set $\d{\V u}^{n+\frac{1}{2}}=\V 0$, and obtain a first-order
algorithm that does not require any time lagging and has improved
stability for very large time step sizes. We compare the three options
(\ref{eq:delta_u_nph}), (\ref{eq:delta_u_nph_visc}), and $\d{\V u}^{n+\frac{1}{2}}=\V 0$
numerically in Section \ref{sub:Deterministic}. The remainder of
the algorithm is not affected by the choice of the slip correction
$\d{\V u}^{n+\frac{1}{2}}$.

\subsection{\label{sub:Deterministic}Efficiency, Stability and Accuracy}

With periodic boundary conditions the velocity and the pressure linear
systems in the incompressible formulation decouple and Fast Fourier
Transforms can be used to solve the system (\ref{eq:provisional_step-1})
efficiently, see Ref. \cite{IBM_Viscoelasticity} for additional details.
We first solve the velocity equation (\ref{eq:provisional_step-1})
without the gradient of pressure term (this is a Helmholtz equation)
using a Fourier transform to diagonalize the discrete Laplacian. Then,
we project the solution onto the space of divergence free vector fields
by subtracting a pressure gradient term. The pressure is a solution
of a discrete Poisson equation, which can also efficiently be computed
using Fourier transforms. Note that it is possible to generalize our
algorithm to non-periodic systems by using the fluid solver developed
by one of us \cite{NonProjection_Griffith} and employed in Ref. \cite{LLNS_Staggered},
at least for the case of neutrally buoyant particles, $m_{e}=0$.
For $m_{e}\neq0$ new iterative solvers for the Stokes subproblem
need to be developed.

We have parallelized the algorithm to run efficiently on Graphics
Processing Units (GPUs), as explained in more detail in Ref. \cite{DirectForcing_Balboa}.
Our public domain implementation \cite{DirectForcing_Balboa} is written
in the CUDA programming environment, and is three-dimensional with
the special case of $N_{z}=1$ cell along the $z$ axes corresponding
to a quasi two-dimensional system. In our implementation we create
one thread per cell, and each thread only writes to the memory address
associated with its cell and only accesses the memory associated with
its own and neighboring cells. This avoids concurrent writes and costly
synchronizations between threads, facilitating efficient execution
on the GPU. For incompressible flow, our present GPU implementation
is specific to periodic systems, and uses the NVIDIA FFT library as
a Poisson/Helmholtz solver.

The stability and accuracy of our spatio-temporal discretization is
controlled by the dimensionless advective and viscous CFL numbers
\begin{equation}
\alpha=\frac{V\D t}{\D x},\mbox{\quad}\beta=\frac{\nu\D t}{\D x^{2}},\label{eq:alpha_beta}
\end{equation}
where $V$ is a typical advection speed, which may be dominated by
the thermal velocity fluctuations or by a deterministic background
flow. Here we always use the same grid spacing along all dimensions,
$\D x=\D y=\D z$. The strength of advection relative to dissipation
is measured by the cell Reynolds number $r=\alpha/\beta=V\D x/\nu$.
Note that for compressible flow (see Ref. \cite{LLNS_Staggered} and
Appendix \ref{AppendixCompressible}) there is a sonic CFL number
$\alpha_{s}=c\D t/\D x$, where $c$ is the speed of sound.

The explicit handling of the advective terms places a stability condition
$\alpha\lesssim1$, in fact, for $\alpha>1$ a particle can move more
than a hydrodynamic cell during a single time step and this causes
not only stability but also implementation difficulties. It is not
hard to see that in the absence of advection our semi-implicit discretization
of viscosity is stable for any value of $\beta$, however, it is only
by keeping $\beta\lesssim1$ that we can ensure the dynamics of all
or at least most fluid modes is resolved \cite{DFDB}. We consider
a temporal integrator to be ``good'' if it produces reasonably-accurate
results with a time step for which at least one of $\alpha$ or $\beta$
is close to $1/2$. Typically, flows at small scales are viscous dominated
($r\ll1$) so that the time step is primarily limited by $\beta$
and not by $\alpha$.

Next we numerically check the deterministic order of accuracy of the
temporal integrator. Based on local truncation error analysis we expect
that the temporal integrator summarized in Section \ref{sub:SummaryIncompressible}
is formally second-order accurate. For small Reynolds numbers, we
expect to see nearly second-order accuracy in practice even if we
use the slip correction (\ref{eq:delta_u_nph_visc}) instead of (\ref{eq:delta_u_nph}).
We also recall that for $m_{e}\neq0$ we made an uncontrolled approximation
in assuming that $\M J\M{\P}\M S$ is translationally-invariant, which
is only accurate to about a percent for the three-point Peskin local
averaging and spreading operators. This approximation leads to another
error in imposing the no-slip condition, which we expect to lead to
first-order accuracy for very small time step sizes.

As a test of the temporal accuracy, we study the deterministic motion
of a particle in an centrally-symmetric harmonic potential $V(r)=kr^{2}/2$,
where $r$ is the distance from the origin and $k$ is a spring constant.
In these tests we keep the spatial discretization (and thus the blob
particle shape) fixed and only change the time step size $\D t$.
We start the particle from rest at a certain distance $r_{0}$ from
the origin and then release it. The particle will perform damped oscillations
under the influence of the spring and viscous friction. We look at
the error in the position of the particle $\V q\left(t\right)$ defined
as the average of the difference between the position for time steps
$\Delta t$ and $\Delta t/2$ over a certain number of time steps
$N_{s}$, from the initial time to a time $T=N_{S}\D t$,
\begin{equation}
E(\D t)=\frac{1}{N_{s}}\sum_{n=1}^{N_{s}}\norm{\mathbf{q}_{\Delta t}\left(n\Delta t\right)-\mathbf{q}_{\Delta t/2}\left(2n\frac{\Delta t}{2}\right)}.\label{eq:E_dt_definition}
\end{equation}
For a numerical scheme with order of accuracy $p$ this error should
behave as $E=O\left(\D t^{p}\right)$ for sufficiently small $\D t$.

\begin{figure*}
\centering{}\includegraphics[width=0.49\columnwidth]{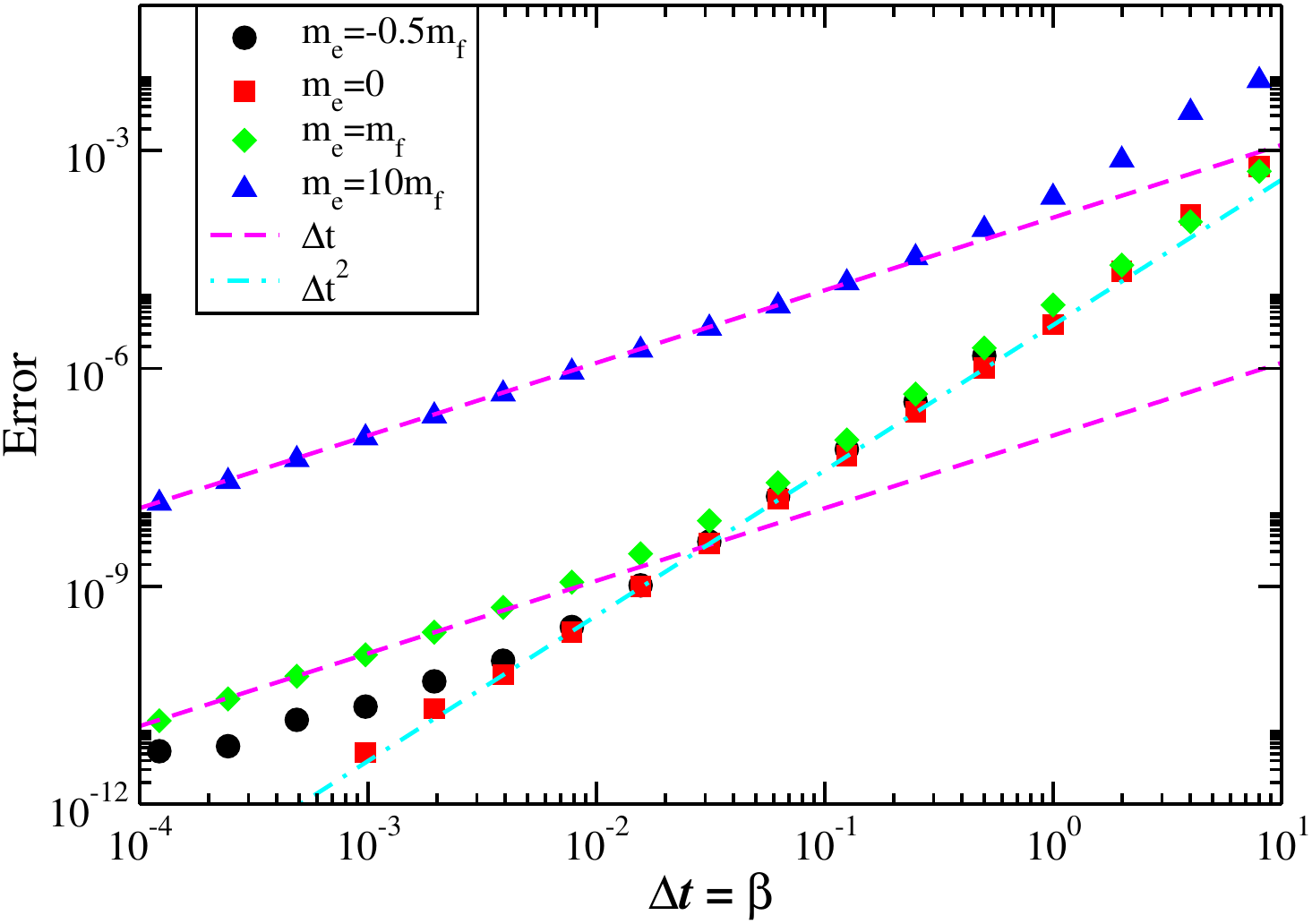}\includegraphics[width=0.49\columnwidth]{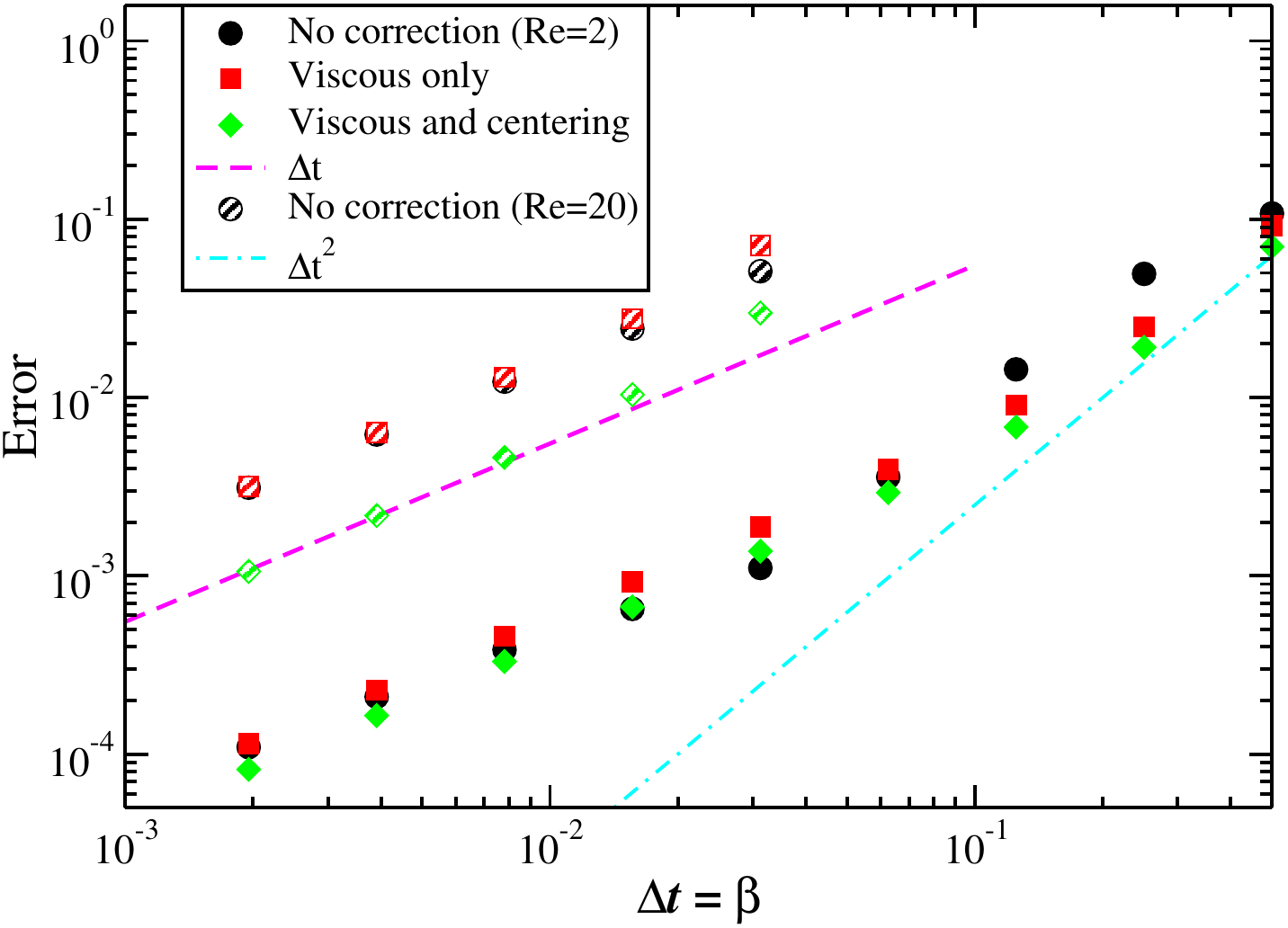}
\caption{\label{fig:convergenceShort}Average error (\ref{eq:E_dt_definition})
over a short deterministic trajectory of a particle initially at rest
and subsequently moving under the action of a harmonic potential.
(\emph{Left panel}) Small Reynolds number, $\mbox{Re}\approx0.02$,
with no-slip correction (\ref{eq:delta_u_nph_visc}). Several values
of the excess particle mass $m_{e}$ relative to the mass of the dragged
fluid $m_{f}$ (symbols, see legend) are shown, including neutrally-buoyant
particles ($m_{e}=0$). Expected error decay rates for a first and
second-order scheme are shown with lines. (\emph{Right panel}) Comparison
of the three choices for the slip correction, (\ref{eq:delta_u_nph})
(viscous and centering corrections), (\ref{eq:delta_u_nph_visc})
(only viscous correction), and (\ref{eq:delta_u_nph}) (no correction),
for Reynolds numbers $\mbox{Re}\approx2$ (full symbols) and $\mbox{Re}\approx20$
(shaded symbols). The excess mass is $m_{e}=10m_{f}$.}
\end{figure*}

We perform this test for several choices of the density of the particle
relative to the fluid, as measured via the ratio $m_{e}/m_{f}$, including
particles less dense than the surrounding fluid (negative excess mass
$m_{e}$). The tests are performed for a periodic system of $16^{3}$
hydrodynamic cells of size $\D x=1$, with fluid density $\rho=1$
and shear viscosity $\eta=1$ (in arbitrary units), with the particle
started at position $x_{0}=r_{0}=10$, and follow the motion of the
particles to a time $T=80$. In the left panel of Fig. \ref{fig:convergenceShort}
we show the error (\ref{eq:E_dt_definition}) for four different values
of the excess mass for a spring constant $k=0.01$, which implies
a small Reynolds number $Re\approx r=u_{\max}\D x/\nu\approx0.02$,
where $u_{\max}$ is the maximal speed of the particle. As expected,
the figure shows clear second-order convergence for neutrally-buoyant
particles ($m_{e}=0$), and a transition from essentially second-order
(at larger $\D t$) to first-order (at smaller $\D t$) accuracy in
the other cases. Notably, for the larger time steps, which are of
more practical interest, we see second-order convergence. As expected,
the transition from second to first order of accuracy occurs at a
larger $\D t$ for particles that are far from neutrally-buoyant,
and for $m_{e}/m_{f}=10$ we see first-order accuracy over a broader
range of time step sizes.

In order to compare the three choices for the the slip correction,
(\ref{eq:delta_u_nph}), (\ref{eq:delta_u_nph_visc}) and $\d{\V u}^{n+\frac{1}{2}}=\V 0$,
in the right panel of Fig. \ref{fig:convergenceShort} we compare
the error for the three choices for the case of a large particle excess
mass $m_{e}=10m_{f}$ and a larger Reynolds number, $\mbox{Re}\approx2$
for $k=1$, and $\mbox{Re}\approx20$ for $k=500$. We see qualitatively
similar behavior as for the small Reynolds number case $k=0.01$.
Compared to the simple choice $\d{\V u}^{n+\frac{1}{2}}=\V 0$, we
see a modest improvement in the error when we use (\ref{eq:delta_u_nph})
for the largest Reynolds number, and we see a small improvement when
we use (\ref{eq:delta_u_nph_visc}) for intermediate time step sizes.
Note that for our choice of parameters the viscous CFL number is $\beta=\D t$
and the advective CFL is $\alpha\approx\beta\,\mbox{Re}$. For large
Reynolds numbers the time step is limited by the requirement $\alpha\lesssim1$. 

Since the majority of the tests presented here are at small Reynolds
numbers, we use (\ref{eq:delta_u_nph_visc}) instead of the more expensive
(\ref{eq:delta_u_nph}). For several of the tests we have also tried
$\d{\V u}^{n+\frac{1}{2}}=\V 0$ and observed similar results.

\section{\label{sec:Results}Results}

In this section we validate and test the performance of the algorithm
summarized in Section \ref{sub:SummaryIncompressible} on a variety
of standard problems from soft-condensed matter applications. We also
examine the behavior of the minimally-resolved blob in a large Reynolds
number flow.

In the first few tests we examine the performance of the algorithm
for thermal systems. For a system at thermodynamic equilibrium at
rest the typical value of the advection velocity to be used in the
definition of the advective CFL number $\alpha$ (\ref{eq:alpha_beta})
is the equilibrium magnitude of the thermal velocity fluctuations,
\[
V\approx\sqrt{\frac{k_{B}T}{\rho\D x^{3}}}.
\]
The temporal integrator we employ here is designed to accurately resolve
the short time dynamics of the blobs when the time step size $\D t$
is reasonably small. We consider a time step size $\D t$ large if
at least one of the advective ($\alpha$) or the viscous ($\beta$)
CFL numbers defined in (\ref{eq:alpha_beta}) becomes comparable to
unity. It is important to emphasize that because of the semi-implicit
second-order nature of the temporal integrator, the algorithm is robust
over a broad range of time step sizes, which would be well beyond
the stability limit of explicit integrators for compressible flow.

\subsection{\label{sub:RDF}Equilibrium Properties}

One of the most important requirements on any scheme that couples
fluctuating hydrodynamics to immersed particles is to reproduce the
Gibbs-Boltzmann distribution at thermodynamic equilibrium. In particular,
the probability distribution of the positions $\V Q=\left\{ \V q_{1},\V q_{2},\dots,\V q_{N}\right\} $
of a collection of $N$ particles interacting with a conservative
potential $U\left(\V Q\right)$ should be
\begin{equation}
P\left(\V Q\right)\sim\exp\left[-\frac{U\left(\V Q\right)}{k_{B}T}\right],\label{eq:P_Q}
\end{equation}
independent of any dynamical parameters such as viscosity or particle
inertia. This follows from the balance between the dissipative and
stochastic forcing terms and requires consistently including thermal
fluctuations in the momentum equation.

\begin{table}
\centering{}%
\begin{tabular}{|c|c|}
\hline 
grid spacing $\D x$  &
1 \tabularnewline
\hline 
grid size &
$32^{3}$\tabularnewline
\hline 
fluid density $\rho$  &
1 \tabularnewline
\hline 
shear viscosity $\eta$  &
1 \tabularnewline
\hline 
time step size $\Delta t$  &
1 \tabularnewline
\hline 
temperature $k_{B}T$  &
0.001 \tabularnewline
\hline 
LJ strength $\epsilon$  &
0.001 \tabularnewline
\hline 
LJ diameter $\sigma$  &
2 \tabularnewline
\hline 
number of particles $N$ &
1000 \tabularnewline
\hline 
\end{tabular}\hspace{1cm}%
\begin{tabular}{|c|c|}
\hline 
grid spacing $\D x$  &
1 \tabularnewline
\hline 
grid size  &
$128^{3}$ \tabularnewline
\hline 
fluid density  &
1 \tabularnewline
\hline 
viscosity  &
1 \tabularnewline
\hline 
advective CFL $\alpha$  &
$0.01$, $0.1$, or $0.25$ \tabularnewline
\hline 
viscous CFL $\beta$  &
$9.2$, $92$, or $230$\tabularnewline
\hline 
excess mass $m_{e}$  &
$m_{f}$ \tabularnewline
\hline 
\end{tabular} \caption{(\emph{Left}) \label{tableRDF}Parameters used in the RDF simulations
shown in the left panel of Fig. \ref{fig:RDF}. (\emph{Right}) \label{tableLubrication}Simulation
parameters for the hydrodynamic interaction simulations presented
in the right panel of Fig. \ref{fig:RDF}.}
\end{table}

\begin{figure}
\centering{}\includegraphics[width=0.49\columnwidth]{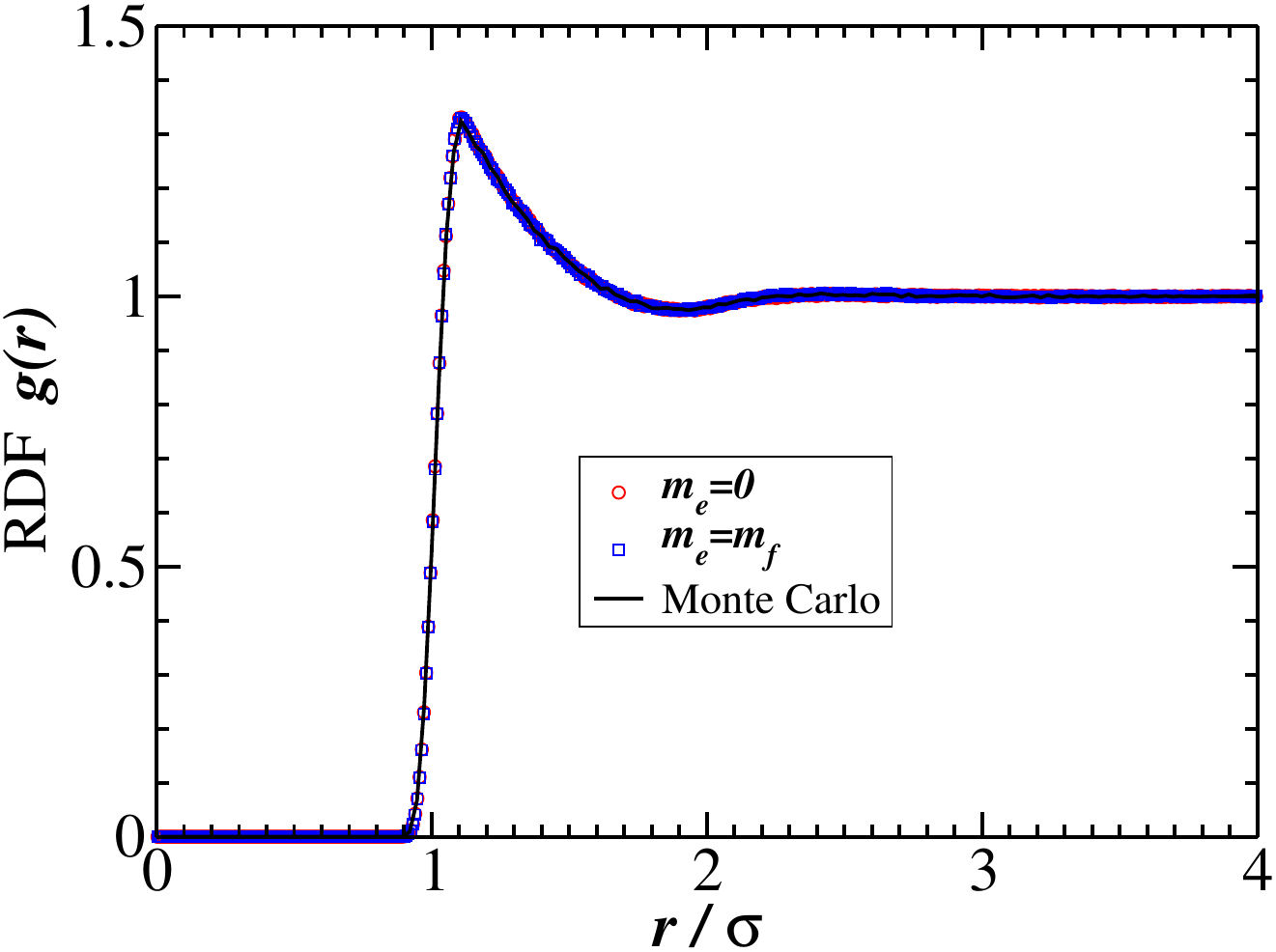}\includegraphics[width=0.49\columnwidth]{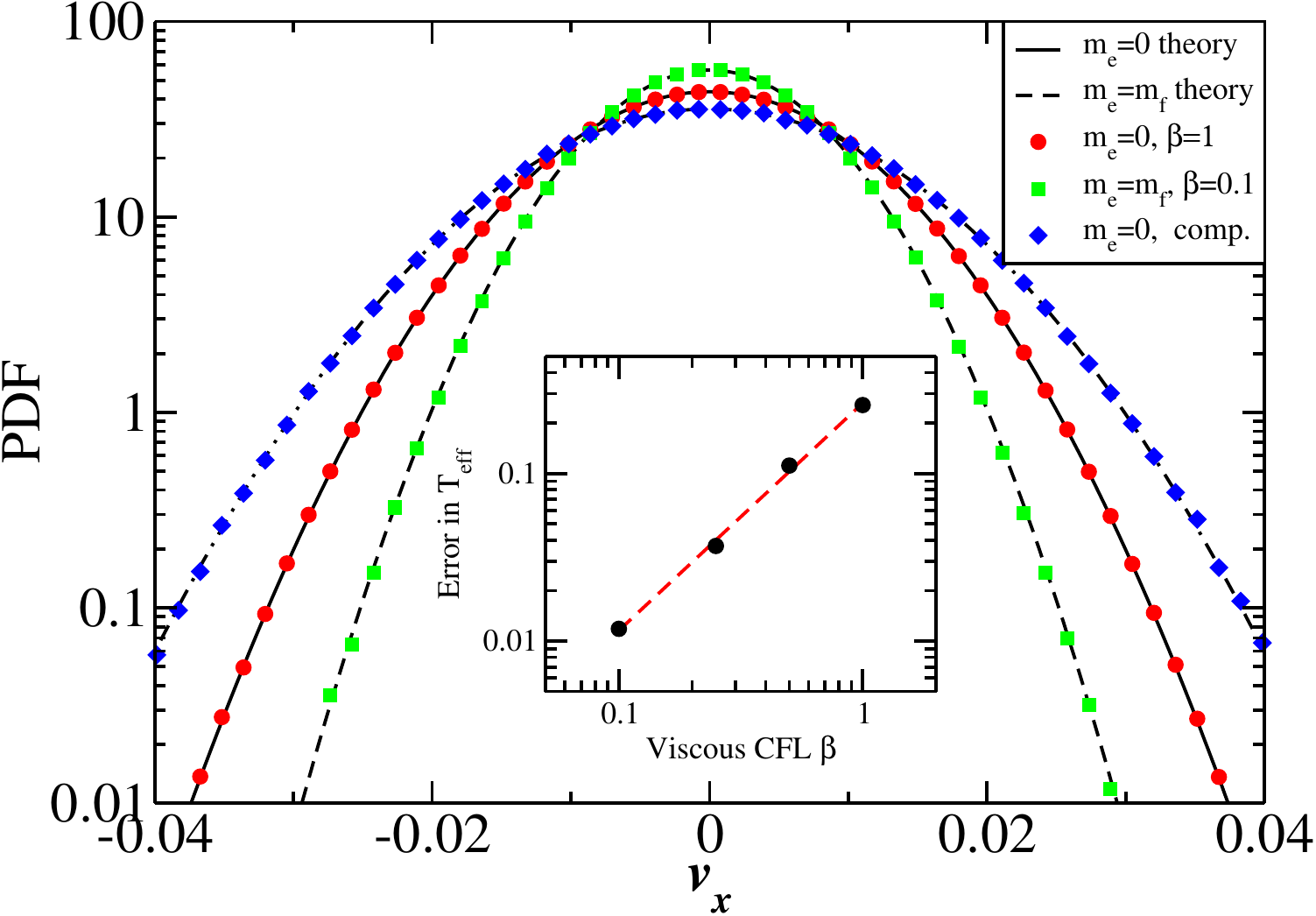}
\caption{\label{fig:RDF} (\emph{Left panel}) The equilibrium radial distribution
function $g(r)$ for a suspension of particles interacting with WCA
potential (\ref{eq:WCA}) with cutoff radius $\sigma$. The results
for two different particle inertias are compared to a Monte Carlo
sampling of the Gibbs-Boltzmann distribution. (\emph{Right panel})
The equilibrium probability distribution for the $x$ component of
the velocity of the particles in the suspension for different excess
masses, for both a compressible and an incompressible fluid (symbols).
The Gaussian distribution dictated by the equipartition principle
is shown for comparison (lines). The inset in the figure shows the
error $\epsilon_{T}$ in the effective temperature of the particles
as a function of the time step size, for $m_{e}=m_{f}$.}
\end{figure}

We verify that our incompressible inertial coupling algorithm gives
the correct equilibrium distribution $P\left(\V Q\right)$ by computing
the radial (pair) distribution function (RDF) $g(r)$ for a collection
of colloidal particles interacting with a pairwise potential $V(r)$,
$U\left(\V Q\right)=\sum_{i,j=1}^{N}\, V\left(\norm{\V q_{i}-\V q_{j}}\right)$.
We use the purely repulsive truncated Lennard-Jones (WCA) potential
\begin{eqnarray}
V(r) & = & \begin{cases}
4\epsilon\left(\left(\frac{\sigma}{r}\right)^{12}-\left(\frac{\sigma}{r}\right)^{6}\right)+\epsilon, & \quad r<2^{1/6}\sigma\\
0, & \quad r>2^{1/6}\sigma
\end{cases}\label{eq:WCA}
\end{eqnarray}
In Fig. \ref{fig:RDF} we compare $g(r)$ between a simulation where
the particles are immersed in an incompressible viscous solvent, and
a standard computation of the equilibrium RDF using a Monte Carlo
algorithm to sample the equilibrium distribution (\ref{eq:P_Q}).
The parameters for these simulations are given in Table \ref{tableRDF}.
For both $m_{e}=0$ and $m_{e}=m_{f}$ we obtain excellent agreement
with the Monte Carlo calculations, even for the rather large time
step size $\beta=1$. 

It is important to observe that the correct equilibrium structure
for the positional degrees of freedom is obtained for both the compressible
and the incompressible formulations, even though the blob velocities
have very different distributions. Specifically, as explained in Section
\ref{sub:Equipartition}, in the incompressible case the variance
of a component of the velocity is not $k_{B}T/m$ as for a compressible
fluid, but rather, $k_{B}T/\tilde{m}$, where the effective blob mass
$\tilde{m}>m$ includes an added mass due to the incompressible fluid
dragged with the blob. In the right panel of Fig. \ref{fig:RDF} we
show the probability distribution for the velocities of the blobs
in the colloidal suspension, and compare them to the theoretical predictions.
The numerical variance of the velocity is sensitive to the time step
size for $m_{e}\neq0$, and a small time step size ($\alpha,\beta<0.25$)
is required to obtain a reasonably accurate variance. This is shown
in the inset of the right panel of Fig. \ref{fig:RDF}, where the
relative error in the effective ``temperature'' of the blob $\epsilon_{T}=\left(\av{u^{2}}-k_{B}T/\tilde{m}\right)/\left(k_{B}T/\tilde{m}\right)$
is shown as a function of the viscous CFL number $\beta$.

\subsection{\label{sub:VACF}\label{sub:VACF_short}Velocity Autocorrelation
Function}

In this section we apply our scheme to a standard test for the coupling
of spherical particle of hydrodynamic radius $R_{H}$ to a compressible
\cite{VACF_Ladd,VACF_LBM_Chineese,DirectForcing_LBM,SmoothedInterface_Compressible,DirectForcing_Balboa}
or incompressible \cite{VACF_IncompressibleSmoothed,BrownianParticle_SIBM,FluctuatingHydro_FluidOnly,ImmersedFEM_Patankar}
fluid solver. The velocity autocorrelation function (VACF) 
\begin{equation}
C(t)=\left\langle v_{x}(0)v_{x}(t)\right\rangle =\frac{1}{d}\av{\V v\left(0\right)\cdot\V v\left(t\right)},\label{eq:VACF_def}
\end{equation}
of a single free Brownian particle diffusing through a periodic fluid
is a non-trivial quantity that contains crucial information at both
short and long times. The integral of the VACF determines the diffusion
coefficient and gets contributions from three distinct stages. Firstly,
at molecular times equipartition dictates that $C(0)=k_{B}T/m$, an
important signature of fluctuation-dissipation balance that has proven
challenging for several fluid-particle coupling methods \cite{VACF_Ladd,VACF_IncompressibleSmoothed,VACF_LBM_Chineese,BrownianParticle_SIBM}.
We recall that for our particle the effective particle mass $m=m_{e}+m_{f}$
includes the mass of the fluid dragged with the particle $m_{f}$,
as well as the excess mass $m_{e}$. The compressible inertial coupling
method is able to reproduce the intercept $k_{B}T/m$ very accurately
even for relatively large sound CFL numbers \cite{DirectForcing_Balboa}.

\begin{figure}
\centering{}\includegraphics[width=0.5\columnwidth]{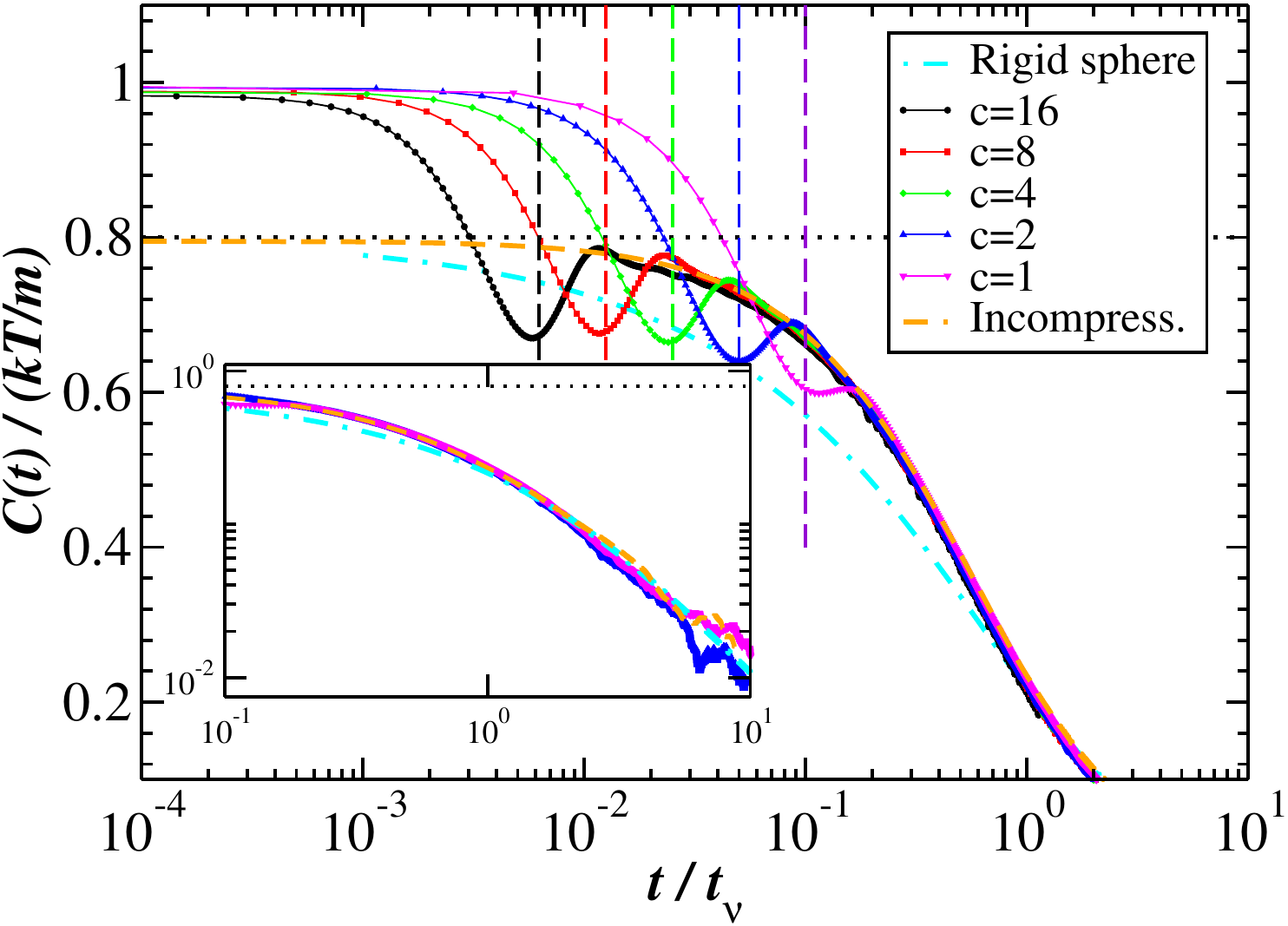}\includegraphics[width=0.5\columnwidth]{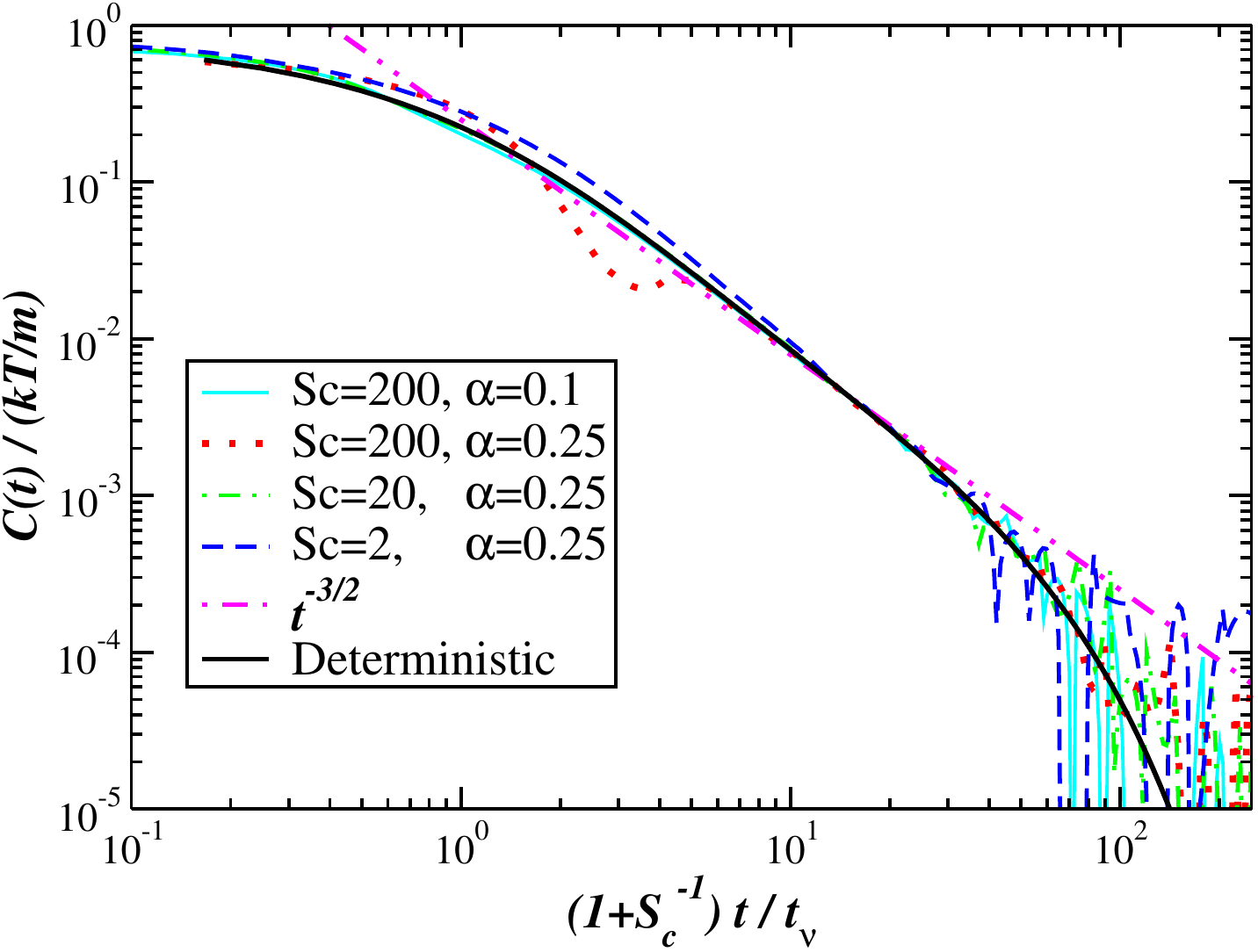}
\caption{\label{fig:VACF}Velocity autocorrelation function (VACF) (\ref{eq:VACF_def})
of a single particle with excess mass $m_{e}=m_{f}$, normalized by
$k_{B}T/m$ so that it should be unity at the origin for a compressible
fluid. (\emph{Left panel}) Comparison between a compressible fluid
for several different speeds of sound $c$ (compressibilities), as
well as an incompressible fluid ($c\rightarrow\infty$). Vertical
lines indicate the sound time scale $t_{c}=2R_{H}/c$, and the asymptotic
power-law tail $(t/t_{\nu})^{-3/2}$ is emphasized in the inset, where
$t_{\nu}=R_{H}^{2}/\nu$ is the viscous time scale. The tail matches
the theoretical predictions for a rigid sphere with the same effective
mass immersed in an incompressible fluid \cite{VACF_Langevin}. All
runs use a small time step size so the dynamics is well-resolved at
short times. (\emph{Right panel}) Comparison between different Schmidt
numbers $S_{c}=\nu/\chi$ for an incompressible fluid. A deterministic
calculation, corresponding to the limit $S_{c}\rightarrow\infty$,
is also shown. In the legend, the time step size is expressed in terms
of the advective CFL number $\alpha$. The scaling of the time axes
is adjusted to overlap the power-law tails (see text).}
\end{figure}

On the time scale of sound waves, $t<t_{c}=2R_{H}/c$, the major effect
of compressibility is that sound waves carry away a fraction of the
particle momentum with the sound speed $c$. The VACF quickly decays
from its initial value to $C(t_{c})\approx k_{B}T/\tilde{m}$, where
$\tilde{m}=m_{e}+d\, m_{f}/\left(d-1\right)$ includes an ``added
mass'' $m_{f}/(d-1)$ that comes from the fluid around the particle
that has to move with the particle because of incompressibility \cite{Landau:Fluid,VACF_Langevin,BrownianCompressibility_Zwanzig}.
The initial decay of the VACF due to sound waves will appear to be
instantaneous (discontinuous) if one increases the speed of sound
to infinity. The incompressible inertial coupling method should produce
an intercept $C(0^{+})=k_{B}T/\tilde{m}$ in agreement with (\ref{eq:m_eff_incomp}),
rather than the equipartition result valid for a compressible fluid.
For example, for $m_{e}=m_{f}$ and $d=3$, we expect $C(0^{+})=0.8\, k_{B}T/m$.
This illustrates the subtle complexity of coupling a fluctuating fluid
solver to immersed particles, even in the absence of external and
interparticle forces. In the left panel of Fig. \ref{fig:VACF} we
show numerical results for the VACF for several different speeds of
sound, obtained using the algorithm summarized in Appendix \ref{AppendixCompressible}.
The approach to the incompressible limit $c\rightarrow\infty$ is
evident in the figure, and it is clear that our incompressible inertia
coupling method correctly reproduces the limiting behavior (without
however suffering from the severe time step limitation of compressible
flow solvers). The parameters for these simulations are shown in Table
\ref{tableVacf}.

\begin{table}
\centering{}%
\begin{tabular}{|c|c|c|}
\hline 
Parameter  &
Fixed $S_{c}$ runs  &
Variable $S_{c}$ runs \tabularnewline
\hline 
excess mass $m_{e}$  &
$m_{e}=m_{f}$  &
$m_{e}=m_{f}$ \tabularnewline
\hline 
grid spacing $\D x$  &
10.543 &
1\tabularnewline
\hline 
grid size $N$ &
$41^{3}$ &
$32^{3}$\tabularnewline
\hline 
fluid density $\rho$  &
1  &
1\tabularnewline
\hline 
shear viscosity $\eta$  &
0.5  &
variable \tabularnewline
\hline 
bulk viscosity $\zeta$  &
0.5  &
not relevant \tabularnewline
\hline 
speed of sound $c$  &
$1-\infty$  &
$\infty$ \tabularnewline
\hline 
Temperature $k_{B}T$  &
$1$  &
$0.1$\tabularnewline
\hline 
viscous CFL $\beta$ &
$10^{-5}-10^{-3}$ &
$\beta=\alpha\sqrt{S_{c}/\left(6\pi\right)}$\tabularnewline
\hline 
Sound CFL $\alpha_{s}$ &
$0.05-0.1$ &
not relevant\tabularnewline
\hline 
Schmidt number $S_{c}$  &
48.2  &
variable \tabularnewline
\hline 
\end{tabular} \caption{\label{tableVacf}Parameters used in the compressible and incompressible
simulations shown in the left panel of Fig. \ref{fig:VACF} (middle
column), as well as the incompressible simulations shown in the right
panel of Fig. \ref{fig:VACF} (right column).}
\end{table}

At the viscous time scale, $t>t_{\nu}=\rho R_{H}^{2}/\eta$, conservation
of momentum (hydrodynamics) in the fluid introduces a memory in the
motion of the particle and the VACF decays with a well-known asymptotic
power-law tail $\sim(t/t_{\nu})^{-d/2}$ \cite{VACF_Alder}. Any numerical
method that solves the time-dependent Navier-Stokes equation (as opposed
to the steady or time-independent Stokes equation, as Brownian or
Stokesian dynamics do) ought to reproduce this power-law decay. The
amplitude of the decay depends on the shape of the particle and is
well-known for the case of a rigid sphere with stick boundaries \cite{VACF_Langevin}.
We expect that the VACF for our blob particle will have the same value
at the origin and the same power-law tail as a rigid sphere of radius
$R_{H}$ and the same ratio $m_{e}/m_{f}$. For the ``equivalent''
rigid sphere we take $m_{f}=\rho V_{s}$ where $V_{s}$ is the volume
of the sphere, and $m_{e}=\left(\rho_{s}-\rho\right)V_{s}$, where
$\rho_{s}$ is the density of the sphere.\textcolor{black}{{} The exact
shape of $C(t)$ will in general be different between a blob and a
rigid sphere. In the inset of the left panel of Fig. }\ref{fig:VACF}
we compare the long-time behavior of the VACF of a blob particle to
that of a rigid sphere, and see the same power-law behavior. Note
that the rigid sphere theory shown here does not account for finite-size
effects. In these tests we use a small time step size in order to
study the properties of the spatial discretization in the absence
of temporal truncation errors.

At long times, the motion of the particle is diffusive with a diffusion
coefficient predicted by the Stokes-Einstein relation to be
\begin{equation}
\chi\approx\chi_{\text{SE}}=\frac{k_{B}T}{6\pi\eta R_{H}}\label{eq:Stokes_Einstein_single}
\end{equation}
in three dimensions, where we recall that for the particular spatial
discretization we employ $R_{H}\sim\D x$ in three dimensions. The
exact coefficient depends on the system size, and for the system size
we use here an excellent approximation is $R_{H}\approx\D x$. We
define here a dimensionless Schmidt number $S_{c}$ based on the Stokes-Einstein
diffusion coefficient (\ref{eq:Stokes_Einstein_single}), 
\[
\frac{\nu}{\chi}\approx S_{c}=\frac{\nu}{\chi_{\text{SE}}}=\frac{6\pi\eta^{2}R_{H}}{\rho k_{B}T}\approx\frac{6\pi\beta^{2}}{\alpha^{2}}.
\]
The Schmidt number is an important quantity that measures how fast
momentum diffuses relative to the particles. In many cases of interest
$S_{c}\gg1$, which means that the dynamics of the particles approaches
the Brownian (overdamped) limit \cite{SELM}. Note that the limit
$S_{c}\rightarrow\infty$ is the same as the deterministic limit $k_{B}T\rightarrow0$,
in which fluctuations become a very weak perturbation to the deterministic
dynamics. Another important dimensionless number is the ``thermal''
Peclet number
\[
\text{Pe}=\frac{V\, R_{H}}{\chi_{SE}}\approx\sqrt{6\pi S_{c}},
\]
which measures the relative importance of advection by the thermal
velocity fluctuations to diffusion. We see that $\text{Pe}\sim S_{c}^{1/2}$
is directly related to the Schmidt number. A similar calculation also
shows that the ``thermal'' Reynolds number is $\text{Re}\sim S_{c}^{-1/2}$.
Therefore $S_{c}$ is the only relevant dimensionless number for a
particle diffusing in a fluid at thermodynamic equilibrium.

It is important to test how well our algorithm works for a range of
Schmidt numbers. We expect the case of small $S_{c}$ to be the most
difficult in terms of accuracy, since the particle can move a substantial
distance (compared to the grid spacing) during a single time step,
$\alpha=O(1)$, and the thermal and cell Reynolds number is also $r=O(1)$.
The case of large $S_{c}$, on the other hand, is the most demanding
in terms of computational effort since particles barely move during
a single time step, $\alpha\ll1$, and $O(S_{c})$ fluid time steps
may be required to reach the diffusive time scale for $\beta=O(1)$.
In order to investigate the long-time behavior we try to maximize
the time step, but always keeping $\alpha<1$, specifically, here
we set $\alpha=0.25$. For the largest Schmidt number we investigate,
$S_{c}\approx200$, this value of $\alpha$ corresponds to a relatively
large $\beta\approx0.81$, and therefore we also try a smaller time
step, corresponding to $\alpha=0.1$ and $\beta\approx0.33$. The
temporal integrator developed here cannot be used for $\beta\gtrsim1$
because the Crank-Nicolson temporal integrator we use for the velocity
equation does not accurately resolve the dynamics of the small wavelength
fluid modes \cite{StochasticImmersedBoundary,DFDB}.

In the right panel of Fig. \ref{fig:VACF} we show the VACF for several
viscosities and thus Schmidt numbers. The parameters for the runs
are given in Table \ref{tableVacf}. The standard theory for the tail
of the VACF (long-time behavior) \cite{VACF_Langevin} implicitly
assumes that $S_{c}\gg1$, and leads to the conclusion that for an
isolated particle in an infinite fluid asymptotically $C(t)\approx\left(t/t_{\nu}\right)^{-d/2}\sim\left(\nu t\right)^{-d/2}$.
A more complete self-consistent mode coupling theory \cite{ModeModeCoupling}
corrects this to account for the fact that while momentum diffuses
around the particle the particle itself diffuses, and predicts that
$C(t)\sim\left[\left(\chi+\nu\right)t\right]^{-d/2}$ \cite{VACF_Alder}.
This means that we expect the tails of the VACFs for different $S_{c}$
values to collapse on one master curve if we plot them as a function
not of $\left(t/t_{\nu}\right)$ but rather of $\left(1+S_{c}^{-1}\right)\left(t/t_{\nu}\right)$.
This is confirmed in the right panel of Fig. \ref{fig:VACF}.

It is evident in Fig. \ref{fig:VACF} that these fluctuating calculations
lead to noisy results for the tail of the VACF, making it difficult
to see the behavior of the long-time behavior. Many researchers have
chosen to calculate the VACF by performing a \emph{deterministic}
calculation, in which the particle is given a small initial kick in
velocity, and then the deterministic algorithm is used to track the
subsequent decay of the velocity. This is sometimes done because thermal
fluctuations are not consistently included in the algorithm, or because
the deterministic calculation is much faster and more accurate, not
requiring as much statistical averaging. In the right panel of Fig.
\ref{fig:VACF} we show the VACF obtained from a deterministic test,
which can be thought of as the VACF in the limit of vanishing fluctuations,
$k_{B}T\rightarrow0$ (equivalently, $S_{c}\rightarrow\infty$). Note
that in the deterministic test the magnitude of the initial velocity
$\V u_{0}$ of the particle has to be chosen to match the thermal
kick, $\norm{\V u_{0}}^{2}=d\, k_{B}T/\tilde{m}$, which in practice
means that the deterministic VACF has to be scaled so that it agrees
with statistical mechanics at the origin. Due to the slight anisotropy
and imperfect translational invariance of the spatial discretization,
in principle even the deterministic result should be averaged over
many initial positions and velocities of the particle. The VACF in
the limit $S_{c}\rightarrow\infty$ shown in Fig. \ref{fig:VACF}
matches the fluctuating runs for the larger Schmidt numbers. Due to
the lack of noise, it also clearly shows the long-time exponential
decay in the VACF at times $t\sim L^{2}/\nu$ \cite{BrownianParticle_SIBM},
where finite-size effects become important.

\subsection{Small Reynolds Number}

In this section we focus on the ability of the blob model to reproduce
some important features of the interaction of rigid spheres with deterministic
low Reynolds number flow. Maxey and collaborators have performed detailed
investigations of the low Reynolds number hydrodynamics for Gaussian
blobs \cite{ForceCoupling_Monopole,ForceCoupling_Stokes,Lomholt2001}
in the context of the Force Coupling Method (FCM). They have already
demonstrated that a blob model can model the behavior of hard sphere
colloidal suspensions with remarkable fidelity given its minimal resolution.
Because our blob is not Gaussian and our spatial discretization is
very different from that employed in the FCM, we examine briefly the
flow around a single particle and the hydrodynamic interactions between
a pair of particles at very small Reynolds number.

\subsubsection{Stokes flow around a blob}

An important property of the blob is its hydrodynamic radius, which
is defined in three dimensions from Stokes law for the drag force
$\V F_{d}=6\pi\eta R_{H}\V v$ experienced by the particle as it moves
slowly through an unbounded fluid at rest far away from the particle.
One can also compare the steady Stokes flow around the blob with the
well-known analytical solution for the flow around a rigid sphere.
These types of calculations were performed in detail for a compressible
fluid in Ref. \cite{DirectForcing_Balboa}, and since the Mach number
was kept small, very similar results are to be expected for an incompressible
fluid. Here we briefly examine the steady Stokes flow around a particle
as a validation of the incompressible formulation and implementation.

\begin{figure}
\centering{}\includegraphics[width=0.49\columnwidth]{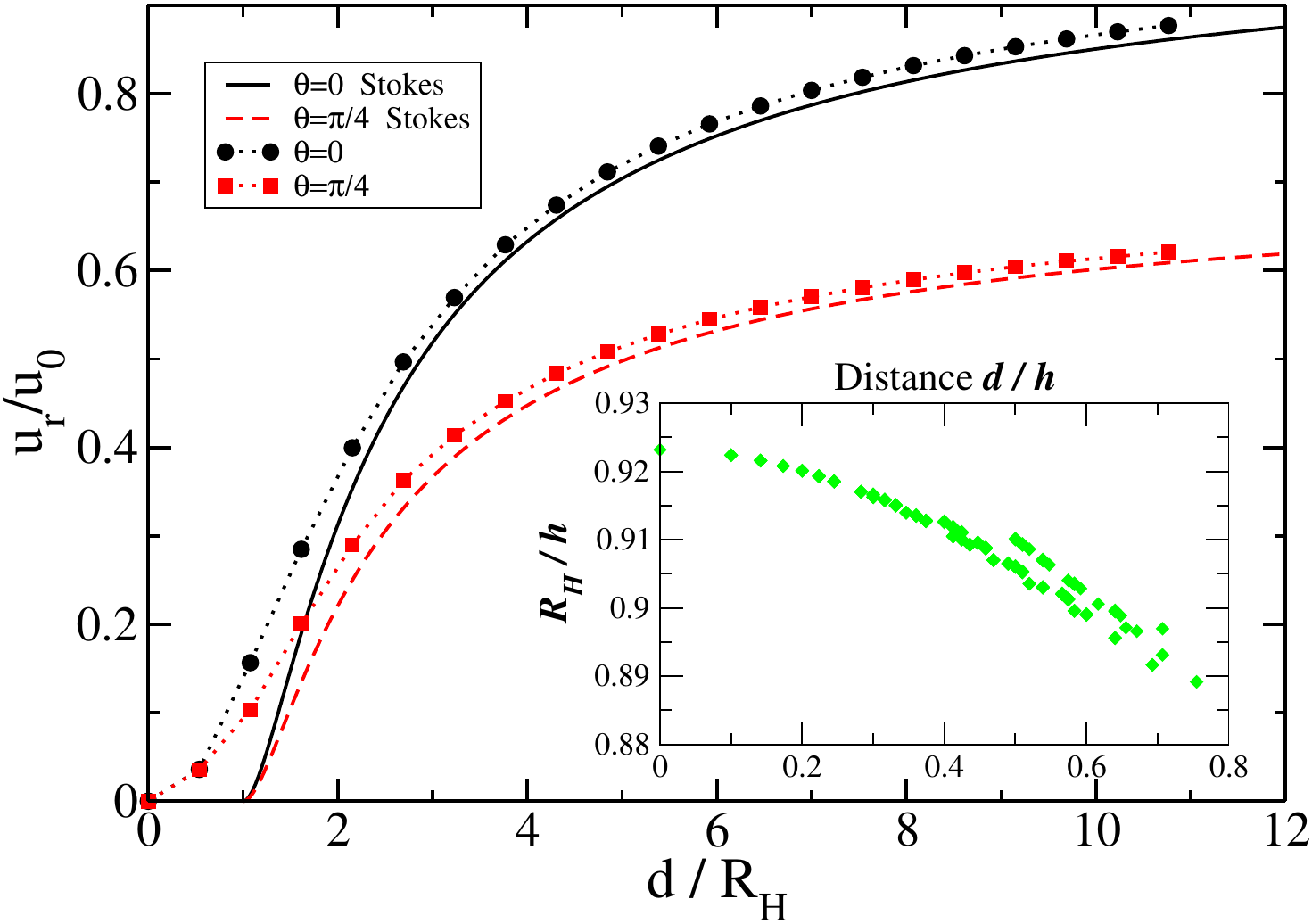}\includegraphics[width=0.49\columnwidth]{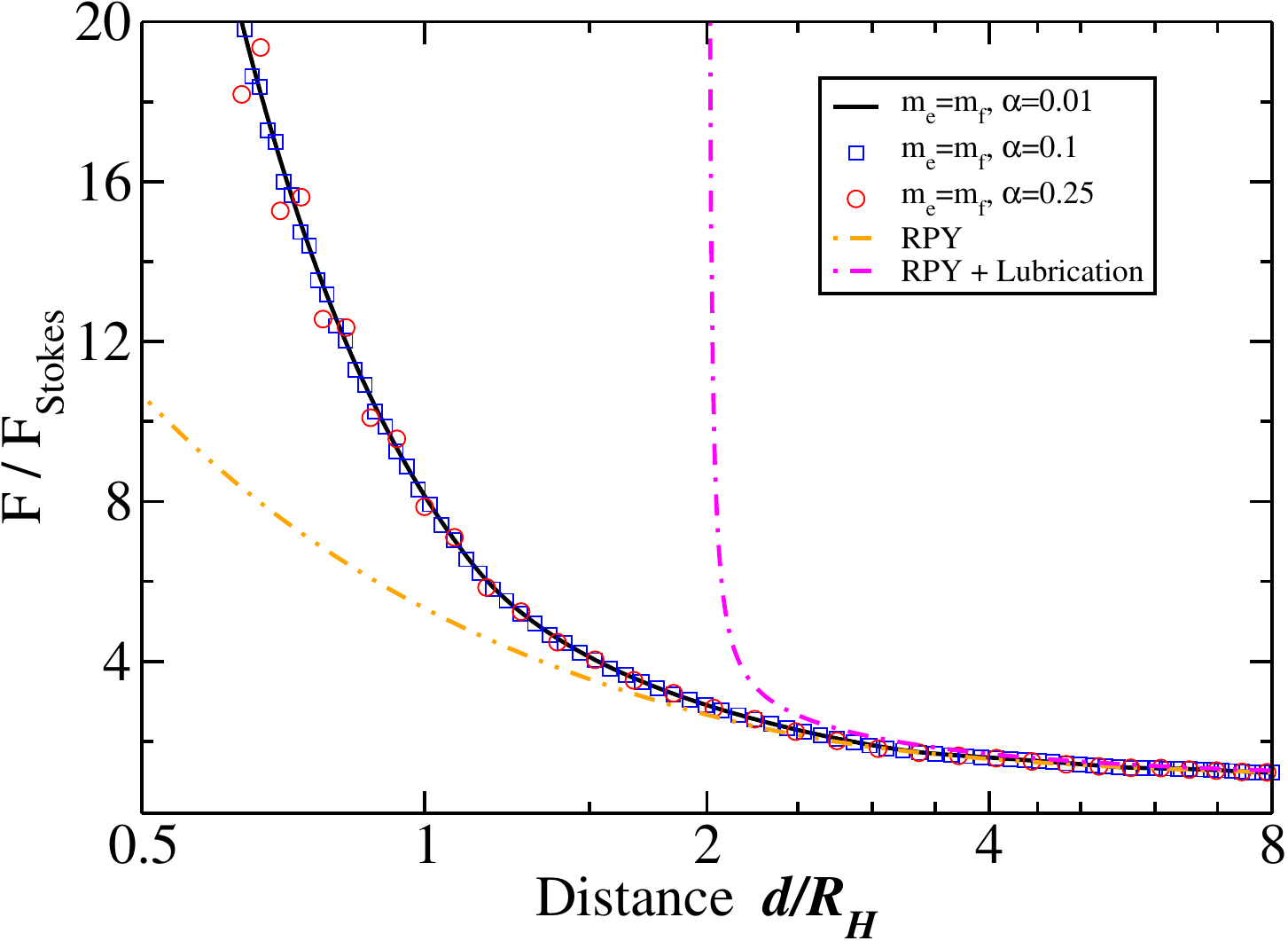}
\caption{\label{fig:Lubrication}(\emph{Left panel}) Comparison of the radial
component of the velocity in steady flow around a fixed blob to the
Stokes solution for a rigid sphere in an unbounded domain. The velocity
at a given point $\V q^{\prime}$ is obtained by using the local averaging
operator, $\V v\left(\V q^{\prime}\right)\approx\M J\left(\V q^{\prime}\right)\V v.$
The inset shows the hydrodynamic radius of the particle for flow along
the $x$ axes, as a function of the distance between it and the center
of the nearest $x$ face of the grid (location of the $x$ component
of velocity). (\emph{Right panel}) Hydrodynamic interaction force
between two particles as a function of the interparticle distance
for several time step sizes. For large distances the Stokes mobility
is recovered, and for moderate distances the next-order correction
(Rotne-Prager mobility) is recovered, independent of dynamical parameters.
At close distances a large increase in the hydrodynamic force is observed,
as in the Rotne-Prager-Yamakawa (RPY) mobility (also shown). Note
that the sharp increase in the hydrodynamic interaction at short distances
is qualitatively similar but distinct from the divergent lubrication
force observed between two rigid spheres (dashed-dotted line).}
\end{figure}

We exert a constant force density (pressure gradient) on the fluid
in a periodic domain of $128^{2}$ hydrodynamic cells, and attach
a single blob to a tether point via a stiff elastic spring. After
an initial transient, a steady state is reached in which the drag
force on the particle is balanced by the spring force. The forcing
is chosen so that the Reynolds number is kept small, $Re=2\D x\, v_{max}/\nu<0.002$.
Because of the long-ranged $r^{-1}$ decay of the flow away from the
particle, there are strong finite-size corrections that are well-known
\cite{DirectForcing_Balboa}. Taking into account these corrections
we estimate $R_{H}\approx0.91\D x$, consistent with the more careful
estimates obtained in Ref. \cite{DirectForcing_Balboa} using a compressible
flow simulation and a non-periodic domain. We emphasize again that
$R_{H}$ is not perfectly translational invariant and changes by a
couple of percent as the particle moves relative to the underlying
fluid solver grid. This is illustrated in the inset in the left panel
of Fig.\ref{fig:Lubrication}. If the Peskin four-point interpolation
function \cite{IBM_PeskinReview} is used instead of the three-point
function, a smaller variance in the hydrodynamic radius (i.e., improved
translational invariance) would be observed \cite{LB_SoftMatter_Review}.

In the left panel of Fig. \ref{fig:Lubrication} we compare the radial
component of the fluid velocity $u_{r}(r)$ along $\theta=0$ (direction
of motion of the incoming flow) and along $\theta=\pi/4$ with the
analytical solution for a solid sphere with no-slip surface in a infinite
system, as a function of the distance $d$ from the particle center.
A surprisingly good agreement is observed even for distances as small
as $d=2R_{H}$, in agreement with previous investigations for Gaussian
blobs \cite{ForceCoupling_Monopole,ForceCoupling_Stokes,Lomholt2001}.
Note, however, that there is flow penetrating the blob at distances
$d<R_{H}$, unlike a rigid sphere. Also note that similar calculations
performed using a frictional coupling \cite{LB_SoftMatter_Review}
in Ref. \cite{DirectForcing_Balboa} clearly reveal much larger penetration
of the flow into the blob, unless a rather large friction constant
is employed.

\subsubsection{Hydrodynamic Interactions}

In this section we investigate the hydrodynamic interaction force
between two particles in the deterministic setting, as done for a
compressible fluid in Ref. \cite{DirectForcing_Balboa}. In this test,
we apply a force $\V F_{0}$ on one particle toward the other particle,
and the opposite force $-\V F_{0}$ on the other particle, so that
the center of mass remains at rest. The applied force is weak so that
the Reynolds number $Re\le10^{-3}$ and the flow is in the Stokes
(steady-state) limit. As the particles approach, we measure the relative
speed of approach $v_{r}$ and compare it to the prediction of Stokes's
law, 
\begin{equation}
\frac{F}{F_{\text{Stokes}}}=-\frac{2F_{0}}{6\pi\eta R_{H}v_{r}}.
\end{equation}

In the right panel of Fig. \ref{fig:RDF} we show results for $F/F_{\text{Stokes}}$
as a function of the interparticle distance $l$. The simulation parameters
are reported in Table \ref{tableLubrication}. We performed several
simulation to verify that the results presented here are independent
of the excess mass and time step size. In the figure we compare the
results of our calculations to a theoretical calculation based on
the Rotne-Prager-Yamakawa (RPY) tensor for the mobility of a pair
of particles in a periodic system \cite{RotnePrager_Periodic}. We
see that the RPY correction correctly captures the behavior for the
blobs for distances $d\gtrsim3$. At short distances there is a strong
repulsive force between the blobs similar to the well-known ``lubrication
force'' that develops as an incompressible fluid is squeezed out
between two approaching rigid spheres. However, unlike the lubrication
force between two rigid spheres, the hydrodynamic interaction force
between two blobs does not diverge like $\left(d-2R_{H}\right)^{-1}$.
This is expected because blobs do not have a well-defined surface;
however, they are not point particles either, and they do squeeze
the incompressible fluid in-between them as they approach each other.

In fact, at short distances the hydrodynamic interaction between blobs
is similar to that for the RPY tensor (see Eq. (8) in Ref. \cite{BD_LB_Ladd}),
also shown in the right panel of Fig. \ref{fig:RDF}. An examination
of the derivation of the RPY mobility (see Eqs. (6) and (7) in Ref.
\cite{BD_LB_Ladd}) reveals that the RPY correction arises due the
Faxen term in (\ref{eq:Faxen_blob}). Therefore, the fact that we
get such good agreement between RPY and the numerical results for
blobs at larger interparticle distances shows that the Faxen radius
of our blob is very close to its hydrodynamic radius, $a_{F}\approx R_{H}\approx0.9\D x$.
Numerical investigations show that the Faxen radius of the blob (which
can be expressed in terms of the second moment of the discrete kernel
function) is translationally-invariant to within about $5\%$. Note
that using kernel functions that try to approximate a Dirac delta
function to higher accuracy will not give the correct Faxen term since
the Faxen radius of a true point particle is zero. In particular,
for the Peskin 6pt function \cite{IBM_PeskinReview} the second moment
and thus the Faxen radius is identically zero. By contrast, the 3pt
function used here gives an excellent approximation to a rigid sphere
at intermediate and large distances.

\subsection{\label{sub:Reynolds}Large Reynolds Number}

As discussed in the Introduction, we believe that the blob particle
model can be successfully used for simulations of particle-laden flows
at moderate and large particle Reynolds number. In the $\pRe\rightarrow0$
limit, we showed that the perturbative flow created by the blob particle
agrees with the Stokes solution for the steady flow around a no-slip
sphere of hydrodynamic radius $R_{H}=0.91\, h$ for the three point
kernel and a staggered grid solver \cite{DirectForcing_Balboa}. This
is consistent with other calculations utilizing a four-point kernel
and a non-staggered grid \cite{IBM_Sphere}. However it is not \emph{a
priori} clear how consistently the blob hydrodynamic behavior will
be at large $\pRe$. The drag force provides a non-trivial test because
it captures the average effect of the perturbative flow. In a previous
study \cite{DirectForcing_Balboa} we observed that the drag on the
blob is consistent with that of a rigid sphere up to $\pRe<10$. This
study was performed using a compressible flow solver, and ensuring
a low Mach number was prohibitively expensive at larger $\pRe$. Using
the incompressible (zero Mach number) solver developed here we now
study larger $\pRe$.

\begin{figure}
\begin{centering}
\includegraphics[height=7cm]{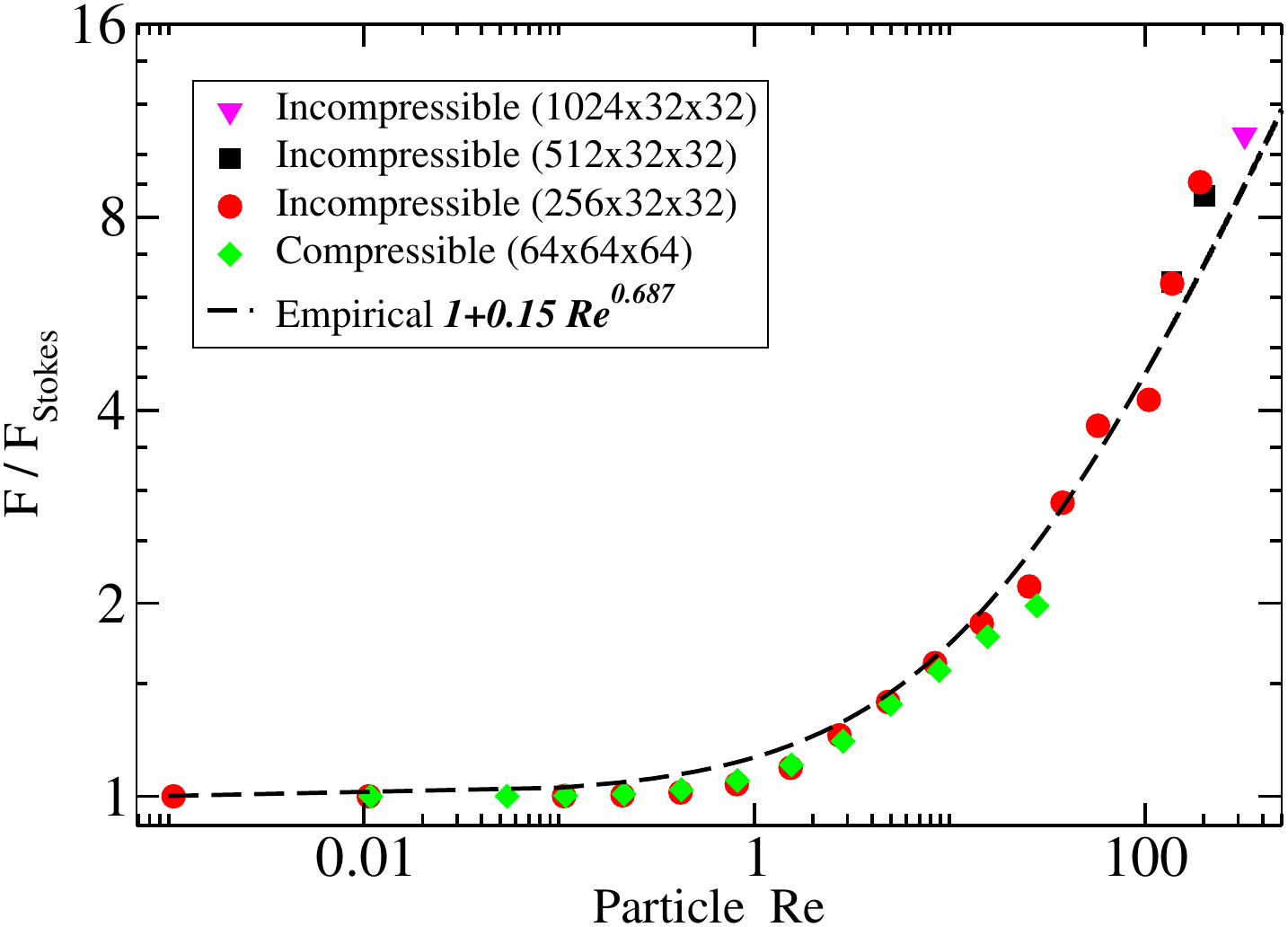}\hspace{1.25cm}\includegraphics[width=4.5cm]{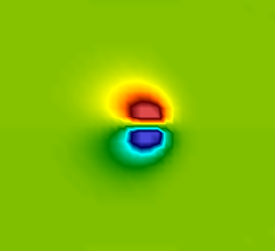}
\par\end{centering}

\begin{centering}
\includegraphics[height=4.5cm]{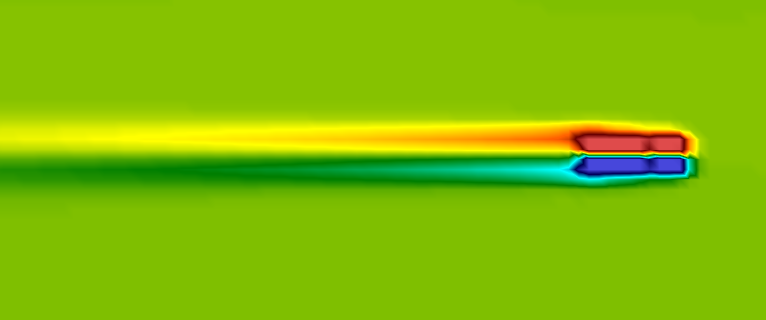}\hspace{0.5cm}\includegraphics[width=4.5cm]{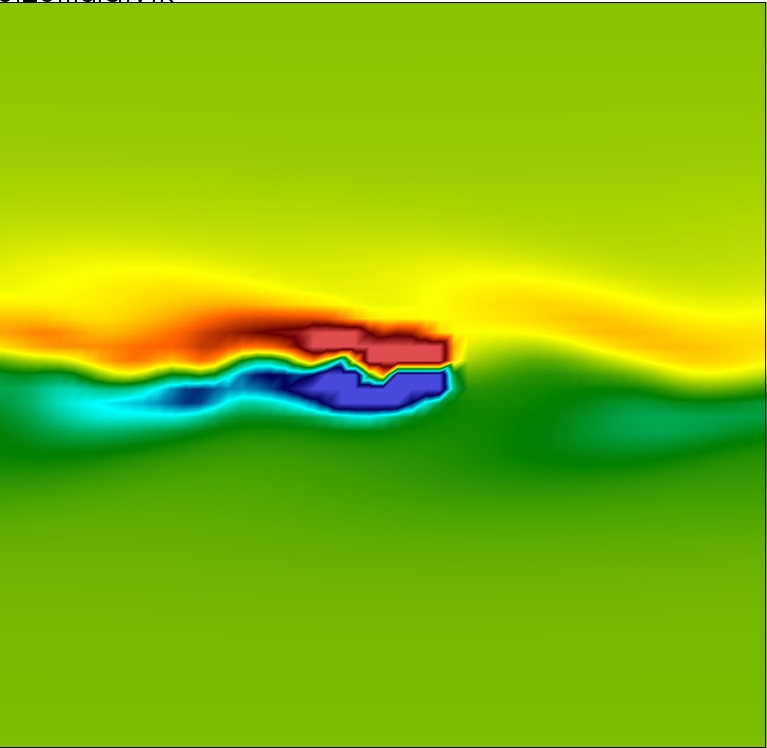}
\par\end{centering}

\centering{}\caption{\label{fig:forceReynolds}(\emph{Top left}) Drag force on a blob particle
in a periodic domain as a function of the particle Reynolds number
$\pRe=2R_{H}\,\av u/\nu$, normalized by the Stokes drag ($\pRe\rightarrow0$)
(symbols). Results for the incompressible and compressible solvers
(see Appendix \ref{AppendixCompressible} and Ref. \cite{DirectForcing_Balboa})
are compared with the empirical law for rigid no-slip spheres (line)
\cite{Clift1978}. The size of the domain box in cells is indicated
in the legend. (\emph{Top right}) Out-of-plane vorticity iso-contours
at $\pRe=1.5$ for a box of $32^{3}$ cells, and (\emph{Bottom left})
at $\pRe=137$ in a long box of $512\times32\times32$ cells. (\emph{Bottom
right}) A snapshot of an unsteady (nearly oscillatory) flow in a box
of $32^{3}$ cells at $\pRe=70$ where the particle sheds vortices
due to the interaction with the wake from its image. The simulation
parameters are as in Table \ref{tableLubrication}, but $m_{e}=0$
and advective CFL $\alpha=0.2$. Error bars are estimated to be less
than 5\%.}
\end{figure}

In the top left panel of Fig. \ref{fig:forceReynolds} we show the
drag force on a blob particle in a periodic domain as a function of
$Re$, normalized by the Stokes limit drag ($Re\rightarrow0$). We
estimate the drag coefficient by dragging the blob with a constant
applied force and measuring the average velocity $v_{0}=\av u$ along
the direction of the applied force at long times. The particle interacts
with its periodic images approximately after time $\tau_{L}=L/v_{0}$
(where $L$ is the box length in flow direction), while viscous transport
around the blob requires a time longer than $\tau_{\nu}=R_{H}^{2}/\nu$
to settle down. In order to mimic the behavior of an isolated blob
in an infinite medium, we must have $\tau_{L}/\tau_{\nu}>1$, or,
equivalently, $L/R_{H}>\mathrm{Re}_{p}$. We therefore performed the
calculations in a box of $2^{n}\times32\times32$ cells, with $2^{n}>3\pRe$;
the size of the simulation is indicated in the legend in the figure.
In the top left panel of Fig. \ref{fig:forceReynolds} we compare
the numerical results with the empirical law for the drag on a rigid
sphere with a no-slip surface \cite{Clift1978}. The agreement is
remarkably good over the studied range $0\leq\pRe\leq324$.

Encouraged by this result we now briefly analyze the structure of
the perturbative flow around the blob, and defer a more detailed study
for future work. In the remaining panels in Fig. \ref{fig:forceReynolds}
we show iso-contours of the vorticity perpendicular to the snapshots'
plane at a few values of $\pRe$. For $\pRe\simeq0$ the fully symmetric
pattern observed around the blob agrees with the Stokes solution at
distances larger than $2R_{H}$ \cite{DirectForcing_Balboa}. The
fore-and-aft symmetry of the Stokes flow becomes distorted by advection
at moderate Reynolds $\pRe\sim1$ leading to the so-called Oseen flow
\cite{Lomholt2001}, as observed in the top right panel of Fig.\ref{fig:forceReynolds}
for $\pRe=1.5$. For $20<\pRe<270$ a transition with symmetry breaking
leads to a steady axisymmetric ``double-thread'' structure \cite{Ormieres1999,Johnson1999a}.
Although we have not studied the transition in detail, our blob model
qualitatively reproduces a steady axisymmetric wake with a bifid vortex
trail, as illustrated in the bottom left panel of Fig. \ref{fig:forceReynolds}
for $\pRe=137$. This type of wake is observed above $\pRe>10$ and
its topology is maintained at least up to $\pRe\simeq300$. For flow
around a rigid sphere a Hopf transition to oscillatory flow leading
to vortex shedding takes place at $\pRe\approx270$ \cite{Ormieres1999,Johnson1999a}.
For the blob we observe small oscillations of the wake, without vortex
shedding at $\pRe>300$, indicating that some possible transition
to unsteady flow could be induced by a small perturbation. In fact,
in particle-laden flows typically $\pRe<100$ \cite{Botto2012}, and
the induction of vortex shedding due to perturbations from the wake
of other particle is a more relevant vorticity source. Using smaller
boxes, where the particle interacts with its image, we frequently
observed induced vortex shedding for $\pRe>70$, as illustrated in
the bottom right panel of Fig. \ref{fig:forceReynolds}.

It is a remarkable fact that the ICM blob minimal-resolution model
can produce wakes containing many of the features of realistic flows.
The thickness of the viscous (Oseen) layer around a sphere decreases
like $R/\pRe$ \cite{Lomholt2001}; therefore, this layer is completely
unresolved for $\pRe>1$ in our model. However, the ``local'' no-slip
constraint is able to capture the non-linear velocity-pressure coupling
which dominates the drag at large $\pRe$. A somewhat similar scenario
was observed in simulations of ultrasound-particle interaction \cite{DirectForcing_Balboa}.
In the inviscid regime (sound frequency faster than $\nu/R_{H}^{2}$),
an excellent agreement with the theory was observed even in cases
where the sound-viscous boundary layer around the blob was unresolved.

\section{\label{sec:Conclusions}Conclusions}

In this work we described a bidirectional coupling between a point-like
``blob'' particle and an incompressible fluctuating fluid, building
on prior work by some of us in the compressible setting \cite{DirectForcing_Balboa}.
At the continuum level, the proposed model includes inertial and stochastic
effects in a consistent manner, ensuring fluctuation-dissipation balance
and independence of equilibrium thermodynamic properties on dynamical
parameters. We constructed a second-order spatio-temporal discretization
that tries to preserve the properties of the continuum model as well
as possible.

Through numerical experiments, we showed that the proposed \emph{inertial
coupling method} (ICM) can consistently describe the dynamics of ``blob''
particles in fluid flow over a broad range of particle and fluid Reynolds
numbers, from Brownian motion to convective regimes. We demonstrated
that the method reproduces well-known non-trivial effects of particle
inertia on the short-time dynamics of Brownian particles, while also
capturing the long-time (Brownian) diffusive dynamics and the associated
equilibrium distribution correctly. Remarkably, we found the minimally-resolved
blob model to reproduce non-trivial features of turbulent flow around
a rigid sphere, including the non-Stokes drag at moderately large
Reynolds numbers $\mathrm{Re}_{P}<300$ and interactions between particle
wakes. As such we believe that the method presented here can be applied
to model the dynamics of dilute and semi-dilute colloidal suspensions
and polymeric fluids over a broader range of conditions than existing
methods \cite{Ermak1978,BrownianDynamics_OrderN,BrownianDynamics_OrderNlogN,LB_SoftMatter_Review,SELM}.

The algorithm we described here was specifically optimized for periodic
boundary conditions. In particular, we constructed a semi-implicit
temporal integrator that only requires a few applications of the FFT
algorithm per time step. This enabled us to implement the algorithm
on GPU platforms, achieving excellent performance with little development
effort. However, this simplicity was not without a cost. Firstly,
we had to make several approximations in order to avoid iterative
linear solvers and use a fixed number of FFTs per time step. Notably,
we had to assume that particles were far away from other particles
compared to their size (dilute conditions). Secondly, in many problems
of interest the fluid is confined in non-periodic geometries such
as channels and periodic boundary conditions are not appropriate.
It is not difficult, at least in principle, to adopt our algorithm
to non-periodic geometries and to dense collections of blobs. This
requires developing specialized preconditioned Krylov linear solvers
for solving the ``inertial'' Stokes problem (\ref{eq:v_implicit_system})
in non-periodic geometries, similar to the fluid-only solver developed
by one of us \cite{NonProjection_Griffith} and implemented in the
IBAMR framework \cite{IBAMR}. In future work we will explore effective
preconditioners for solving (\ref{eq:v_implicit_system}) using a
Krylov method.

The temporal integrator algorithm we developed in this work can accurately
resolve the inertial, viscous, and fluctuating dynamics of particles
immersed in an incompressible fluid if the viscous and advective CFL
numbers are less than unity. Some further work is required to tackle
the large \emph{separation of time scales} present in many situations
of practical interest. For example, in the Brownian dynamics limit
there is an increasing separation of scales between the particle movement,
the viscous damping, and the inertial dynamics. For the second-order
temporal integrator that we presented here, the case of large Schmidt
number $S_{c}\gg1$ is the most demanding in terms of computational
effort. This is because particles barely move during a single time
step, $\alpha\ll1$, and $O(S_{c})$ fluid time steps may be required
to reach the diffusive time scale for $\beta=O(1)$. For the case
of a neutrally-buoyant particle ($m_{e}=0$) and periodic boundary
conditions Atzberger \emph{et al.} \cite{StochasticImmersedBoundary}
have developed a specialized exponential integrator that tackles the
large separation of time scales that arise when $S_{c}\gg1$. Future
work will consider extending such techniques to non-periodic systems
and to the case $m_{e}\neq0$.

In ICM the fluid-particle coupling is expressed as a no-slip constraint
equating the translational particle velocity with the local fluid
velocity. This implies that only the \emph{monopole} term (\emph{stokeslet})
is included in the fluid-particle force. As a consequence the present
approach can only accurately resolve the fluid flow at distances larger
than the typical size of the particles. However, it is a remarkable
fact that with such minimal resolution permits to capture so many
hydrodynamic effects in a qualitatively, and, in some cases, quantitatively
correct way (see Sec. \ref{sec:Results} and also Ref. \cite{DirectForcing_Balboa}).
Even though lubrication flows (at low $\mathrm{Re}_{P}$) or viscous
boundary layers (at large $\mathrm{Re}_{P}$ and/or high forcing rates)
are unresolved, the locally averaged no-slip constraint proves to
be remarkably robust in capturing their essential hydrodynamics and
permits to go beyond the Stokes limit. Crucial to this success is
the fact that we employ a carefully-constructed spatial discretization,
combining nearly grid-invariant Lagrangian-Eulerian transformations,
local energy and momentum conservation, and a staggered discretization
of the incompressibility condition and the fluctuating stresses.

It is not difficult to extend our approach to also include the anti-symmetric
component of the dipole (\emph{rotlet}) stress \cite{ForceCoupling_Stokes}.
Firstly, particle rotational degrees of freedom would need to be added
to the blob description, along with an angular velocity $\V{\omega}$
and an associated excess moment of inertia. We would need to impose
an additional \emph{rotational no-slip} constraint, requiring that
the particle rotate with the locally-averaged angular velocity of
the fluid, $\V{\omega}=\M J\left(\grad\times\V v\right)/2,$ and distribute
the fluid-particle torque $\V{\tau}$ (Lagrange multiplier) as a force
density $\M f_{\V{\tau}}=-\grad\times\left(\M S\V{\tau}\right)/2$
in the fluid momentum equation. This type of approach has already
been employed in the deterministic context to model suspensions of
neutrally-buoyant semi-rigid rods \cite{IBM_Generalized,IBM_TwistBend}.
To our knowledge, such an approach has not yet been applied to fluctuating
hydrodynamics and the resulting rotational diffusion has not been
investigated.

An additional \emph{rigidity constraint} would be required to also
constrain the locally-averaged deformation tensor, and thus consistently
include the \emph{symmetric} components of the dipole (\emph{stresslet})
force terms, as proposed by Maxey and collaborators in the context
of the deterministic Force Coupling Method (FCM) \cite{ForceCoupling_Stokes}.
Unlike the smooth Gaussian kernels used in the FCM \cite{ForceCoupling_Stokes},
the existing Peskin kernels do not have well-behaved derivatives.
Therefore, it appears necessary to generalize the IBM kernels to enable
the local averaging (``interpolation'') of spatial derivatives in
a reasonably translationally-invariant manner. The inclusion of thermal
fluctuations of the stresslet requires careful consideration even
at the continuum level, and will be the subject of future research.

In a different spirit, the approach used here for single ``blobs''
can be extended to account for the finite extent and shape of arbitrarily-shaped
rigid bodies immersed in fluid flow. One approach that has been successfully
employed in the deterministic setting is to construct the immersed
body out of a collection of blobs constrained to move rigidly \cite{RigidIBAMR}.
Future work will consider the inclusion of thermal fluctuations in
this type of approach, and the minimal amount of resolution required
to capture the geometry of immersed particles.

At the other extreme, our method can be used to provide a coarse-grained
model for hydrodynamics at very small scales, for example, for the
Brownian motion of a small molecule suspended in a simple solvent.
At such small scales, the suspended particles do not have a fixed,
or even a well-defined shape, and there are many competing and sometimes
canceling effects: normal and tangential slip \cite{StokesEinstein_SlipBC},
breakdown of Navier-Stokes hydrodynamics, non-Gaussian fluctuations,
non-Markovian effects, etc. Our simple fluid-particle coupling model
can be used to isolate hydrodynamic from non-hydrodynamic effects
and study basic physics questions about the importance of inertia
and fluctuations on Brownian motion, going beyond the uncontrolled
approximations required by existing theoretical approaches. Notably,
preliminary investigations have shown that for small Schmidt numbers
nonlinear effects become important and lead to a non-trivial contribution
of the thermal fluctuations to the mean fluid-particle force. Specifically,
the mobility of a particle in a fluctuating fluid was found to differ
from that in a deterministic (Stokes) fluid, thus leading to a deviation
of the diffusion coefficient from the standard Stokes-Einstein prediction.
More careful investigations are required to assess how well self-consistent
mode-coupling theories can model this effect, as well as to study
the influence of particle inertia (density contrast), random slip
(see Appendix \ref{sec:AppendixLangevin}), and spatial dimensionality.
\begin{acknowledgments}
We thank Alejandro Garcia, Charles Peskin, Paul Atzberger, Eric Vanden-Eijnden,
Martin Maxey, Tony Ladd and Burkhard Dünweg for informative and inspiring
discussions. A. Donev was supported in part by the Air Force Office
of Scientific Research under grant number FA9550-12-1-0356. B. Griffith
acknowledges research support from the National Science Foundation
under awards OCI 1047734 and DMS 1016554. R. Delgado-Buscalioni and
F. Balboa acknowledge funding from the Spanish government FIS2010-22047-C05
and from the Comunidad de Madrid MODELICO-CM (S2009/ESP-1691). Collaboration
between A. Donev and R. Delgado-Buscalioni was fostered at the Kavli
Institute for Theoretical Physics in Santa Barbara, California, and
supported in part by the National Science Foundation under Grant No.
NSF PHY05-51164.
\end{acknowledgments}
\begin{appendix}

\part*{Appendix}

\section{\label{AppendixPeriodicBCs}Periodic Boundary Conditions}

The pressure-free fluid-only equation (\ref{eq:fluid_only_no_p})
can formally be written in a form suitable for direct application
of standard numerical methods for integrating initial value problems,
\begin{equation}
\partial_{t}\V v=\M{\rho}_{\text{eff}}^{-1}\left(\M{\mathcal{P}}\V f+\M{\mathcal{P}}\M S\M F\right),\label{eq:dt_v_rho_inv}
\end{equation}
although this form is only useful if one can actually compute the
action of the inverse inertia tensor $\M{\rho}_{\text{eff}}^{-1}$.
It turns out that this is possible for periodic systems.

To see this, we expand $\M{\rho}_{\text{eff}}^{-1}$ into a formal
series,
\begin{equation}
\M{\rho}_{\text{eff}}^{-1}=\rho^{-1}\left[\M I-m_{e}\rho^{-1}\M{\mathcal{P}}\M S\V J\M{\mathcal{P}}+\left(m_{e}\rho^{-1}\M{\mathcal{P}}\M S\V J\M{\mathcal{P}}\right)^{2}-\left(m_{e}\rho^{-1}\M{\mathcal{P}}\M S\V J\M{\mathcal{P}}\right)^{3}+\dots\right].\label{eq:rho_inv_series}
\end{equation}
Observe that $\left(\M{\mathcal{P}}\M S\V J\M{\mathcal{P}}\right)^{n}=\M{\mathcal{P}}\M S\left(\V J\M{\mathcal{P}}\M S\right)^{n-1}\V J\M{\mathcal{P}}$
involves the $\left(n-1\right)$-st power of the $d\times d$ matrix
$\D{\widetilde{\V V}}^{-1}=\M J\M{\mathcal{P}}\M S$, where we made
use of the fact that $\M{\mathcal{P}}^{2}=\M{\mathcal{P}}$. Recall
from (\ref{eq:dV_JS}) that $\M J\M S=\D V^{-1}\M I$ is a related
to the inverse of the particle volume, which is independent of the
position of the particle for the particular kernel function used herein.
In principle, the matrix $\D{\widetilde{\V V}}^{-1}$ could depend
on the position of the particle because of the appearance of $\M{\mathcal{P}}$,
which implicitly encodes the boundary conditions. However, periodic
systems are translationally invariant and therefore $\D{\widetilde{\V V}}^{-1}$
cannot depend on the position of the particle and is simply a constant
$d\times d$ matrix.

By performing a relatively-straightforward calculation in Fourier
space it is possible to show that for periodic systems much larger
than the kernel extent $\D{\widetilde{\V V}}^{-1}$ is simply a multiple
of the identity matrix,
\begin{equation}
\D{\widetilde{\V V}}^{-1}=\M J\M{\mathcal{P}}\M S=\frac{d-1}{d}\D V^{-1}\,\M I.\label{eq:JPS}
\end{equation}
The prefactor $(d-1)/d$ in the relation accounts for the elimination
of the longitudinal (compressible) velocity mode by the projection
operator. We can therefore define the \emph{effective} particle volume
accounting for incompressibility to be
\[
\D{\widetilde{V}}=\frac{d}{d-1}\D V,
\]
and accordingly, define the effective mass of the fluid dragged with
the particle to be $\tilde{m}_{f}=\rho\D{\widetilde{V}}=d\, m_{f}/\left(d-1\right)$,
and the effective particle inertia to be $\tilde{m}=\tilde{m}_{f}+m_{e}$.
In three dimensions, the added fluid mass due to incompressibility
is $m_{f}/2$, which is a well-known result for rigid spheres immersed
in an incompressible fluid \cite{Landau:Fluid,VACF_Langevin}.

Note that for periodic systems where some of the dimensions of the
unit cell are comparable to the kernel width (\ref{eq:JPS}) is only
an approximation. Notably, for a three-dimensional periodic box of
shape $L_{x}\times L_{y}\times L_{z}$, if $L_{z}\ll a$ the value
of $\D{\widetilde{\V V}}^{-1}$ converges to the two-dimensional result,
\begin{equation}
\D{\widetilde{\V V}}_{\text{2D}}^{-1}=\D V^{-1}\left[\begin{array}{ccc}
\frac{1}{2} & 0 & 0\\
0 & \frac{1}{2} & 0\\
0 & 0 & 1
\end{array}\right],\label{eq:dV_2D}
\end{equation}
rather than the three-dimensional $\D{\widetilde{\V V}}_{\text{3D}}^{-1}=\left(2\D V^{-1}/3\right)\,\M I$.

Using (\ref{eq:JPS}) it is possible to simplify the infinite series
(\ref{eq:rho_inv_series}) and obtain the important result for a single
particle immersed in a periodic fluid,
\begin{equation}
\M{\rho}_{\text{eff}}^{-1}=\left(\rho\M I+m_{e}\M{\mathcal{P}}\M S\V J\M{\mathcal{P}}\right)^{-1}=\rho^{-1}\left(\M I-\frac{m_{e}\D{\widetilde{V}}}{\tilde{m}}\M{\mathcal{P}}\M S\V J\M{\mathcal{P}}\right).\label{eq:rho_eff_inv_incomp}
\end{equation}
Using this result we can rewrite (\ref{eq:dt_v_rho_inv}) in the simple
form
\begin{equation}
\rho\partial_{t}\V v=\M{\mathcal{P}}\left(\M I-\frac{m_{e}\D{\widetilde{V}}}{\tilde{m}}\M S\V J\right)\M{\mathcal{P}}\V f+\frac{\tilde{m}_{f}}{\tilde{m}}\M{\mathcal{P}}\M S\M F,\label{eq:fluid_only_periodic}
\end{equation}
which is useful for analysis and for numerical approximations.

It is important to point out, however, that when there are many particles
present, there is no simple formula for $\M{\rho}_{\text{eff}}^{-1}$.
That is, we cannot just add a summation over all particles in (\ref{eq:rho_eff_inv_incomp}).
This is because there are cross-terms between two particles $i$ and
$j$ in the infinite series (\ref{eq:rho_inv_series}) involving the
operator $\V J_{i}\M{\mathcal{P}}\M S_{j}$. If the particles are
\emph{not} overlapping, meaning that the kernel functions of the different
particles have disjoint support, then $\V J_{i}\M S_{j}=\M 0$, however,
this is not true for incompressible flow because the projection $\M{\mathcal{P}}$
is a non-local operator involving the inverse Laplacian. Nevertheless,
the cross terms decay fast as particles become well-separated from
each other. Specifically, theoretical calculations suggest that to
leading order $\V J_{i}\M{\mathcal{P}}\M S_{j}$ decays with the distance
between the two particles like dipole and quadrupole terms, and is
thus expected to be very small in semi-dilute suspensions \cite{VACF_Suspension}.
In many problems of practical interest there are repulsive forces
between the particles that will keep them from coming close to each
other, and the approximation $\V J_{i}\M{\mathcal{P}}\M S_{j}\approx\left(\D{\widetilde{V}}\right)^{-1}\delta_{ij}\,\M I$,
\begin{equation}
\M{\rho}_{\text{eff}}^{-1}\approx\rho^{-1}\left[\M I-\sum_{i}\frac{\left(m_{e}\right)_{i}\D{\widetilde{V}}_{i}}{\tilde{m}_{i}}\M{\mathcal{P}}\M S_{i}\V J_{i}\M{\mathcal{P}}\right]\label{eq:rho_eff_inv_approx}
\end{equation}
will be quite accurate even for multiparticle systems. We investigate
the accuracy of the approximation $\V J_{i}\M{\mathcal{P}}\M S_{j}=\M 0$
for $i\neq j$ numerically in Section \ref{sub:TranslationalInvariance}.

\section{\label{sec:AppendixLangevin}Fluctuation-Dissipation Balance}

In this Appendix we demonstrate that the coupled fluid-particle equations
written in the form (\ref{eq:fluid_only_no_p})
\begin{eqnarray}
\partial_{t}\V v & = & \M{\rho}_{\text{eff}}^{-1}\M{\mathcal{P}}\left\{ -\left[\rho\left(\V v\cdot\grad\right)+m_{e}\M S\M J\left(\V v\cdot\frac{\partial}{\partial\V q}\M J\right)\right]\V v+\M S\M F\right.\nonumber \\
 & + & \left.\eta\grad^{2}\V v+\grad\cdot\left[\left(k_{B}T\eta\right)^{\frac{1}{2}}\left(\M{\mathcal{W}}+\M{\mathcal{W}}^{T}\right)\right]\right\} \label{eq:dv_t_FDB}\\
\frac{d\V q}{dt} & = & \M J\V v,\label{eq:dq_t_FDB}
\end{eqnarray}
obey fluctuation-dissipation balance with respect to the Gibbs-Boltzmann
distribution with coarse-grained free energy
\[
H\left(\V v,\V q\right)=\frac{1}{2}\int\M v^{T}\M{\rho}_{\text{eff}}\V v\, d\V r+U\left(\V q\right),
\]
where the effective fluid inertia operator $\M{\rho}_{\text{eff}}$
is given in (\ref{eq:rho_eff_one_particle}). The calculations here
will follow the techniques described in detail in Ref. \cite{OttingerBook}
(ignoring boundary terms), and are purely formal in the continuum
(infinite-dimensional) setting. More precisely, the equations we write
should really be interpreted as a short-hand notation for a spatially-discretized
system in which there is a finite number of degrees of freedom \cite{SELM}.

It is well-known \cite{GrabertBook,OttingerBook} that within an isothermal
and Markovian approximation the generic form of evolution equations
for a set of macroscopic variables $\V x$ has the Ito Langevin form
\begin{equation}
\partial_{t}\V x=-\M N\left(\V x\right)\frac{\partial H}{\partial\V x}+\left(2k_{B}T\right)^{\frac{1}{2}}\M B\left(\V x\right)\M W(t)+\left(k_{B}T\right)\frac{\partial}{\partial\V x}\cdot\M N^{\star}\left(\V x\right),\label{eq:x_t_general}
\end{equation}
where $\M W(t)$ denotes a collection of independent white noise processes,
star denotes an adjoint, and $\left(\partial_{\V x}\cdot\M N^{\star}\right)_{k}=\partial N_{kj}/\partial x_{j}$
in indicial notation. We will suppress the explicit dependence on
$\V x$ and usually write the \emph{mobility operator} as $\M N\equiv\M N\left(\V x\right)$.
The fluctuation dissipation balance is contained in the relation
\[
\M B\M B^{\star}=\M M=\frac{1}{2}\left(\M N+\M N^{\star}\right),
\]
and the last term in (\ref{eq:x_t_general}) is a ``thermal'' drift
which ensures that the dynamics obeys detailed-balance (time-reversibility)
at equilibrium with respect the Gibbs-Boltzmann distribution $Z^{-1}\exp\left[-H\left(\V x\right)/k_{B}T\right]$.

In our case, the coarse-grained variables are $\V x=\left(\V v,\V q\right)$
and the thermodynamic driving forces are given by the functional and
partial derivatives

\begin{eqnarray}
\frac{\partial H}{\partial\V v} & = & \M{\rho}_{\text{eff}}\V v\label{eq:dH_dv}\\
\frac{\partial H}{\partial\V q} & = & -\V F\left(\V q\right)+m_{e}\M J\left(\V v\cdot\frac{\partial}{\partial\V q}\M J\right)\V v.\label{eq:dH_dq}
\end{eqnarray}
Note that the kinetic term $m_{e}\M S\M J\left(\V v\cdot\frac{\partial}{\partial\V q}\M J\right)\V v$
appearing in (\ref{eq:dv_t_FDB}) term follows from $\partial H/\partial\V q$,
and therefore the approximation $\M a_{\M J}\approx\V 0$ in (\ref{eq:a_J_def})
is not consistent in general.

\subsection{Mobility Operator}

The mobility operator that gives (\ref{eq:dv_t_FDB},\ref{eq:dq_t_FDB})
from the thermodynamic driving forces (\ref{eq:dH_dv},\ref{eq:dH_dq})
can be symbolically written in the block form
\[
\M N=\left[\begin{array}{cc}
\M{\rho}_{\text{eff}}^{-1}\M N_{\text{NS}}\M{\rho}_{\text{eff}}^{-1} & \M{\rho}_{\text{eff}}^{-1}\M{\mathcal{P}}\M S\\
-\M J\M{\mathcal{P}}\M{\rho}_{\text{eff}}^{-1} & \M N_{\text{BD}}
\end{array}\right],\quad\M B=\left[\begin{array}{c}
\M{\rho}_{\text{eff}}^{-1}\M B_{\text{NS}}\\
\M B_{\text{BD}}
\end{array}\right]
\]
where the subscript $\text{NS}$ denotes the corresponding operators
for the fluctuating Navier-Stokes equations without any immersed particles
\cite{HamiltonianFluid}. In the form of the equations that we employ
$\M N_{\text{BD}}=\M B_{\text{BD}}=\M 0$. Fluctuation-dissipation
balance then follows from the corresponding property of the pure fluid
equations, $\M B_{\text{NS}}\M B_{\text{NS}}^{\star}=\left(\M N_{\text{NS}}+\M N_{\text{NS}}^{\star}\right)/2$,
since there is no dissipative terms in the particle equation. The
fluid-particle coupling contribution is non-dissipative or skew-adjoint,
which follows from the antisymmetry between to the lower left and
upper right blocks in $\M N$.

It is, however, consistent with the general framework of augmented
Langevin descriptions and fluctuation-dissipation balance to allow
for a nonzero $\M B_{\text{BD}}$ with $\M N_{\text{BD}}=\M B_{\text{BD}}\M B_{\text{BD}}^{\star}$.
This does not affect the fluid equation but changes the particle equation.
For example, the simple choice
\[
\M B_{\text{BD}}=\sqrt{\zeta}\,\M I,\quad\M N_{\text{BD}}=\zeta\,\M I,
\]
leads to a modified equation of motion for the particle,
\[
\frac{d\V q}{dt}=\M J\left[\M I-\zeta m_{e}\left(\V v\cdot\frac{\partial}{\partial\V q}\M J\right)\right]\V v+\zeta\V F\left(\V q\right)+\sqrt{2\zeta k_{B}T}\,\V{\mathcal{W}}_{\V q},
\]
where $\V{\mathcal{W}}_{\V q}\left(t\right)$ is a white-noise ``random
slip''. If $\V v=\V 0$ the above equation is recognized as the usual
equation of Brownian dynamics with friction coefficient $\zeta^{-1}$
and mobility $\zeta$. It is therefore expected that adding a random
slip component would increase the diffusion coefficient of the particle
by $\zeta k_{B}T$. This can be used to tune the diffusion coefficient
to some target (e.g., experimental) value. In this work, we fix $\zeta=0$,
and in future work we will consider second-order algorithms for $\zeta>0$
for the case of a neutrally buoyant particle, $m_{e}=0$.

\subsection{\label{sub:ThermalDrift}Thermal Drift}

Complications arise because of the presence of the last term in (\ref{eq:x_t_general})
(proportional to $k_{B}T$), which is non-zero because $\M J$ and
$\M S$ depend on $\V q$. This ``thermal drift'' term arises because
$\V u$ is eliminated from the description as a variable adiabatically-slaved
to $\V v$, neglecting the fact that the particle position fluctuates
rapidly due to the fluctuations in $\V u$. This term was identified
in Ref. \cite{SELM} as missing from the formulation of the Stochastic
Immersed Boundary Method \cite{StochasticImmersedBoundary}, which
is equivalent to our formulation with $m_{e}=0$. The missing contribution
is an extra term in the velocity equation (\ref{eq:dv_t_FDB}) of
the form
\[
\V f_{\text{th}}=\left(k_{B}T\right)\frac{\partial}{\partial\V q}\cdot\left(\M J\M{\mathcal{P}}\M{\rho}_{\text{eff}}^{-1}\right).
\]
One can, in principle, numerically evaluate this term without requiring
any derivatives by using ``random finite-differences'' \cite{MultiscaleIntegrators}
via the identity%
{} 
\[
\frac{\partial}{\partial\V q}\cdot\left(\M J\M{\mathcal{P}}\M{\rho}_{\text{eff}}^{-1}\right)=\lim_{\epsilon\rightarrow0}\,\epsilon^{-2}\av{\left[\M{\rho}_{\text{eff}}^{-1}\left(\V q+\D{\V q}\right)\M{\mathcal{P}}\M S\left(\V q+\D{\V q}\right)\right]\D{\V q}-\left[\M{\rho}_{\text{eff}}^{-1}\left(\V q\right)\M{\mathcal{P}}\M S\left(\V q\right)\right]\D{\V q}}_{\D{\V q}}.
\]
where the expectation value is with respect to the small random particle
displacement $\D{\V q}=\epsilon\M W_{\V q}$, and the components of
$\M W_{\V q}$ are independent standard normal variates.

For periodic boundaries, using (\ref{eq:rho_eff_inv_incomp}) we can
simplify, for any $\V q$,
\[
\left[\M{\rho}_{\text{eff}}^{-1}\left(\V q\right)\M{\mathcal{P}}\M S\left(\V q\right)\right]\D{\V q}=\frac{\tilde{m}_{f}}{\tilde{m}}\M{\mathcal{P}}\left[\M S\left(\V q\right)\right]\D{\V q},
\]
and therefore the thermal drift%
\begin{equation}
\V f_{\text{th}}=\left(k_{B}T\right)\frac{\tilde{m}_{f}}{\tilde{m}}\M{\mathcal{P}}\,\lim_{\epsilon\rightarrow0}\,\epsilon^{-2}\av{\left[\M S\left(\V q+\D{\V q}\right)-\M S\left(\V q\right)\right]\D{\V q}}_{\D{\V q}}=-\frac{\tilde{m}_{f}}{\tilde{m}}\M{\mathcal{P}}\grad\M S\left(k_{B}T\right)=\V 0\label{eq:f_th}
\end{equation}
vanishes as the projection of a gradient of a scalar for incompressible
flow.  Observe that we can interpret $\V f_{\text{th}}$ as the divergence
of an additional ``thermal'' contribution to the kinetic stress,
\begin{equation}
\M{\sigma}_{\text{kin}}+\M{\sigma}_{\text{th}}=-\rho\V v\V v^{T}-\M S\left(m_{e}\V u\V u^{T}+\frac{\tilde{m}_{f}}{\tilde{m}}k_{B}T\right)\M I.\label{eq:sigma_kin-1}
\end{equation}
In the limit of large Schmidt number, the particle positions evolves
much slower than the momenta (and thus also $\V u$), and therefore
the temporal average of the second term gives an average ``themal''
contribution to the pressure
\[
\pi_{\text{th}}=\M S\left(\frac{m_{e}}{d}\av{\V u^{T}\V u}+\frac{\tilde{m}_{f}}{\tilde{m}}k_{B}T\right)=\left(\frac{m_{e}}{\tilde{m}}+\frac{\tilde{m}_{f}}{\tilde{m}}\right)\M S\left(k_{B}T\right)=\M S\left(k_{B}T\right),
\]
where we used the equipartition result (\ref{eq:m_eff_incomp}). We
see that $\pi_{\text{th}}$ has the physical interpretation as an
osmotic pressure contribution arising from the thermal motion of the
suspended particle, even when excess particle inertia is present.

Note that some of the mathematical manipulations used above rely on
continuum identities which are not necessarily obeyed by the discrete
operators. In particular, for the spatial discretization that we employ,
the discrete vector field $\V w=\frac{\partial}{\partial\V q}\M S$
is not equal to a discrete gradient of a scalar. However, numerical
observations suggest that it is very close to a discrete gradient
of a scalar field, in the sense that application of the discrete projection
reduces the magnitude by several orders, $\norm{\M{\P}\V w}\ll\norm{\V w}$,
regardless of the position of the particle relative to the fluid grid.
This is one more case in which we find that the Peskin discrete local
averaging and local spreading operators closely mimic the properties
of the continuum operators even though they were not specifically
designed with the staggered-grid discrete projection $\M{\P}$ in
mind.

Nevertheless, maintaining strict discrete fluctuation-dissipation
balance in the semi-discrete equations requires keeping $\V f_{\text{th}}$
in the momentum balance equation. Note that the gradient in $\grad\M S$
is a \emph{continuum} rather than a discrete gradient. There are two
ways to implement this gradient numerically. The first is to differentiate
the Peskin kernel and spread a force $-\left(k_{B}T\right)\tilde{m}_{f}/\tilde{m}$
using the derivative of the kernel instead of the kernel itself, as
we do in the calculations reported below. The second is to use a random
finite difference \cite{MultiscaleIntegrators} and numerically obtain
the required derivative \emph{in expectation} by adding a term
\begin{equation}
\frac{\tilde{m}_{f}}{\tilde{m}}\,\frac{k_{B}T}{\epsilon}\left[\M S\left(\V q+\epsilon\M W_{\V q}\right)-\M S\left(\V q\right)\right]\M W_{\V q}\quad\mbox{ or }\quad-\left(k_{B}T\right)\frac{\tilde{m}_{f}}{\tilde{m}}\V{\partial}\delta_{a}\left(\V q-\V r\right)\label{eq:thermal_drift}
\end{equation}
to the momentum balance equation, where $\M W_{\V q}$ is generated
independently at each time step, and $\V{\partial}\delta_{a}$ is
the gradient of the kernel.

\begin{figure*}
\begin{centering}
\includegraphics[width=0.49\textwidth]{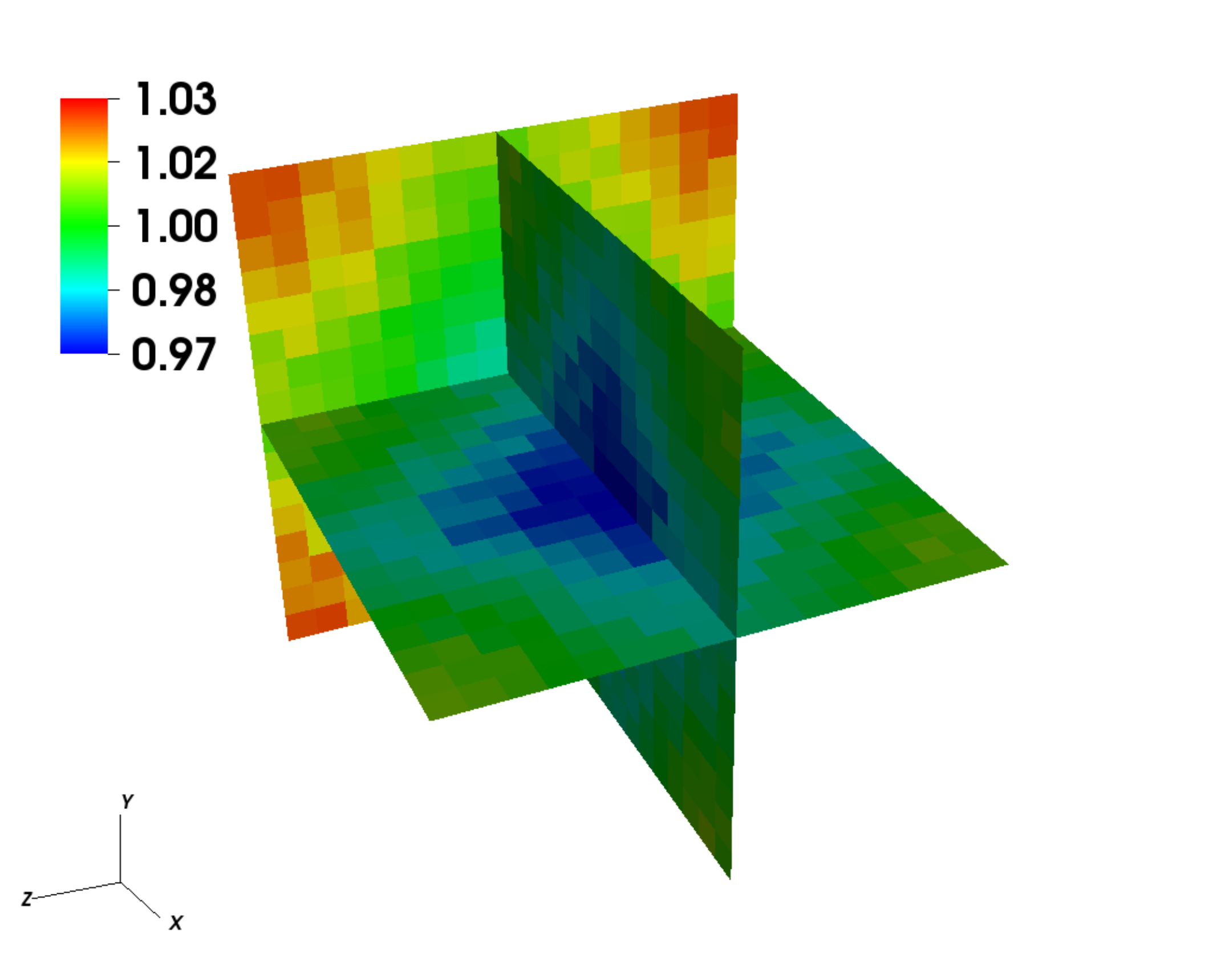}\includegraphics[width=0.49\textwidth]{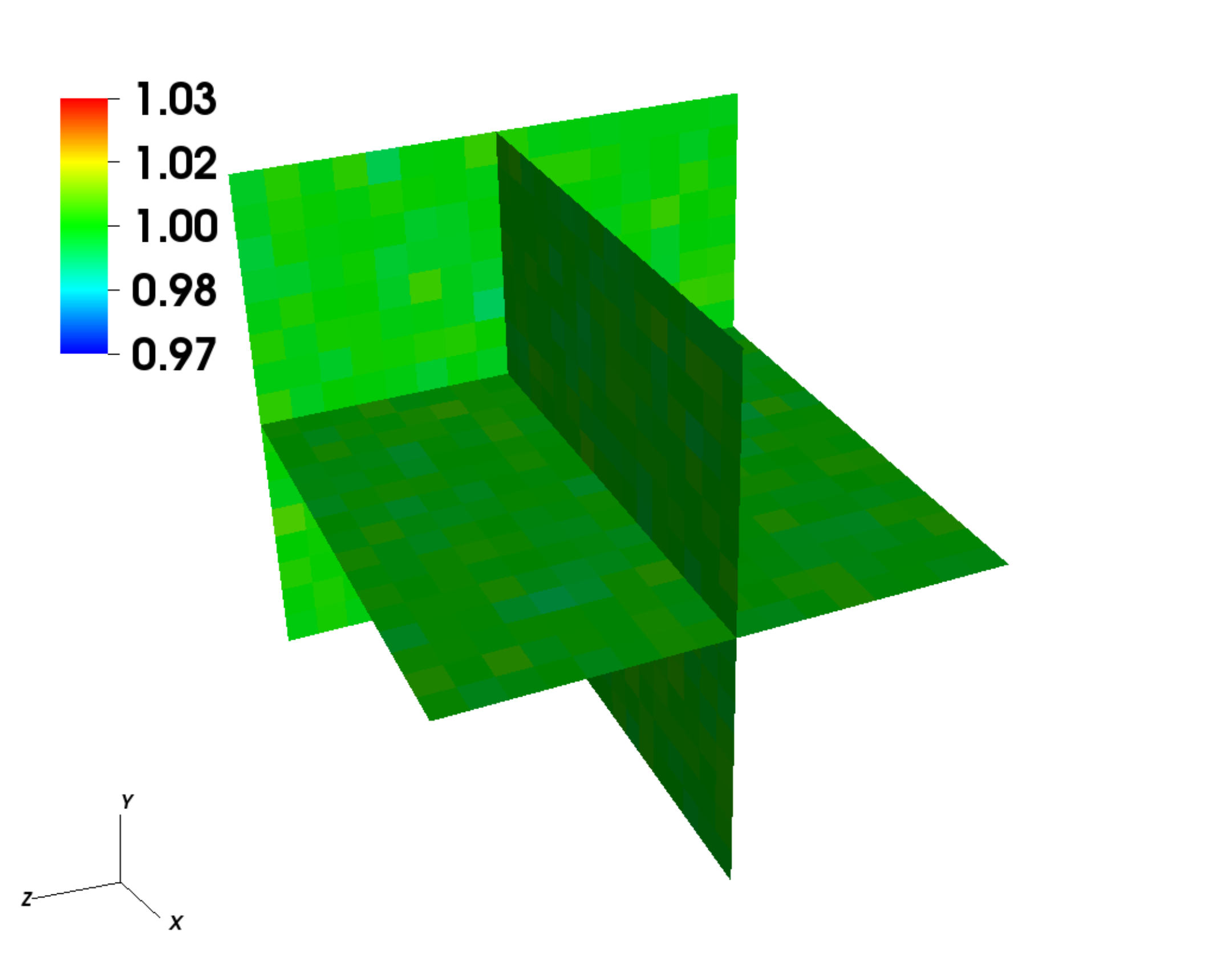}
\par\end{centering}

\caption{\label{fig:DriftBias}Histogram of the equilibrium distribution of
the position a particle freely diffusing at thermodynamic equilibrium
in a translationally-invariant system, without including the drift
$\V f_{\text{th}}$ (left) and including it (right). The distribution
is projected on one grid cell by averaging over all cells in a periodic
grid of $16^{3}$ cells, and averaging over 100 non-interacting neutrally-buyuoant
particles ($m_{e}=0$). A small time step size corresponding to viscous
CFL number $\beta=0.05$ is used in order to minimize time discretization
artifacts.}

\end{figure*}

To demonstrate that it is necessary to include (\ref{eq:thermal_drift})
in the discrete setting, we investigate the equilibrium distribution
of the positions of a number of non-interacting particles (i.e., an
ideal gas of particles) in a periodic domain. Due to the translational
invariance of the problem we know that the particle positions should
be uniformly distributed through space. However, the presence of the
fluid grid breaks translational invariance and if discrete fluctuation-dissipation
balance is not strictly obeyed grid artifacts can appear in the solution.
In Fig. \ref{fig:DriftBias} we show a histogram of the positions
of a freely-diffusing particle relative to one grid cell (by grid
translational invariance all cells and particles are equivalent).
We clearly see that when $\V f_{\text{th}}$ is omitted the particle
spends more time near the corners of the grid cell (nodes of the grid)
than near the center of the grid cell, by about $6\%$ for these parameters,
Including the thermal drift forcing in the momentum equation eliminates
these artifacts to within statistical and temporal integration errors.
In this test we use neutrally-buyuoant particles because the particles
do not interact via a steric repulsion so the approximation (\ref{eq:rho_eff_inv_approx})
fails and the no-slip condition is not enforced to sufficient accuracy
by the algorithm developed here. We note that similar results are
obtained for the compressible fluid equations, solved using the algorithm
described in Appendix \ref{AppendixCompressible}.

\section{\label{sec:VelocityCorrection}Semi-Implicit Inertial Velocity Correction}

In this Appendix we explain how we solve the linear system (\ref{eq:unperturbed_v},\ref{eq:sys_dv_eq},\ref{eq:u_np1_eq},\ref{eq:no_slip_split})
using only FFT-based linear subsolvers.

First, we solve (\ref{eq:unperturbed_v}) for $\tilde{\V v}^{n+1}$
and $\tilde{\pi}^{n+\frac{1}{2}}$ using a projection algorithm and
FFTs as explained, for example, in Ref. \cite{IBM_Viscoelasticity}.
Then, we eliminate $\V u^{n+1}$ in (\ref{eq:u_np1_eq}) using the
no-slip constraint (\ref{eq:no_slip_split}) to obtain 
\begin{equation}
\D t\,\left(\V{\lambda}^{n+\frac{1}{2}}+\V F^{n+\frac{1}{2}}\right)=m_{e}\left[\M J\left(\tilde{\V v}^{n+1}+\D{\V v}^{n+\frac{1}{2}}\right)+\D{\V u}^{n+\frac{1}{2}}-\V u^{n}\right].\label{eq:lambda_semi_implicit}
\end{equation}
Substituting this expression into (\ref{eq:sys_dv_eq}), we get the
fluid correction equation
\begin{equation}
\left[\left(\rho\V I-\frac{\D t}{2}\eta\M L\right)+m_{e}\M S\M J\right]\D{\V v}^{n+\frac{1}{2}}+\D t\,\grad\left(\D{\pi}^{n+\frac{1}{2}}\right)=m_{e}\M S\left(\V u^{n}-\M J\tilde{\V v}^{n+1}-\D{\V u}^{n+\frac{1}{2}}\right),\label{eq:dv_semi_implicit}
\end{equation}
where $\M L\equiv\grad^{2}$ denotes the discrete Laplacian operator.
This is nothing but a rewriting of (\ref{eq:v_implicit_system}) and
just as difficult to solve, so we need to make further approximations.

We begin solving (\ref{eq:dv_semi_implicit}) by writing 
\begin{equation}
\rho\V I-\frac{\D t}{2}\eta\M L+m_{e}\M S\M J=\left(\rho\V I+m_{e}\M S\M J\right)\left(\V I-\frac{\D t}{2}\nu\M L\right)+\frac{m_{e}\nu\D t}{2}\M S\M J\M L.\label{eq:Florens_trick}
\end{equation}
If we ignore the last term we obtain an approximation of order $O\left(m_{e}\D t\right)$.
We can improve the accuracy for viscous-dominated flows if we time
lag the last term, that is, evaluate it during the previous time step.
Using (\ref{eq:Florens_trick}) in (\ref{eq:dv_semi_implicit}) along
with time lagging we get the modified fluid correction equation
\begin{equation}
\left(\rho\V I+m_{e}\M S\M J\right)\left(\V I-\frac{\D t}{2}\nu\M L\right)\D{\V v}^{n+\frac{1}{2}}+\D t\,\grad\left(\D{\pi}^{n+\frac{1}{2}}\right)=\M S\left[m_{e}\left(\V u^{n}-\M J\tilde{\V v}^{n+1}-\d{\V u}^{n+\frac{1}{2}}\right)\right]=\M S\D{\V p},\label{eq:dv_modified}
\end{equation}
where we have time-lagged the last term in (\ref{eq:Florens_trick})
and denoted 
\[
\d{\V u}^{n+\frac{1}{2}}=\D{\V u}^{n+\frac{1}{2}}+\frac{\nu\D t}{2}\M J^{n-\frac{1}{2}}\M L\D{\V v}^{n-\frac{1}{2}}.
\]
Note that if one approximates $\d{\V u}^{n+\frac{1}{2}}=\V 0$ (lowering
the order of accuracy), then as $\D t\rightarrow\infty$ we get $\D{\V v}^{n+\frac{1}{2}}\rightarrow\V 0$
since both $\rho\V I+m_{e}\M S\M J$ and $\V I-\D t\nu\M L/2$ are
positive semi-definite matrices. This is consistent with physical
intuition that inertia should play a negligible role at long time
scales, and implies that the algorithm will be unconditionally stable
in the absence of advection.

Denote $\D{\tilde{\V v}}=\left(\V I-\D t\nu\M L/2\right)\D{\V v}^{n+\frac{1}{2}}$,
and note that $\grad\cdot\left(\D{\tilde{\V v}}\right)=\grad\cdot\left(\D{\V v}^{n+\frac{1}{2}}\right)=0$
with periodic BCs since the divergence and the Laplacian commute.
Therefore, we can solve (\ref{eq:dv_modified}) by first solving
\begin{equation}
\left(\rho\V I+m_{e}\M S\M J\right)\D{\tilde{\V v}}+\D t\,\grad\left(\D{\pi}^{n+\frac{1}{2}}\right)=\M S\D{\V p},\label{eq:dv_tilde_eq}
\end{equation}
subject to $\grad\cdot\left(\D{\tilde{\V v}}\right)=0$, and then
solving
\begin{equation}
\left(\V I-\frac{\D t}{2}\nu\M L\right)\D{\V v}^{n+\frac{1}{2}}=\D{\tilde{\V v}},\label{eq:dv_eq}
\end{equation}
using FFTs to diagonalize the Laplacian \cite{IBM_Viscoelasticity}.

We solve (\ref{eq:dv_tilde_eq}) using the pressure-elimination procedure
described in Appendix \ref{AppendixPeriodicBCs}, to obtain the equivalent
pressure-free equation
\[
\left(\rho\V I+m_{e}\M{\P}\M S\M J\M{\P}\right)\D{\tilde{\V v}}=\M{\P}\M S\D{\V p},
\]
where $\M{\P}$ denotes the discrete projection operator \cite{LLNS_Staggered}.
By assuming the particles are sufficiently far away from each other,
we can employ the approximation (\ref{eq:rho_eff_inv_discrete}) to
approximately solve (\ref{eq:dv_tilde_eq}) {[}compare to (\ref{eq:fluid_only_periodic})
with $\V f=\V 0${]},
\begin{equation}
\D{\tilde{\V v}}=\frac{\tilde{m}_{f}}{\rho\tilde{m}}\M{\P}\M S\D{\V p}=\frac{m_{e}\tilde{m}_{f}}{\rho\left(\tilde{m}_{f}+m_{e}\right)}\M{\P}\M S\left(\V u^{n}-\M J\tilde{\V v}^{n+1}-\d{\V u}^{n+\frac{1}{2}}\right),\label{eq:dv_tilde_sol}
\end{equation}
where we recall that $\tilde{m}_{f}=d\, m_{f}/\left(d-1\right)$.

In order to ensure strict momentum conservation even in the presence
of the approximation (\ref{eq:rho_eff_inv_discrete}), we rewrite
(\ref{eq:dv_eq}) using (\ref{eq:dv_tilde_eq}) in the form
\begin{equation}
\left(\rho\V I-\frac{\D t}{2}\eta\M L\right)\D{\V v}^{n+\frac{1}{2}}+\D t\,\grad\left(\D{\pi}^{n+\frac{1}{2}}\right)=\M S\left(\D{\V p}-m_{e}\M J\D{\tilde{\V v}}\right),\label{eq:dv_eq_modified}
\end{equation}
subject to $\grad\cdot\left(\D{\V v}^{n+\frac{1}{2}}\right)=0$. This
is consistent with (\ref{eq:sys_dv_eq}) if we set
\begin{equation}
\D t\,\left(\V{\lambda}^{n+\frac{1}{2}}+\V F^{n+\frac{1}{2}}\right)=m_{e}\left[\M J\left(\tilde{\V v}^{n+1}+\D{\tilde{\V v}}\right)+\d{\V u}^{n+\frac{1}{2}}-\V u^{n}\right],\label{eq:lambda_sol}
\end{equation}
which is to be compared to (\ref{eq:lambda_semi_implicit}). The linear
system (\ref{eq:dv_eq_modified}) can be solved using the same techniques
as used to solve (\ref{eq:unperturbed_v}).

Having determined $\D{\V v}^{n+\frac{1}{2}}$ we can update the fluid
velocity (momentum), $\V v^{n+1}=\tilde{\V v}^{n+1}+\D{\V v}^{n+\frac{1}{2}}$.
If desired, we can update the particle momentum in a conservative
manner by substituting (\ref{eq:lambda_sol}) into (\ref{eq:u_np1_eq})
to get
\begin{equation}
\V u^{n+1}=\M J\left(\tilde{\V v}^{n+1}+\D{\tilde{\V v}}\right)+\d{\V u}^{n+\frac{1}{2}}.\label{eq:u_np1_modified}
\end{equation}
From (\ref{eq:u_np1_modified}) we see that the approximations we
used lead to a violation of the no-slip constraint (\ref{eq:no_slip_split})
\[
\D{\V u^{n+1}}=\V u^{n+1}-\M J\left(\tilde{\V v}^{n+1}+\D{\V v}^{n+\frac{1}{2}}\right)-\D{\V u}^{n+\frac{1}{2}}\approx-\frac{\nu\D t}{2}\left(\M J^{n+\frac{1}{2}}\M L\D{\V v}^{n+\frac{1}{2}}-\M J^{n-\frac{1}{2}}\M L\D{\V v}^{n-\frac{1}{2}}\right),
\]
which is of $O\left(m_{e}\D t^{2}\right)$ if $\D{\V v}$ is smooth
in time, $\D{\V v}^{n+\frac{1}{2}}=\D{\V v}^{n-\frac{1}{2}}+O\left(\D t\right)$.
This means that our procedure approximates the solution of (\ref{eq:unperturbed_v},\ref{eq:sys_dv_eq},\ref{eq:u_np1_eq},\ref{eq:no_slip_split})
with slip $\D{\V u^{n+1}}=O\left(m_{e}\D t^{2}\right)$. Therefore,
in the deterministic setting, for sufficiently smooth flows, the temporal
integrator summarized in Section \ref{sub:SummaryIncompressible}
is expected to be second-order accurate.

\section{\label{AppendixCompressible}Compressible Algorithm}

In this Appendix we propose a modification of the first-order temporal
integrator developed in Ref. \cite{DirectForcing_Balboa}, following
a similar approach to the incompressible algorithm summarized in Section
\ref{sub:SummaryIncompressible}. The algorithm summarized below is
still only first-order accurate if $m_{e}\neq0$, however, it is second-order
accurate for $m_{e}=0$ unlike the algorithm used in Ref. \cite{DirectForcing_Balboa}.
Numerical results show that the modified algorithm below is a substantial
improvement over that used previously \cite{DirectForcing_Balboa}.
\begin{enumerate}
\item Estimate the position of the particle at the midpoint,
\begin{equation}
\V q^{n+\frac{1}{2}}=\V q^{n}+\frac{\D t}{2}\M J^{n}\V v^{n},\label{eq:particle_pred_1}
\end{equation}
and evaluate the external or interparticle forces $\V F^{n+\frac{1}{2}}\left(\V q^{n+\frac{1}{2}}\right)$.
\item Solve the coupled density and unperturbed momentum equations \cite{StagerredFluctHydro}
\begin{eqnarray*}
D_{t}\rho & = & -\rho\left(\grad\cdot\V v\right)\\
\rho\left(D_{t}\V v\right) & = & -c_{T}^{2}\grad\rho-\grad\cdot\M{\sigma}+\M S^{n+\frac{1}{2}}\V F^{n+\frac{1}{2}}-\left(k_{B}T\right)\sum_{i}\frac{\left(\tilde{m}_{f}\right)_{i}}{\tilde{m}_{i}}\V{\partial}\delta_{a}\left(\V q_{i}^{n+\frac{1}{2}}-\V r\right)
\end{eqnarray*}
using a step of the third-order Runge-Kutta algorithm described in
Ref. \cite{LLNS_Staggered}, to obtain the density $\rho^{n+1}$ and
the unperturbed fluid velocity $\tilde{\V v}^{n+1}$. During this
step we treat the force density $\M S^{n+\frac{1}{2}}\V F^{n+\frac{1}{2}}$
in the momentum equation as a constant external forcing. Here we have
included the thermal drift term (\ref{eq:f_th}) as the last term
on the right hand side of the momentum equation, denoting the gradient
of the kernel with $\V{\partial}\delta_{a}$; omitting this term violates
discrete fluctuation-dissipation balance and leads to small but measurable
unphysical grid artifacts in the particle dynamics.
\item If $m_{e}=0$, set $\V v^{n+1}=\tilde{\V v}^{n+1}$ and skip to step
\ref{enu:update_q-1}.
\item Calculate the momentum exchange between the particle and the fluid
during the time step \cite{DirectForcing_Balboa},
\begin{equation}
\D t\,\left(\V{\lambda}^{n+\frac{1}{2}}+\V F^{n+\frac{1}{2}}\right)=\D{\V p}=\frac{m_{e}m_{f}}{m_{e}+m_{f}}\left(\M J^{n+\frac{1}{2}}\tilde{\V v}^{n+1}-\V u^{n}\right),\label{eq:lambda_semi_implicit-1}
\end{equation}
where the mass of the dragged fluid is estimated from the local density
as 
\[
m_{f}=\D V\,\M J^{n+\frac{1}{2}}\rho^{n+1}.
\]

\item Update the particle momentum,
\begin{equation}
\V u^{n+1}=\V u^{n}+\frac{\D{\V p}}{m_{e}}=\V u^{n}+\frac{m_{f}}{m_{e}+m_{f}}\left(\M J^{n+\frac{1}{2}}\tilde{\V v}^{n+1}-\V u^{n}\right).\label{eq:update_u-1}
\end{equation}

\item Update the fluid velocity to enforce the no-slip condition $\V u^{n+1}=\M J^{n+\frac{1}{2}}\V v^{n+1}$,
\begin{equation}
\V v^{n+1}=\tilde{\V v}^{n+1}-\frac{\D V}{m_{f}}\M S^{n+\frac{1}{2}}\D{\V p}=\tilde{\V v}^{n+1}+\D V\,\M S^{n+\frac{1}{2}}\left(\V u^{n+1}-\M J^{n+\frac{1}{2}}\tilde{\V v}^{n+1}\right)\label{eq:update_v-1}
\end{equation}
Note that this conserves the total momentum since
\[
\rho^{n+1}\V v^{n+1}-\rho^{n+1}\tilde{\V v}^{n+1}=-\left(\M J^{n+\frac{1}{2}}\rho^{n+1}\right)^{-1}\int\rho^{n+1}\M S^{n+\frac{1}{2}}\D{\V p}\, d\V r=-\D{\V p}.
\]

\item \label{enu:update_q-1}Update the particle position,
\begin{equation}
\V q^{n+1}=\V q^{n}+\frac{\D t}{2}\M J^{n+\frac{1}{2}}\left(\V v^{n+1}+\V v^{n}\right).\label{eq:update_q-1}
\end{equation}

\end{enumerate}
There are some differences between this algorithm and the incompressible
algorithm summarized in Section \ref{sub:SummaryIncompressible}.
Notably, viscosity is handled explicitly in the compressible algorithm
to avoid costly linear solvers. In the incompressible algorithm the
no-slip condition is violated slightly while here the velocity correction
$\V v^{n+1}-\tilde{\V v}^{n+1}$ is designed to enforce the no-slip
condition exactly even in the presence of density variations. Another
important difference is that the compressible algorithm presented
here is not second-order accurate even for small Reynolds number because
viscous corrections of $O\left(\nu\D t\right)$ are not taken into
account when computing $\V{\lambda}^{n+\frac{1}{2}}$. Since the time
step is typically strongly limited by propagation of sound, typically
$\beta\ll1$ and there is little need for higher-order handling of
the viscous terms.

\end{appendix}


\begin{thebibliography}{100}

\bibitem{KimKarrila}
S.~Kim and S.J. Karrila.
\newblock {\em Microhydrodynamics:Principles and Selected Applications}.
\newblock Butterworth Heinemann, Boston, 1991.

\bibitem{SRD_Review}
R.~Kapral.
\newblock {Multiparticle collision dynamics: simulation of complex systems on
  mesoscales}.
\newblock {\em Adv. Chem. Phys}, 140:89, 2008.

\bibitem{LB_SoftMatter_Review}
B.~D{\"u}nweg and A.J.C. Ladd.
\newblock {Lattice Boltzmann simulations of soft matter systems}.
\newblock {\em Adv. Comp. Sim. for Soft Matter Sciences III}, pages 89--166,
  2009.

\bibitem{MigrationFractination}
C.W. Hsu and Y.L. Chen.
\newblock Migration and fractionation of deformable particles in microchannel.
\newblock {\em The Journal of chemical physics}, 133:034906, 2010.

\bibitem{Eaton2009}
J.~K. Eaton.
\newblock {Two-way coupled turbulence simulations of gas-particle flows using
  point-particle tracking}.
\newblock {\em International Journal of Multiphase Flow}, 35(9):792--800, 2009.

\bibitem{Tanaka2010}
T.~Tanaka and J.~K. Eaton.
\newblock {Sub-Kolmogorov resolution partical image velocimetry measurements of
  particle-laden forced turbulence}.
\newblock {\em Journal of Fluid Mechanics}, 643:177, 2010.

\bibitem{Rhodes_book}
M.~Rhodes.
\newblock {\em Introduction to particle technology}.
\newblock John Wiley and Sons, 2008.

\bibitem{Bian2012}
X.~Bian, S.~Litvinov, R.~Qian, M.~Ellero, and N.~A. Adams.
\newblock {Multiscale modeling of particle in suspension with smoothed
  dissipative particle dynamics}.
\newblock {\em Physics of Fluids}, 24(1):012002, 2012.

\bibitem{SDPD_Scaling}
A.~V{\'a}zquez-Quesada, M.~Ellero, and P.~Espa{\~n}ol.
\newblock {Consistent scaling of thermal fluctuations in smoothed dissipative
  particle dynamics}.
\newblock {\em J. Chem. Phys.}, 130:034901, 2009.

\bibitem{Horbach2006}
J.~Horbach and S.~Succi.
\newblock {Lattice Boltzmann versus Molecular Dynamics Simulation of Nanoscale
  Hydrodynamic Flows}.
\newblock {\em Physical Review Letters}, 96(22):1--4, 2006.

\bibitem{Melchionna2010}
Simone Melchionna.
\newblock {Incorporation of smooth spherical bodies in the Lattice-Boltzmann
  method}.
\newblock {\em J. Comp. Phys}, 230(10):3966, 2011.

\bibitem{Hu2001}
H.~H. Hu, Patankar~N. A., and M.~Y. Zhu.
\newblock {Direct numerical simulation of fluid-solid systems using the
  arbitrary Lagrangian-Eulerian technique}.
\newblock {\em J. Comp. Phys.}, 169:427, 2001.

\bibitem{Yamamoto2007}
R.~Yamamoto, K.~Kim, and Y.~Nakayama.
\newblock {Strict simulations of non-equilibrium dynamics of colloids}.
\newblock {\em Colloids and Surfaces A}, 311:42, 2007.

\bibitem{Luo2009}
X.~Luo, M.~R. Maxey, and G.~Karniadakis.
\newblock {Smoothed profile method for particulate flows: Error analysis and
  simulations}.
\newblock {\em Journal of Computational Physics}, 228:1750, 2009.

\bibitem{Maxey1983}
M.~R. Maxey and J.~J. Riley.
\newblock {Equation of motion for a small rigid sphere in a nonuniform flow}.
\newblock {\em Physics of Fluids}, 26:883, 1983.

\bibitem{Lomholt2001}
S.~Lomholt.
\newblock {\em {Numerical Investigations of Macroscopic Particle Dynamics in
  Microflows}}.
\newblock PhD thesis, Ris\o National Laboratory, Roskilde, Denmark, 2001.

\bibitem{Monchaux2012}
R.~Monchaux, M.~Bourgoin, and A.~Cartellier.
\newblock {Analyzing preferential concentration and clustering of inertial
  particles in turbulence}.
\newblock {\em International Journal of Multiphase Flow}, 40:1--18, 2012.

\bibitem{Falck2004}
E.~Falck, J.~M. Lahtinen, I.~Vattulainen, and T.~Ala-Nissila.
\newblock {Influence of hydrodynamics on many-particle diffusion in 2D
  colloidal suspensions.}
\newblock {\em The European physical journal. E, Soft matter}, 13(3):267--75,
  2004.

\bibitem{Ripoll2005}
M.~Ripoll, K.~Mussawisade, R.~Winkler, and G.~Gompper.
\newblock {Dynamic regimes of fluids simulated by multiparticle-collision
  dynamics}.
\newblock {\em Physical Review E}, 72(1):016701, July 2005.

\bibitem{Balbin2009}
M.~Balvin, E.~Sohn, T.~Iracki, G.~Drazer, and J.~Frechette.
\newblock Directional locking and the role of irreversible interactions in
  deterministic hydrodynamics separations in microfluidic devices.
\newblock {\em Phys. Rev. Lett.}, 103:078301, 2009.

\bibitem{Ermak1978}
D.~L. Ermak and J.~A. McCammon.
\newblock {Brownian dynamics with hydrodynamic interactions}.
\newblock {\em The Journal of Chemical Physics}, 69(4):1352, 1978.

\bibitem{BrownianDynamics_DNA}
R.~M. Jendrejack, J.~J. de~Pablo, and M.~D. Graham.
\newblock {Stochastic simulations of DNA in flow: Dynamics and the effects of
  hydrodynamic interactions}.
\newblock {\em J. Chem. Phys.}, 116(17):7752--7759, 2002.

\bibitem{Banchio2003}
Adolfo~J. Banchio and John~F. Brady.
\newblock {Accelerated Stokesian dynamics: Brownian motion}.
\newblock {\em The Journal of Chemical Physics}, 118(22):10323, 2003.

\bibitem{BrownianDynamics_OrderNlogN}
A.~Sierou and J.~F. Brady.
\newblock {Accelerated Stokesian Dynamics simulations}.
\newblock {\em J. Fluid Mech.}, 448:115--146, 2001.

\bibitem{BrownianDynamics_OrderN}
J.~P. Hernandez-Ortiz, J.~J. de~Pablo, and M.~D. Graham.
\newblock {Fast Computation of Many-Particle Hydrodynamic and Electrostatic
  Interactions in a Confined Geometry}.
\newblock {\em Phys. Rev. Lett.}, 98(14):140602, 2007.

\bibitem{FluctuatingHydroMD_Coveney}
G.~Giupponi, G.~De Fabritiis, and P.~V. Coveney.
\newblock {Hybrid method coupling fluctuating hydrodynamics and molecular
  dynamics for the simulation of macromolecules}.
\newblock {\em J. Chem. Phys.}, 126(15):154903, 2007.

\bibitem{SELM}
P.~J. Atzberger.
\newblock {Stochastic Eulerian-Lagrangian Methods for Fluid-Structure
  Interactions with Thermal Fluctuations}.
\newblock {\em J. Comp. Phys.}, 230:2821--2837, 2011.

\bibitem{StochasticImmersedBoundary}
P.~J. Atzberger, P.~R. Kramer, and C.~S. Peskin.
\newblock {A stochastic immersed boundary method for fluid-structure dynamics
  at microscopic length scales}.
\newblock {\em J. Comp. Phys.}, 224:1255--1292, 2007.

\bibitem{ForceCoupling_Stokes}
S.~Lomholt and M.R. Maxey.
\newblock {Force-coupling method for particulate two-phase flow: Stokes flow}.
\newblock {\em J. Comp. Phys.}, 184(2):381--405, 2003.

\bibitem{ForceCoupling_Monopole}
M.~R. Maxey and B.~K. Patel.
\newblock {Localized force representations for particles sedimenting in Stokes
  flow}.
\newblock {\em International journal of multiphase flow}, 27(9):1603--1626,
  2001.

\bibitem{ForceCoupling_Fluctuations}
Eric~E. Keaveny.
\newblock Fluctuating force-coupling method for simulations of colloidal
  suspensions.
\newblock {\em J. Comp. Phys.}, 269(0):61 -- 79, 2014.

\bibitem{IBM_PeskinReview}
C.S. Peskin.
\newblock {The immersed boundary method}.
\newblock {\em Acta Numerica}, 11:479--517, 2002.

\bibitem{Lovalenti1993}
P.~M. Lovalenti and J.~F. Brady.
\newblock {The force on a bubble, drop, or particle in arbitrary time-dependent
  motion at small Reynolds number}.
\newblock {\em Physics of Fluids}, 5(5):2104--2116, 1993.

\bibitem{Botto2012}
L.~Botto and A.~Prosperetti.
\newblock {A fully resolved numerical simulation of turbulent flow past one or
  several spherical particles}.
\newblock {\em Physics of Fluids}, 24(1):013303, 2012.

\bibitem{Uhlmann2008}
M.~Uhlmann.
\newblock Interface-resolved direct numerical simulation of vertical
  particulate channel flow in the turbulent regime.
\newblock {\em Physics of Fluids}, 20(5):053305, 2008.

\bibitem{FluctuatingHydro_FluidOnly}
N.~Sharma and N.~A. Patankar.
\newblock {Direct numerical simulation of the Brownian motion of particles by
  using fluctuating hydrodynamic equations}.
\newblock {\em J. Comput. Phys.}, 201:466--486, 2004.

\bibitem{FIMAT_Patankar}
Y.~Chen, N.~Sharma, and N.~Patankar.
\newblock {Fluctuating Immersed Material (FIMAT) dynamics for the direct
  simulation of the Brownian motion of particles}.
\newblock In {\em IUTAM Symposium on Computational Approaches to Multiphase
  Flow}, pages 119--129. Springer, 2006.

\bibitem{ImmersedFEM_Patankar}
Adrian~M Kopacz, Neelesh~A Patankar, and Wing~K Liu.
\newblock The immersed molecular finite element method.
\newblock {\em Computer Methods in Applied Mechanics and Engineering},
  233:28--39, 2012.

\bibitem{ParticleLaden_Proteus}
Zhi-Gang Feng and Efstathios~E Michaelides.
\newblock < i> proteus</i>: a direct forcing method in the simulations of
  particulate flows.
\newblock {\em Journal of Computational Physics}, 202(1):20--51, 2005.

\bibitem{Lucci2010}
F.~Lucci, A.~Ferrante, and S.~Elghobashi.
\newblock {Modulation of isotropic turbulence by particles of Taylor
  length-scale size}.
\newblock {\em Journal of Fluid Mechanics}, 650:5, 2010.

\bibitem{ParticleLaden_Heat}
Zhi-Gang Feng and Efstathios~E Michaelides.
\newblock Inclusion of heat transfer computations for particle laden flows.
\newblock {\em Physics of Fluids}, 20:040604, 2008.

\bibitem{Dejoan2011}
A.~Dejoan.
\newblock {DNS experiments on the settling of heavy particles in homogeneous
  turbulence: two-way coupling and Reynolds number effects}.
\newblock {\em Journal of Physics: Conference Series}, 333:012006, December
  2011.

\bibitem{DirectForcing_Balboa}
F.~Balboa Usabiaga, I.~Pagonabarraga, and R.~Delgado-Buscalioni.
\newblock {Inertial coupling for point particle fluctuating hydrodynamics}.
\newblock {\em J. Comp. Phys.}, 235:701--722, 2013.

\bibitem{LLNS_Staggered}
F.~Balboa Usabiaga, J.~B. Bell, R.~Delgado-Buscalioni, A.~Donev, T.~G. Fai,
  B.~E. Griffith, and C.~S. Peskin.
\newblock {Staggered Schemes for Fluctuating Hydrodynamics}.
\newblock {\em SIAM J. Multiscale Modeling and Simulation}, 10(4):1369--1408,
  2012.

\bibitem{Wang2011}
J.~Wang and J.~Dual.
\newblock {Theoretical and numerical calculations for the time-averaged
  acoustic force and torque acting on a rigid cylinder of arbitrary size in a
  low viscosity fluid.}
\newblock {\em The Journal of the Acoustical Society of America},
  129(6):3490--501, 2011.

\bibitem{SELM_Reduction}
G.~Tabak and P.J. Atzberger.
\newblock Systematic stochastic reduction of inertial fluid-structure
  interactions subject to thermal fluctuations.
\newblock {\em arXiv preprint arXiv:1211.3798}, 2015.

\bibitem{DirectForcing_Uhlmann}
M.~Uhlmann.
\newblock {An immersed boundary method with direct forcing for the simulation
  of particulate flows}.
\newblock {\em J. Comp. Phys.}, 209(2):448--476, 2005.

\bibitem{bellColellaGlaz:1989}
J.~B. Bell, P.~Colella, and H.~M. Glaz.
\newblock A second order projection method for the incompressible
  {N}avier-{S}tokes equations.
\newblock {\em J. Comp. Phys.}, 85(2):257--283, 1989.

\bibitem{ReactiveBlobs}
A.~Pal~Singh Bhalla, B.~E. Griffith, N.~A. Patankar, and A.~Donev.
\newblock {A Minimally-Resolved Immersed Boundary Model for Reaction-Diffusion
  Problems}.
\newblock {\em J. Chem. Phys.}, 139(21):214112, 2013.

\bibitem{CompressibleBlobs}
F.~Balboa~Usabiaga and R.~Delgado-Buscalioni.
\newblock {A minimal model for acoustic forces on Brownian particles}.
\newblock {\em Phys. Rev. E}, 88:063304, 2013.

\bibitem{FluctHydroNonEq_Book}
J.~M.~O.~De Zarate and J.~V. Sengers.
\newblock {\em {Hydrodynamic fluctuations in fluids and fluid mixtures}}.
\newblock Elsevier Science Ltd, 2006.

\bibitem{DFDB}
S.~Delong, B.~E. Griffith, E.~Vanden-Eijnden, and A.~Donev.
\newblock {Temporal Integrators for Fluctuating Hydrodynamics}.
\newblock {\em Phys. Rev. E}, 87(3):033302, 2013.

\bibitem{DiscreteLLNS_Espanol}
P.~Espa{\~n}ol, J.G. Anero, and I.~Z{\'u}{\~n}iga.
\newblock {Microscopic derivation of discrete hydrodynamics}.
\newblock {\em J. Chem. Phys.}, 131:244117, 2009.

\bibitem{alphaNS_Titi}
C.~Foias, D.~D Holm, and E.~S Titi.
\newblock {The Navier--Stokes-alpha model of fluid turbulence}.
\newblock {\em Physica D: Nonlinear Phenomena}, 152:505--519, 2001.

\bibitem{alphaNS_BCs}
M.~van Reeuwijk, H.~J.~J. Jonker, and K.~HanjalicÌ.
\newblock {Incompressibility of the Leray-$\alpha$ model for wall-bounded
  flows}.
\newblock {\em Physics of Fluids}, 18(1):018103, January 2006.

\bibitem{StokesLaw}
A.~Donev, T.~G. Fai, and E.~Vanden-Eijnden.
\newblock {Reversible Diffusion by Thermal Fluctuations}.
\newblock Arxiv preprint 1306.3158, 2013.

\bibitem{LowMachExplicit}
A.~Donev, A.~J. Nonaka, Y.~Sun, T.~G. Fai, A.~L. Garcia, and J.~B. Bell.
\newblock {Low Mach Number Fluctuating Hydrodynamics of Diffusively Mixing
  Fluids}.
\newblock {\em Communications in Applied Mathematics and Computational
  Science}, 9(1):47--105, 2014.

\bibitem{DaPratoBook}
G.~Da Prato.
\newblock {\em {Kolmogorov equations for stochastic PDEs}}.
\newblock Birkhauser, 2004.

\bibitem{Ladd:93}
A.J.C. Ladd.
\newblock Short-time motion of colloidal particles: Numerical simulation via a
  fluctuating lattice-{B}oltzmann equation.
\newblock {\em Phys. Rev. Lett.}, 70(9):1339--1342, Mar 1993.

\bibitem{VACF_LBM_Chineese}
D.~Nie and J.~Lin.
\newblock {A fluctuating lattice-Boltzmann model for direct numerical
  simulation of particle Brownian motion}.
\newblock {\em Particuology}, 7(6):501--506, 2009.

\bibitem{FaxenRelations}
J.M. Rallison.
\newblock Note on the faxen relations for a particle in stokes flow.
\newblock {\em J. Fluid Mech}, 88(part 3):529--533, 1978.

\bibitem{IBM_Implicit_Peskin}
Y.~Mori and C.S. Peskin.
\newblock {Implicit second-order immersed boundary methods with boundary mass}.
\newblock {\em Computer Methods in Applied Mechanics and Engineering},
  197(25-28):2049--2067, 2008.

\bibitem{InertialIBM_nonuniform}
L.~Zhu and C.S. Peskin.
\newblock Simulation of a flapping flexible filament in a flowing soap film by
  the immersed boundary method.
\newblock {\em J. Comp. Phys.}, 179(2):452--468, 2002.

\bibitem{InertialIBM_Penalty}
Y.~Kim and C.S. Peskin.
\newblock {Penalty immersed boundary method for an elastic boundary with mass}.
\newblock {\em Physics of Fluids}, 19:053103, 2007.

\bibitem{InertialIBM_Density}
Y.~Kim and C.S. Peskin.
\newblock {Numerical study of incompressible fluid dynamics with nonuniform
  density by the immersed boundary method}.
\newblock {\em Physics of Fluids}, 20:062101, 2008.

\bibitem{balboa2011}
R.~Delgado-Buscalioni and F.~B. Usabiaga.
\newblock Particle hydrodynamics using hybrid models: from molecular to
  colloidal fluids.
\newblock In E.~O{\~n}ate and R.~Owen, editors, {\em Particle-Based Methods II:
  Fundamentals and Applications}. International Center for Numerical Methods in
  Engineering, 2011.

\bibitem{Chorin68}
A.~J. Chorin.
\newblock {Numerical Solution of the Navier-Stokes Equations}.
\newblock {\em J. Math. Comp.}, 22:745--762, 1968.

\bibitem{BrownianCompressibility_Zwanzig}
R.~Zwanzig and M.~Bixon.
\newblock {Compressibility effects in the hydrodynamic theory of Brownian
  motion}.
\newblock {\em J. Fluid Mech.}, 69:21--25, 1975.

\bibitem{GrabertBook}
H.~Grabert.
\newblock {\em Projection operator techniques in nonequilibrium statistical
  mechanics}.
\newblock Springer-Verlag, 1982.

\bibitem{CoarseGraining_Pep}
P.~Espa{\~n}ol.
\newblock Statistical mechanics of coarse-graining.
\newblock {\em Novel Methods in Soft Matter Simulations}, pages 69--115, 2004.

\bibitem{HamiltonianFluid}
P.J. Morrison.
\newblock Hamiltonian description of the ideal fluid.
\newblock {\em Rev. Mod. Phys.}, 70(2):467, 1998.

\bibitem{OttingerBook}
H.~C. {\"O}ttinger.
\newblock {\em Beyond equilibrium thermodynamics}.
\newblock Wiley Online Library, 2005.

\bibitem{VACF_Suspension}
B.~Cichocki and B.~U. Felderhof.
\newblock {Velocity autocorrelation function of interacting Brownian
  particles}.
\newblock {\em Phys. Rev. E}, 51:5549--5555, 1995.

\bibitem{IBAMR}
B.E. Griffith, R.D. Hornung, D.M. McQueen, and C.S. Peskin.
\newblock {An adaptive, formally second order accurate version of the immersed
  boundary method}.
\newblock {\em J. Comput. Phys.}, 223(1):10--49, 2007.
\newblock Software available at \url{http://ibamr.googlecode.com}.

\bibitem{NonProjection_Griffith}
B.E. Griffith.
\newblock {An accurate and efficient method for the incompressible
  Navier-Stokes equations using the projection method as a preconditioner}.
\newblock {\em J. Comp. Phys.}, 228(20):7565--7595, 2009.

\bibitem{DiscreteDiffusion_Espanol}
P.~Espa{\~n}ol and I.~Z{\'u}{\~n}iga.
\newblock On the definition of discrete hydrodynamic variables.
\newblock {\em J. Chem. Phys}, 131:164106, 2009.

\bibitem{IBM_Sphere}
T.T. Bringley and C.S. Peskin.
\newblock {Validation of a simple method for representing spheres and slender
  bodies in an immersed boundary method for Stokes flow on an unbounded
  domain}.
\newblock {\em J. Comp. Phys.}, 227(11):5397--5425, 2008.

\bibitem{IBMKernels_Mori}
Yang Liu and Yoichiro Mori.
\newblock Properties of discrete delta functions and local convergence of the
  immersed boundary method.
\newblock {\em SIAM Journal on Numerical Analysis}, 50(6):2986--3015, 2012.

\bibitem{IBMDelta_Boundary}
B.E. Griffith, X.~Luo, D.M. McQueen, and C.S. Peskin.
\newblock Simulating the fluid dynamics of natural and prosthetic heart valves
  using the immersed boundary method.
\newblock {\em International Journal of Applied Mechanics}, 1(01):137--177,
  2009.

\bibitem{Chorin:1968}
A.J. Chorin.
\newblock Numerical simulation of the {N}avier-{S}tokes equations.
\newblock {\em Mathematics of Computation}, 22(104):745--762, 1968.

\bibitem{IBM_Viscoelasticity}
D.~Devendran and C.~S. Peskin.
\newblock An immersed boundary energy-based method for incompressible
  viscoelasticity.
\newblock {\em J. Comp. Phys.}, 231(14):4613--4642, 2012.

\bibitem{VACF_Ladd}
A.J.C. Ladd.
\newblock {Numerical simulations of particulate suspensions via a discretized
  Boltzmann equation. II. Numerical results}.
\newblock {\em Journal of Fluid Mechanics}, 271(1):311--339, 1994.

\bibitem{DirectForcing_LBM}
S.~Melchionna.
\newblock {Incorporation of smooth spherical bodies in the Lattice Boltzmann
  method}.
\newblock {\em J. Comp. Phys.}, 230:3966--3976, 2011.

\bibitem{SmoothedInterface_Compressible}
R.~Tatsumi and R.~Yamamoto.
\newblock Direct numerical simulation of dispersed particles in a compressible
  fluid.
\newblock {\em Phys. Rev. E}, 85:066704, 2012.

\bibitem{VACF_IncompressibleSmoothed}
T.~Iwashita, Y.~Nakayama, and R.~Yamamoto.
\newblock {Velocity Autocorrelation Function of Fluctuating Particles in
  Incompressible Fluids: Toward Direct Numerical Simulation of Particle
  Dispersions}.
\newblock {\em Progress of Theoretical Physics Supplement}, 178:86--91, 2009.

\bibitem{BrownianParticle_SIBM}
P.~J. Atzberger.
\newblock {Velocity correlations of a thermally fluctuating Brownian particle:
  A novel model of the hydrodynamic coupling}.
\newblock {\em Physics Letters A}, 351(4-5):225--230, 2006.

\bibitem{VACF_Langevin}
E.~J. Hinch.
\newblock {Application of the Langevin equation to fluid suspensions}.
\newblock {\em J. Fluid Mech.}, 72(03):499--511, 1975.

\bibitem{Landau:Fluid}
L.D. Landau and E.M. Lifshitz.
\newblock {\em Fluid Mechanics}, volume~6 of {\em Course of Theoretical
  Physics}.
\newblock Pergamon Press, Oxford, England, 1959.

\bibitem{VACF_Alder}
B.~J. Alder and T.~E. Wainwright.
\newblock Decay of the velocity autocorrelation function.
\newblock {\em Phys. Rev. A}, 1(1):18--21, 1970.

\bibitem{ModeModeCoupling}
Y.~{Pomeau} and P.~{R{\'e}sibois}.
\newblock {Time dependent correlation functions and mode-mode coupling
  theories}.
\newblock {\em Phys. Rep.}, 19:63--139, June 1975.

\bibitem{RotnePrager_Periodic}
C.~W.~J. Beenakker.
\newblock {Ewald sum of the Rotne-Prager tensor}.
\newblock {\em J. Chem. Phys.}, 85:1581, 1986.

\bibitem{BD_LB_Ladd}
Rahul Kekre, Jason~E. Butler, and Anthony J.~C. Ladd.
\newblock {Comparison of lattice-Boltzmann and Brownian-dynamics simulations of
  polymer migration in confined flows}.
\newblock {\em Phys. Rev. E}, 82:011802, 2010.

\bibitem{Clift1978}
R.~Clift, J.~R. Grace, and M.~E. Weber.
\newblock {\em {Bubbles, drops and Particles}}.
\newblock Academic Press, 1978.

\bibitem{Ormieres1999}
D.~Ormi\`{e}res and M.~Provansal.
\newblock {Transition to Turbulence in the Wake of a Sphere}.
\newblock {\em Physical Review Letters}, 83(1):5--8, 1999.

\bibitem{Johnson1999a}
T.~A. Johnson and V.~C. Patel.
\newblock {Flow past a sphere up to a Reynolds number of 300}.
\newblock {\em Journal of Fluid Mechanics}, 378(-1):19--70, 1999.

\bibitem{IBM_Generalized}
B.E. Griffith and S.~Lim.
\newblock {Simulating an elastic ring with bend and twist by an adaptive
  generalized immersed boundary method}.
\newblock {\em Commun. Comput. Phys.}, 12:433--461, 2012.

\bibitem{IBM_TwistBend}
S.~Lim, A.~Ferent, X.~S. Wang, and C.~S. Peskin.
\newblock Dynamics of a closed rod with twist and bend in fluid.
\newblock {\em SIAM~J~Sci~Comput}, 31(1):273--302, 2008.

\bibitem{RigidIBAMR}
A.~P.~S. Bhalla, R.~Bale, B.~E. Griffith, and N.~A. Patankar.
\newblock {A unified mathematical framework and an adaptive numerical method
  for fluid-structure interaction with rigid, deforming, and elastic bodies}.
\newblock {\em Journal of Computational Physics}, 250:446--476, 2013.

\bibitem{StokesEinstein_SlipBC}
J.~T. Hynes, Raymond K., and Michael W.
\newblock {Molecular theory of translational diffusion: Microscopic
  generalization of the normal velocity boundary condition}.
\newblock {\em J. Chem. Phys.}, 70(3):1456, February 1979.

\bibitem{MultiscaleIntegrators}
S.~Delong, Y.~Sun, B.~E. Griffith, E.~Vanden-Eijnden, and A.~Donev.
\newblock {Multiscale temporal integrators for fluctuating hydrodynamics}.
\newblock {\em Phys. Rev. E}, 90:063312, 2014.
\newblock Software available at
  \url{https://github.com/stochasticHydroTools/MixingIBAMR}.

\bibitem{StagerredFluctHydro}
N.~K. Voulgarakis and J.-W. Chu.
\newblock {Bridging fluctuating hydrodynamics and molecular dynamics
  simulations of fluids}.
\newblock {\em J. Chem. Phys.}, 130(13):134111, 2009.

\end{thebibliography}

\end{document}